\documentclass[11pt,a4paper]{llncs}

\usepackage[usenames,dvipsnames]{color}
\usepackage[francais,english]{babel}
\usepackage[T1]{fontenc}
\usepackage[utf8]{inputenc}
\usepackage{amsmath}
\usepackage{epsfig}
\usepackage{wrapfig}
\usepackage{paralist}
\usepackage{graphics}
\usepackage{stmaryrd}
\usepackage{txfonts}
\usepackage{framed}
\usepackage{makecell}
\usepackage{url}
\usepackage[colorlinks]{hyperref}
\usepackage{proof}

\usepackage{algorithm}
\usepackage{algorithmicx}
\usepackage[noend]{algpseudocode}

\usepackage[draft]{commenting}

\newif\ifLongVersion\LongVersiontrue

\ifLongVersion\else
\newtheorem{remark}{Remark}
\fi

\newcommand{\bigO}{\mathcal{O}}

\newcommand{\pspace}{\textsf{PSPACE}}
\newcommand{\ptime}{\textsf{P}}

\newcommand{\bigospace}[1]{\mathsf{SPACE}({#1})}
\newcommand{\bigonspace}[1]{\mathsf{NSPACE}({#1})}

\newcommand{\set}[1]{\left\{ #1 \right\}}
\newcommand{\tuple}[1]{\left\langle #1 \right\rangle}
\renewcommand{\vec}[1]{\mathbf #1}

\newcommand{\setof}[2]{\left\{#1\,\middle|\:#2\right\}}

\newcommand{\defequal}{\stackrel{\scriptscriptstyle{\mathsf{def}}}{=}}

\newcommand{\len}[1]{{|{#1}|}}
\newcommand{\card}[1]{{||{#1}||}}

\newcommand{\nat}{{\bf \mathbb{N}}}
\newcommand{\zed}{{\bf \mathbb{Z}}}

\renewcommand{\paragraph}[1]{\noindent{\bf #1}}

\renewcommand{\proof}[1]{\ifLongVersion \noindent\emph{Proof}: {#1} \else\fi}
\newcommand{\prop}[2]{\ifLongVersion \begin{proposition}\label{#1} {#2} \end{proposition} \else\fi}

\newcommand{\teq}{\approx}

\newcommand{\I}{\mathcal{I}}

\newcommand{\emp}{\mathsf{emp}}
\newcommand{\wand}{
 \mathrel{\mbox{$\hspace*{-0.03em}\mathord{-}\hspace*{-0.66em}
 \mathord{-}\hspace*{-0.36em}\mathord{*}$\hspace*{-0.005em}}}} % {\multimap}
\newcommand{\seplog}{\mathsf{SL}}
\newcommand{\tinyseplog}{\mathsf{\scriptscriptstyle{sl}}}
\newcommand{\seplogk}[1]{\seplog^{\!\scriptstyle{#1}}}

\newcommand{\finbsrsl}[1]{\bsr^\fincard(\seplog^{#1})}
\newcommand{\infbsrsl}[1]{\bsr^\infcard(\seplog^{#1})}

\newcommand{\infaxiom}[1]{#1_{\infty}} 

\newcommand{\fol}{\mathsf{FO}}
\newcommand{\bsr}{\mathsf{BSR}}

\newcommand{\fv}[1]{\mathsf{var}({#1})}

\newcommand{\dom}{\mathrm{dom}}

\newcommand{\size}[1]{\mathsf{size}(#1)}
\newcommand{\vars}{\mathsf{Var}}

\renewcommand{\int}{\mathsf{\scriptscriptstyle{i}}}

\newcommand{\foltrans}[1]{\tau({#1})}

\newcommand{\pfunc}{\mathfrak{p}}
\newcommand{\aconst}{\mathfrak{a}}
\newcommand{\bconst}{\mathfrak{b}}
\newcommand{\cconst}{\mathfrak{c}}

\newcommand{\uconst}{\mathfrak{u}}
\newcommand{\vconst}{\mathfrak{v}}
\newcommand{\wconst}{\mathfrak{w}}

\newcommand{\maxn}[1]{\mathcal{N}({#1})}

\newcommand{\axioms}[1]{\mathcal{A}({#1})}

\newcommand{\minconj}[2]{\mathrm{minh}({#1},{#2})}
\newcommand{\maxconj}[2]{\mathrm{maxh}({#1},{#2})}

\ifLongVersion
\newcommand{\U}{\mathfrak{U}}
\else
\renewcommand{\U}{\mathfrak{U}}
\fi

\newcommand{\D}{\mathfrak{D}}

\newcommand{\afunc}{\mathfrak{i}}
\newcommand{\astruct}{\mathcal{S}}
\newcommand{\astore}{\mathfrak{s}}
\newcommand{\aheap}{\mathfrak{h}}
\renewcommand{\iff}{\Leftrightarrow}
\newcommand{\pto}{\hookrightarrow}
\newcommand{\alloc}{\mathsf{alloc}}
\newcommand{\nalloc}{\mathsf{nalloc}}
\newcommand{\allocno}[1]{\#_a({#1})}
\newcommand{\nallocno}[2]{\#_{n}({#1},{#2})}
\newcommand{\septraction}{\multimap}
\newcommand{\avar}[1]{\mathsf{av}({#1})}
\newcommand{\nvar}[1]{\mathsf{nv}({#1})}
\newcommand{\footprint}[2]{\mathsf{fp}_{#2}({#1})}
\newcommand{\afootprint}[1]{\mathsf{fp}_a({#1})}
\newcommand{\negpto}[2]{\mathsf{npto}({#1},{#2})}
\newcommand{\elim}{\mathsf{elim}}

\newcommand{\finelim}{\mathsf{elim}^{\scriptscriptstyle{\fincard}}}
\newcommand{\infelim}{\mathsf{elim}^{\scriptscriptstyle{\infcard}}}
\newcommand{\genelim}{\mathsf{elim}^{\scriptscriptstyle{\gencard}}}

\newcommand{\finmt}[1]{\mu^{\scriptscriptstyle{\fincard}}({#1})}
\newcommand{\infmt}[1]{\mu^{\scriptscriptstyle{\infcard}}({#1})}
\newcommand{\genmt}[1]{\mu^{\scriptscriptstyle{\gencard}}({#1})}

\newcommand{\gencard}{\dagger}
\newcommand{\fincard}{\mathit{fin}}
\newcommand{\infcard}{\mathit{inf}}

\newcommand{\cclose}[1]{\mathsf{cc}({#1})}
\newcommand{\pclose}[1]{\mathsf{pc}({#1})}
\newcommand{\uclose}[1]{\mathsf{dc}({#1})}
\newcommand{\completion}[2]{({#1})^{\scriptscriptstyle{#2}}}
\newcommand{\eqs}[1]{\mathsf{E}({#1})}
\newcommand{\allocs}[1]{\mathsf{A}({#1})}
\newcommand{\conj}[1]{\left[{#1}\right]}
\newcommand{\dnf}[1]{\left({#1}\right)^{\scriptscriptstyle{\mathsf{dnf}}}}
\newcommand{\minterm}[1]{\left[{#1}\right]^{\scriptscriptstyle{\mathsf{mt}}}}
\newcommand{\minheap}{\mathrm{min}}
\newcommand{\bound}[1]{\mathcal{M}({#1})}
\newcommand{\bounded}{$\mathcal{M}$-bounded}
\newcommand{\datano}[2]{\delta_{#1}({#2})}
\newcommand{\maxheap}{\mathrm{max}}

\newcommand{\Bool}{\mathsf{Bool}}

\newcommand{\allvect}[1]{\mathsf{vect}^k(#1)}
\newcommand{\controlledk}[1]{{#1}-controlled}
\newcommand{\controlled}{$\maxi$-controlled}
\newcommand{\ncontrolled}{non-$\maxi$-controlled}
\newcommand{\maxi}{\alpha}
\newcommand{\Nmaxi}{\beta}
\newcommand{\kconst}{\overline{\mathsf{x}}}

\newcommand{\Tcont}[1]{\gamma(#1)}
\newcommand{\contAx}{{\mathcal C}(\maxi)}
\newcommand{\alocation}{\ell}
\newcommand{\asetbis}{Y}
\newcommand{\aset}{X}
\newcommand{\interval}[2]{[#1,#2]}

\newcommand{\Tnocont}[1]{\delta(#1)}
\newcommand{\isdef}{\defequal}
\newcommand{\sequence}[2]{(#1, \ldots, #2)}

\newcommand{\equivfin}{\equiv^{\fincard}}
\newcommand{\equivinf}{\equiv^{\infcard}}

\declareauthor{ri}{Radu}{blue}
\declareauthor{np}{Nicolas}{red}
\declareauthor{me}{Mnacho}{brown!70!black}

\begin{document}
%%%%%%%%%%%%%%%%%%%%%%%%%%%%%%%%%%%%%%%%%%%%%%%%%%%%%%%%%%%%%%%%%%%%%%%%%%%%%%%

\title{On the Expressive Completeness of Bernays-Sch\"onfinkel-Ramsey Separation Logic}

\author{Mnacho Echenim\inst{1}, Radu Iosif\inst{2} and Nicolas Peltier\inst{1}}

 \institute{Univ. Grenoble Alpes, CNRS, LIG, F-38000 Grenoble France
   \and Univ. Grenoble Alpes, CNRS, VERIMAG, F-38000 Grenoble France}

\maketitle

\begin{abstract}
This paper investigates the satisfiability problem for Separation
Logic, with unrestricted nesting of separating conjunctions and
implications, for prenex formulae with quantifier prefix in the
language $\exists^*\forall^*$, in the cases where the universe of
possible locations is either countably infinite or finite. In analogy with
first-order logic with uninterpreted predicates and equality, we call
this fragment Bernays-Sch\"onfinkel-Ramsey Separation Logic
[$\bsr(\seplogk{k})$]. We %found
show  that, unlike in first-order logic, the
(in)finite satisfiability problem is undecidable for
$\bsr(\seplogk{k})$ and we define two non-trivial subsets thereof,
that are decidable for finite and infinite satisfiability,
respectively, by controlling the occurrences of universally quantified
variables within the scope of separating implications, as well as the
polarity of the occurrences of the latter. The decidability results
are obtained by a controlled elimination of separating connectives,
described as \begin{inparaenum}[(i)]
\item an effective translation of a prenex form Separation Logic
  formula into a combination of a small number of \emph{test
    formulae}, using only first-order connectives, followed by
\item a translation of the latter into an equisatisfiable first-order
  formula.
\end{inparaenum}

\end{abstract}

\section{Introduction}

Separation Logic \cite{IshtiaqOHearn01,Reynolds02} is a logical
framework used in program verification to describe properties of the
dynamically allocated memory, such as topologies of data structures
(lists, trees), (un)reachability between pointers, etc. The quest for
automated push-button program verification methods motivates the need
for understanding the decidability, complexity and expressive power of
various dialects thereof, that are used as assertion languages in
Hoare-style proofs \cite{IshtiaqOHearn01}, or logic-based abstract
domains in static analysis \cite{Infer}.

In a nutshell, given an integer $k\geq1$, the logic $\seplogk{k}$ is
obtained from the first-order theory of a finite functional relation
of arity $k+1$, called a \emph{heap}\footnote{Intuitively, $k$ is the
  number of record fields in each memory cell.}, by adding two
non-classical connectives:~\begin{inparaenum}[(i)]
\item the \emph{separating conjunction} $\phi_1 * \phi_2$, that
  asserts a split of the heap into disjoint heaps satisfying $\phi_1$
  and $\phi_2$ respectively, and
\item the \emph{separating implication} or
\emph{magic wand} $\phi_1 \wand \phi_2$, stating that each extension
of the heap by a heap satisfying $\phi_1$ must satisfy
$\phi_2$. 
\end{inparaenum}
The separating connectives $*$ and $\wand$ allow concise definitions
of program semantics, via weakest precondition calculi
\cite{IshtiaqOHearn01} and easy-to-write specifications of recursive
linked data structures (e.g.\ singly- and doubly-linked lists, trees
with linked leaves and parent pointers, etc.), when higher-order
inductive definitions are added \cite{Reynolds02}.

A typical problem in verification, occurring as a subgoal in a
Hoare-style proof of a program or in an inductive proof of inclusion
between least fixed point models (sets of heaps) of higher-order
predicates, is deciding the validity of entailments between
existentially quantified formulae in the base assertion language. This
problem is reduced to the (un)satisfiability of an $\seplogk{k}$
formula with quantifier prefix in the language
$\exists^*\forall^*$. In analogy with first-order logic with equality
and uninterpreted predicates \cite{Ramsey87}, we call this fragment
Bernays-Sch\"onfinkel-Ramsey Separation Logic [$\bsr(\seplogk{k})$].

Unlike the Bernays-Sch\"onfinkel-Ramsey fragment of first-order logic,
$\bsr(\seplogk{k})$ is difficult to reason about, due to the
unrestricted use of separating connectives. A way to circumvent this
problem is to define a small set of patterns, called \emph{test
  formulae} in the literature
\cite{PhD-lozes,BrocheninDemriLozes11,DemriDeters14,DemriGalmicheWendlingMery14},
that are parametric in their arguments and some integer constants, and
prove that every formula in the fragment is equivalent to a classical
combination of instances of those patterns, bound only with
first-order connectives.

These expressive completeness results are, in some sense, similar to
the elimination of existential quantifiers in some interpreted
theories of first-order logic, such as Presburger arithmetic. In fact,
the existential quantifiers are not completely eliminated, but rather
confined to a small set of modulo constraints, in which they occur in
a controlled fashion. Similarly, in $\seplogk{k}$, it is possible to
confine the separating conjunction $*$ and implication $\wand$ to a
small set of test formulae and convert each $\seplogk{k}$ formula from
a certain fragment into an equivalent boolean combination of test
formulae. As with Presburger arithmetic, this is an argument for
showing decidability of the logical fragment under consideration.

\paragraph{Our contributions} The main contributions of this paper are:
\begin{compactenum}
\item We show that the finite and infinite satisfiability problems are
  undecidable for the logic $\bsr(\seplogk{k})$, interpreted over
  heaps with $k\geq2$ record fields. The main reason for undecidability
  lies in the presence of universally quantified variables within the
  scope of a separating implication, that occurs, moreover, under an
  even number of negations.
\item By disallowing universally quantified variables in the scope of
  positive occurrences of separating implications, and even stronger,
  disallowing positive occurrences thereof, we define two non-trivial
  fragments $\infbsrsl{k}$ and $\finbsrsl{k}$ of $\bsr(\seplogk{k})$,
  for which the infinite and finite satisfiability problems
  are \pspace-complete, respectively. These results establish neat
  decidability frontiers within $\bsr(\seplogk{k})$.
\end{compactenum}

In contrast with the majority of the literature on Separation Logic,
here the universe of \emph{available} memory locations (besides the
ones occurring in the heap, which is finite) is not automatically
assumed to be infinite. In fact, we consider both cases in which the
universe is countably infinite and finite. In particular, the finite
universe hypothesis is useful when dealing with bounded memory issues,
for instance checking that the execution of the program satisfies its
postcondition, provided that there are enough many available memory
cells.

Having different interpretations of the universe is also motivated by
a recent integration of $\seplogk{k}$ within a DPLL($T$)-based SMT solver
\cite{ReynoldsIosifKingSerban16,Vmcai17}, in which the $\seplog$
theory is parameterized by the theory of locations, just like the
theories of arrays and sets are parameterized by theories of values.

Surprisingly, when considering a finite universe, the separating
connectives allow to define bounds also on the cardinality of the
universe and on the number of free locations (not in the heap),
besides specifying the shape and cardinality of the heap. As a
result, the conditions needed for decidability within
$\bsr(\seplogk{k})$ turn out to be stronger for finite universes than
for infinite ones. The argument for decidability relies
on \begin{inparaenum}[(i)]
\item the definition of a restricted set of test formulae capturing
  all properties of heaps, that can be expressed in quantifier-free
  $\seplogk{k}$, together with
\item an equivalence-preserving syntactic translation of a prenex form
  $\seplogk{k}$ formula into a boolean combination of test formulae,
  with the same quantifier prefix. The latter formula is translated
  into first-order logic and decidability is established by tracking
  those formulae of $\bsr(\seplogk{k})$ that translate into the
  classical Bernays-Sch\"onfinkel-Ramsey fragment of first-order logic
  \cite{Ramsey87}.
\end{inparaenum}

\ifLongVersion\else
For space reasons, lemmas and proofs not included in the paper are
given in Appendix \ref{app:proofs}.
\fi

\paragraph{Related Work.}
Expressive completeness results exist for quantifier-free
$\seplogk{1}$ \cite{PhD-lozes,BrocheninDemriLozes11} and for
$\seplogk{1}$ with one and two quantified variables
\cite{DemriGalmicheWendlingMery14,DemriDeters14}. There, the existence
of equivalent boolean combinations of test formulae is showed
implicitly, using a finite enumeration of equivalence classes of
models, instead of an effective transformation. Instead, here we
present an explicit equivalence-preserving transformation of
quantifier-free $\seplogk{k}$ formulae over heaps with $k\geq2$ record
fields into boolean combinations of test formulae, and translate the
latter into first-order logic.

Another translation of quantifier-free $\seplogk{k}$ into first-order
logic with equality has been described in
\cite{CalcagnoGardnerHague05}. There, the small model property of
quantifier-free $\seplogk{k}$ \cite{CalcagnoYangOHearn01} is used to
bound the number of first-order variables to be considered and the
separating connectives are interpreted as first-order quantifiers. The
result is an equisatisfiable first-order formula whose satisfiability
can be checked in \pspace. This translation scheme cannot be, however,
directly applied to $\bsr(\seplogk{k})$, which does not have a small
model property, and is, moreover, undecidable.

Existing decidability and complexity results for various $\seplogk{k}$
fragments
\cite{CalcagnoYangOHearn01,BrocheninDemriLozes11,DemriGalmicheWendlingMery14,DemriDeters14}
always assume the universe of heap locations to be countably
infinite. In this paper we consider, in addition, the case where the
universe is finite. Theory-parameterized versions of
$\bsr(\seplogk{k})$ have been shown to be undecidable, e.g.\ when
integer linear arithmetic is used to reason about locations, and
wrongly claimed to be \pspace-complete for countably infinite and
finite unbounded location sorts, with no relation other than equality
\cite{Vmcai17}. Here we correct the wrong claim of \cite{Vmcai17} and
draw a precise chart of decidability for both infinite and finite
satisfiability of $\bsr(\seplogk{k})$.

\section{Preliminaries}

We denote by $\zed$ the set of integers and by $\nat$ the set of
positive integers including zero. We define $\zed_{\infty} = \zed \cup
\set{\infty}$ and $\nat_{\infty} = \nat \cup \set{\infty}$, where for
each $n \in \zed$ we have $n + \infty = \infty$ and $n<\infty$. For a
countable set $S$ we denote by $\card{S} \in \nat_\infty$ the
cardinality of $S$. A decision problem is in
$\mathsf{(N)}\bigospace{n}$ if it can be decided by a
(nondeterministic) Turing machine in space $\bigO(n)$ and in $\pspace$
if it is in $\bigospace{n^c}$ for some integer $c \geq 1$, independent
of the input.

Let $\vars$ be a countable set of variables, denoted as $x,y,z$ and
$U$ be a sort. A \emph{function symbol} $f$ has $\#(f) \geq 0$
arguments of sort $U$ and a sort $\sigma(f)$, which is either the
boolean sort $\Bool$ or $U$. If $\#(f)=0$, we call $f$ a
\emph{constant}. We use $\bot$ and $\top$ for the boolean constants
false and true, respectively. First-order ($\fol$) terms $t$ and
formulae $\varphi$ are defined by the following grammar:
\[\begin{array}{rcl}
t & := & x \mid f(t_1,\ldots,t_{\#(f)}) \\ \varphi & := & \bot \mid
\top \mid \varphi_1 \wedge \varphi_2 \mid \neg \varphi_1 \mid \exists
x ~.~ \varphi_1 \mid t_1 \teq t_2 \mid p(t_1,\ldots,t_{\#(p)})
\end{array}\]
where $x\in \vars$, $f$ and $p$ are function symbols, $\sigma(f)=U$
and $\sigma(p)=\Bool$. We write $\varphi_1 \vee \varphi_2$ for
$\neg(\neg\varphi_1 \wedge \neg\varphi_2)$, $\varphi_1 \rightarrow
\varphi_2$ for $\neg\varphi_1 \vee \varphi_2$, $\varphi_1
\leftrightarrow \varphi_2$ for $\varphi_1 \rightarrow \varphi_2 \wedge
\varphi_2 \rightarrow \varphi_1$ and $\forall x ~.~ \varphi$ for
$\neg\exists x ~.~ \neg\varphi$.

The size of a formula $\varphi$, denoted as $\size{\varphi}$, is the
number of symbols needed to write it down. Let $\fv{\varphi}$ be the
set of variables that occur free in $\varphi$, i.e.\ not in the scope
of a quantifier. A \emph{sentence} $\varphi$ is a formula where
$\fv{\varphi} = \emptyset$. Given formulae $\varphi$, $\phi$ and
$\psi$, we write $\varphi[\phi]$ when $\phi$ is a subformula of
$\varphi$ and denote by $\varphi[\psi/\phi]$ the formula obtained by
substituting $\psi$ for $\phi$ in $\varphi$.

First-order formulae are interpreted over $\fol$-structures (called
structures, when no confusion arises) $\astruct =
(\U,\astore,\afunc)$, where $\U$ is a countable set, called the
\emph{universe}, the elements of which are called \emph{locations},
$\astore : \vars \rightharpoonup \U$ is a mapping of variables to
locations, called a \emph{store} and $\afunc$ interprets each function
symbol $f$ by a function $f^\afunc : \U^{\#(f)} \rightarrow \U$, if
$\sigma(f) = U$ and $f^\afunc : \U^{\#(f)} \rightarrow
\{\bot^\afunc,\top^\afunc\}$ if $\sigma(f)= \Bool$. A structure
$(\U,\astore,\afunc)$ is \emph{finite} when $\card{\U} \in \nat$ and
\emph{infinite} otherwise.

We write $\astruct \models \varphi$ iff $\varphi$ is true when
interpreted in $\astruct$. This relation is defined recursively on the
structure of $\varphi$, as usual. When $\astruct \models \varphi$, we
say that $\astruct$ is a \emph{model} of $\varphi$.  A formula is
\emph{satisfiable} when it has a model. We write $\varphi_1 \models
\varphi_2$ when every model of $\varphi_1$ is also a model of
$\varphi_2$ and by $\varphi_1 \equiv \varphi_2$ we mean $\varphi_1
\models \varphi_2$ and $\varphi_1 \models \varphi_2$.  The
\emph{(in)finite satisfiability problem} asks, given a formula
$\varphi$, whether a (in)finite model exists for this formula.

The Bernays-Sch\"onfinkel-Ramsey fragment of $\fol$, denoted by
$\bsr(\fol)$, is the set of sentences $\exists x_1 \ldots \exists x_n
\forall y_1 \ldots \forall y_m ~.~ \varphi$, where $\varphi$ is a
quantifier-free formula in which all function symbols $f$ of arity
$\#(f)>0$ have sort $\sigma(f)=\Bool$. It is known that any
satisfiable $\bsr(\fol)$ sentence has a finite model with at most
$\max(1,n)$ locations, where $n$ is the length of the existential
quantifier prefix\footnote{See, e.g., \cite[Proposition
    6.2.17]{BorgerGradelGurevich97}.}.

\subsection{Separation Logic}

Let $k \in \nat$ be a strictly positive integer. The logic $\seplogk{k}$
is the set of formulae generated by the grammar below: 
\[\begin{array}{rcl}
\varphi & := & \bot \mid \top \mid \emp \mid x \teq y \mid x \mapsto
(y_1,\ldots,y_k) \mid \\ && \varphi \wedge \varphi \mid \neg \varphi
\mid \varphi * \varphi \mid \varphi \wand \varphi \mid \exists x ~.~
\varphi
\end{array}\]
where $x,y,y_1,\ldots,y_k \in \vars$. The connectives $*$ and $\wand$
are respectively called the \emph{separating conjunction} and
\emph{separating implication} (\emph{magic wand}). We write $\varphi_1
\septraction \varphi_2$ for $\neg(\varphi_1 \wand \neg \varphi_2)$
(also called \emph{septraction}) and denote by $\vec{y}$, $\vec{y'}$
the tuples $(y_1,\ldots,y_k), (y'_1,\ldots,y'_k) \in \vars^k$,
respectively. The size of an $\seplogk{k}$ formula $\varphi$, denoted
$\size{\varphi}$, is the number of symbols needed to write it down.

Given an $\seplogk{k}$ formula $\phi$ and a subformula $\psi$ of $\phi$,
we say that $\psi$ \emph{occurs at polarity} $p \in \set{-1,0,1}$ iff
one of the following holds: \begin{inparaenum}[(i)]
\item $\phi = \psi$ and $p=1$, 
\item $\phi = \neg\phi_1$ and $\psi$ occurs at polarity $-p$ in
  $\phi_1$,
\item $\phi = \phi_1 \wedge \phi_2$ or $\phi = \phi_1 * \phi_2$, and
  $\psi$ occurs at polarity $p$ in $\phi_i$, for some $i=1,2$, or
\item $\phi = \phi_1 \wand \phi_2$ and either $\psi$ is a subformula
  of $\phi_1$ and $p=0$, or $\psi$ occurs at polarity $p$ in $\phi_2$.
\end{inparaenum}
A polarity of $1,0$ or $-1$ is also referred to as positive, neutral
or negative, respectively. 

$\seplogk{k}$ formulae are interpreted over
$\seplog$-\emph{structures} (called structures when no confusion
arises) $\I = (\U, \astore, \aheap)$, where $\U$ and $\astore$ are as
before and $\aheap : \U \rightharpoonup_{\mathit{fin}} \U^k$ is a
finite partial mapping of locations to $k$-tuples of locations, called
a \emph{heap}. As before, a structure $(\U,\astore,\aheap)$ is finite
when $\card{\U} \in \nat$ and infinite otherwise. We denote by
$\dom(\aheap)$ the domain of the heap $\aheap$ and by $\card{\aheap}
\in \nat$ the cardinality of $\dom(\aheap)$. Two heaps $\aheap_1$ and
$\aheap_2$ are \emph{disjoint} iff $\dom(\aheap_1) \cap \dom(\aheap_2)
= \emptyset$, in which case $\aheap_1 \uplus \aheap_2$ denotes their
union ($\uplus$ is undefined for non-disjoint heaps). A heap $\aheap'$
is an \emph{extension} of $\aheap$ iff $\aheap' = \aheap \uplus
\aheap''$, for some heap $\aheap''$. The relation $(\U,\astore,\aheap)
\models \varphi$ is defined inductively, as follows:
\[\begin{array}{lcl}
(\U,\astore,\aheap) \models \emp & \iff & \aheap = \emptyset \\ 
(\U,\astore,\aheap) \models x \mapsto (y_1,\ldots,y_k) & \iff &
\aheap = \set{\tuple{\astore(x),(\astore(y_1), \ldots, \astore(y_k))}}  \\
(\U,\astore,\aheap) \models \varphi_1 * \varphi_2 & \iff & \text{there exist disjoint heaps} \\ 
&& h_1,h_2 \text{ such that } h=h_1\uplus h_2 \\
&& \text{and } (\U,\astore,\aheap_i) \models \varphi_i \text{, for $i = 1,2$} \\
(\U,\astore,\aheap) \models \varphi_1 \wand \varphi_2 & \iff &
\text{for all heaps $\aheap'$ disjoint from $\aheap$} \\
&& \text{such that } (\U,\astore,\aheap') \models \varphi_1 \text{, we} \\
&& \text{have } (\U,\astore,\aheap'\uplus\aheap) \models \varphi_2
\end{array}\]
The semantics of equality, boolean and first-order connectives is the
usual one. Satisfiability, entailment and equivalence are defined for
$\seplogk{k}$ as for $\fol$ formulae. The (in)finite satisfiability
problem for $\seplogk{k}$ asks whether a (in)finite model exists for a
given formula. We write $\phi \equiv^{\fincard} \psi$ [$\phi
  \equiv^{\infcard} \psi$] whenever $(\U,\astore,\aheap) \models \phi
\iff (\U,\astore,\aheap) \models \psi$ holds for every finite
     [infinite] structure $(\U,\astore,\aheap)$.

The Bernays-Sch\"onfinkel-Ramsey fragment of $\seplogk{k}$, denoted by
$\bsr(\seplogk{k})$, is the set of sentences $\exists x_1 \ldots
\exists x_n \forall y_1 \ldots \forall y_m ~.~ \phi$, where $\phi$ is
a quantifier-free $\seplogk{k}$ formula. Since there are no function
symbols of arity greater than zero in $\seplogk{k}$, there are no
restrictions, other than the form of the quantifier prefix, defining
$\bsr(\seplogk{k})$.

\section{Test Formulae for $\seplogk{k}$}
\label{sec:test-formulae}

We define a small set of $\seplogk{k}$ patterns of formulae, possibly
parameterized by a positive integer, called \emph{test
  formulae}. These patterns capture properties related to allocation,
points-to relations in the heap and cardinality constraints.
\begin{definition}\label{def:test-formulae}
The following patterns are called \emph{test formulae}:
\[\begin{array}{rcl}
x \pto \vec{y} & \defequal & x \mapsto \vec{y} * \top \\
\alloc(x) & \defequal & x \mapsto \underbrace{(x,\ldots,x)}_{k \text{ times}} \wand \bot \\
\len{h} \geq n & \defequal & \left\{\begin{array}{ll}
\len{h} \geq n-1 * \neg\emp, & \text{if $n>0$} \\
\top, & \text{if $n = 0$} \\
\bot, & \text{if $n = \infty$}
\end{array}\right. \\
\len{U} \geq n & \defequal & \top \septraction \len{h} \geq n,~ n \in \nat \\
\len{h} \geq \len{U} - n & \defequal & \len{h} \geq n + 1 \wand \bot, n \in \nat \\
%\finite & \defequal & \top \septraction \len{h} \geq \len{U} - 0 \\
\end{array}\]
and $x \teq y$, where $x,y \in \vars$, $\vec{y} \in \vars^k$ and $n
\in \nat_\infty$ is a positive integer or $\infty$. A \emph{literal}
is either a test formula or its negation.
\end{definition}
The intuitive semantics of test formulae is formally stated below:

\begin{proposition}\label{prop:test-formulae}
  Given an $\seplog$-structure $(\U,\astore,\aheap)$, we have:
  \[\begin{array}{rcl}
  (\U,\astore,\aheap) \models x \pto \vec{y} & \iff & \aheap(\astore(x)) = (\astore(y_1),\ldots, \astore(y_k)) \\
  (\U,\astore,\aheap) \models \len{U} \geq n & \iff & \card{\U} \geq n \\
  (\U,\astore,\aheap) \models \alloc(x) & \iff & \astore(x) \in \dom(\aheap) \\
  (\U,\astore,\aheap) \models \len{h} \geq \len{U} - n & \iff & \card{\aheap} \geq \card{\U} - n \\
  (\U,\astore,\aheap) \models \len{h} \geq n & \iff & \card{\aheap} \geq n \\
%  \I \models \finite & \iff & \card{\U} \in \nat
  \end{array}\]
  for all variables $x, y_1, \ldots, y_k \in \vars$ and  integers $n \in \nat$. 
\end{proposition}
\proof{Let $\I = (\U,\astore,\aheap)$ and, given a set of locations
  $\U$ and a finite set $L\subseteq \U$, we will denote by $\aheap_L$
  the heap with domain $L$, such that for all $\ell\in L$,
  $\aheap_L(\ell) = (\ell, \ldots, \ell)$. It is clear that
  $\card{\aheap_L} = \card{L}$.

  \noindent
  $\boxed{\I \models x \pto \vec{y}\iff \aheap(\astore(x)) =
    (\astore(y_1),\ldots, \astore(y_k))}$ Assume that $\I \models x
  \pto \vec{y}$. Then by definition, there exist disjoint heaps
  $\aheap_1$, $\aheap_2$ such that $(\U, \astore, \aheap_1) \models
  x\mapsto \vec{y}$, $(\U, \astore,\aheap_2)\models \top$ and $\aheap
  = \aheap_1\uplus\aheap_2$. Thus $\astore(x)\in \dom(\aheap_1)
  \subseteq \dom(\aheap)$ and $\aheap(\astore(x)) =
  \aheap_1(\astore(x)) = (\astore(y_1),\ldots,
  \astore(y_k))$. Conversely, assume $\aheap(\astore(x)) =
  (\astore(y_1),\ldots, \astore(y_k))$. Then $\aheap$ is of the form
  $\aheap_1\uplus\aheap_2$, where $\aheap_1$ is the restriction of
  $\aheap$ to $\set{\astore(x)}$ and $\aheap_2$ is the restriction of
  $\aheap$ to $\U\setminus \set{\astore(x)}$. It is straightforward to
  verify that $\aheap_1\models x\mapsto \vec{y}$ and $\aheap_2\models
  \top$.
    
  \noindent
  $\boxed{\I \models \alloc(x) \iff \astore(x) \in \dom(\aheap)}$
  Assume that $\I \models \alloc(x)$. Then there cannot be any heap
  $\aheap_1$ disjoint from $\aheap$, such that $(\U, \astore,\aheap_1)
  \models x \mapsto (x,\ldots,x)$. But for $L = \set{\astore(x)}$, we
  have $(\U, \astore,\aheap_L) \models x \mapsto (x,\ldots,x)$, thus
  $\aheap_L$ is not disjoint from $\aheap$ and necessarily,
  $\astore(x) \in \dom(\aheap)$. Conversely, assume $\astore(x) \in
  \dom(\aheap)$, and let $\aheap_1$ be a heap such that $(\U, \astore,
  \aheap_1)\models x \mapsto (x,\ldots,x)$. Then $\aheap_1$ cannot be
  disjoint from $\aheap$, which proves that $\I\models \alloc(x)$.
  
  \noindent
  $\boxed{\I \models \len{h} \geq n\iff \card{\aheap} \geq n}$ Assume
  that $\I\models \len{h} \geq n$. Then since $\aheap$ has a finite
  domain, it is clear that $\card{\aheap}\geq n$ if $n< 0$ and that no
  such structure exists if $n = \infty$. When $n\geq 0$, we prove the
  result by induction on $n$. The case where $n=0$ is straightforward
  to prove. Otherwise, there exist disjoint heaps $\aheap_1,\aheap_2$
  such that $(\U, \astore, \aheap_1)\models \len{h}\geq n-1$, $(\U,
  \astore, \aheap_2)\models \neg\emp$ and $\aheap =
  \aheap_1\uplus\aheap_2$. By the induction hypothesis
  $\card{\aheap_1} \geq n-1$ and by definition, $\card{\aheap_2} \geq
  1$, so that $\card{\aheap_1\uplus\aheap_2} \geq n$. Conversely,
  assume that $\card{\aheap} \geq n$. This always holds if $n\leq 0$
  and never holds if $n=\infty$. Otherwise, we prove the result by
  induction on $n$. Assume $n>0$, so that $\dom(\aheap) \neq
  \emptyset$. Consider $\ell\in \dom(\aheap)$ and let $\aheap_1$ and
  $\aheap_2$ respectively denote the restrictions of $\aheap$ to
  $\U\setminus\set{\ell}$ and to $\set{\ell}$, so that $\aheap =
  \aheap_1\uplus\aheap_2$. Since $\card{\aheap_1}\geq n-1$, by the
  induction hypothesis $(\U,\astore, \aheap_1)\models \len{h}\geq
  n-1$, and since $\dom(\aheap_2) \neq \emptyset$, we have the result.
  
  \noindent
  $\boxed{\I \models \len{U} \geq n \iff \card{\U} \geq n}$ Assume
  that $\I\models \len{U}\geq n$. Then there exists a heap $\aheap_1$
  disjoint from $\aheap$ such that $(\U,\astore,\aheap\uplus\aheap_1)
  \models \len{h}\geq n$. This entails that
  $\card{\aheap\uplus\aheap_1}\geq n$ and since
  $\dom(\aheap\uplus\aheap_1)\subseteq \U$, necessarily,
  $\card{\U}\geq n$. Conversely, if $\card{\U}\geq n$, then there
  exists a set $L\subseteq \U$ such that $\dom(\aheap)\subseteq L$ and
  $\card{L} = \max \set{n,\,\card{\aheap}}$. Then $(\U,\astore,
  \aheap\uplus\aheap_L) \models \len{h}\geq n$, which proves that
  $\I\models \len{U}\geq n$.

  \noindent
  $\boxed{\I \models \len{h} \geq \len{U} - n \iff \card{\aheap} \geq
    \card{\U} - n}$ Assume that $\I\models \len{h}\geq \len{U} -
  n$. Then there is no heap disjoint from $\aheap$ with a domain of
  cardinality at least $n+1$. In particular, if $L = \U\setminus
  \dom(\aheap)$, then necessarily, $\card{\aheap_1} \leq n$. Since
  $\card{\U} = \card{\aheap} +\card{\aheap_1}$, we deduce that
  $\card{\aheap} \geq \card{\U} - n$. Conversely, if $\card{\aheap}
  \geq \card{\U} - n$ then there is no heap disjoint from $\aheap$
  with a domain of cardinality at least $n+1$, so that $\I\models
  \len{h} \geq \len{U} - n$. \qed
  
%  \noindent
%  $\boxed{\I \models \finite \iff \card{\U} \in \nat}$ Assume that
%  $\I\models \finite$. Then there exists a heap $\aheap_1$ disjoint
%  from $\aheap$ such that $(\U,\astore, \aheap\uplus\aheap_1)\models
%  \len{h}\geq \len{U} - 0$. This means that
%  $\card{\aheap\uplus\aheap_1} \geq \card{\U}$, and since both
%  $\aheap$ and $\aheap_1$ have a finite domain, $\U$ is necessarily
%  finite. Conversely, assume that $\card{\U} \in \nat$. Then the set
%  $L\defequal \U\setminus \dom(\aheap)$ is finite and since
%  $\card{\aheap\uplus\aheap_L} = \card{\U}$, we deduce that
%  $(\U,\astore, \aheap\uplus\aheap_L) \models \len{h}\geq\len{U} -
%  0$. Thus $\I\models \finite$. \qed 
}

Not all atoms of $\seplogk{k}$ are test formulae, for instance $x
\mapsto \vec{y}$ and $\emp$ are not test formulae. However, by
Proposition \ref{prop:test-formulae}, we have the equivalences $x
\mapsto \vec{y} \equiv x \pto \vec{y} \wedge \neg\len{h} \geq 2$ and
$\emp \equiv \neg\len{h} \geq 1$. Moreover, for any $n \in \nat$, the
test formulae $\len{U}\geq n$ and $\len{h}\geq \len{U}-n$ become
trivially true and false, respectively, if we consider the universe to
be infinite.

The integer parameter $n$ occurring in $\len{h} \geq n$, $\len{U} \geq
n$ and $\len{h} \geq \len{U} - n$ is assumed to be written in unary
notation. We write $t < u$ for $\neg(t \geq u)$ and $t \teq u$ for $t
\geq u \wedge t < u + 1$, where $t,u \in
\set{n,\len{h},\len{U},\len{U}-n \mid n \in \nat}$. For technical
convenience, we also define the following linear combinations.

\begin{definition}\label{def:lin-comb}
  Given integers $\alpha,\beta \in \zed$, where
  $\alpha\not\in\set{0,1}$, let $\len{h} \geq \alpha \cdot \len{U} +
  \beta \defequal$
\[\left\{\begin{array}{ll}
\bot & \text{ if $\alpha > 1,~ \beta > 0$} \\
\top & \text{ if $\alpha, \beta < 0$} \\
\begin{array}{l}
\len{U} < \left\lceil\frac{1-\beta}{\alpha-1}\right\rceil ~\wedge \\
\bigwedge_{\scriptscriptstyle{1 \leq n \leq \left\lfloor\frac{-\beta}{\alpha - 1}\right\rfloor}}
(\len{U} \teq n \rightarrow \len{h} \geq \alpha \cdot n + \beta)
\end{array} & \text{ if $\alpha > 1,~ \beta \leq 0$} \\
\bigwedge_{\scriptscriptstyle{1 \leq n < \left\lfloor\frac{-\beta}{\alpha}\right\rfloor}}
	(\len{U} \teq n \rightarrow \len{h} \geq \alpha \cdot n + \beta) & 
\text{ if $\alpha < 0,~ \beta \geq 0$}
\end{array}\right.\]
\end{definition}

\begin{proposition}\label{prop:linear-combination}
  Given an $\seplog$-structure $(\U,\astore,\aheap)$, we have
  $(\U,\astore,\aheap) \models \len{h} \geq \alpha \cdot \len{U} +
  \beta$ iff $\card{\aheap} \geq \alpha \cdot \card{\U} + \beta$, for
  all $\alpha,\beta \in \zed$, $\alpha\not\in\set{0,1}$.
\end{proposition}
\proof{
  We distinguish the four cases below: \begin{compactitem}
  \item If $\alpha > 1$ and $\beta > 0$ then $(\U,\astore,\aheap)
    \models \bot \iff \card{\U} \geq \card{\aheap} \geq \alpha \cdot
    \card{\U} + \beta$, never.
  \item If $\alpha < 0$ and $\beta < 0$ then $(\U,\astore,\aheap)
    \models \top \iff \card{\aheap} \geq 0 \geq \alpha \cdot \card{\U} +
    \beta$, always.
  \item If $\alpha > 1$ and $\beta \geq 0$, assume first that
    $(\U,\astore,\aheap) \models \len{h} \geq \alpha \cdot \len{U} +
    \beta$. Then $(\U,\astore,\aheap) \models \len{U} <
    \left\lceil\frac{1-\beta}{\alpha-1}\right\rceil$ thus $1 \leq
    \card{\U} < \left\lceil\frac{1-\beta}{\alpha-1}\right\rceil$, by
    Proposition \ref{prop:test-formulae}. If $\card{\U} >
    \left\lfloor\frac{-\beta}{\alpha-1}\right\rfloor$ then $\card{\U}
    \geq \left\lceil\frac{1-\beta}{\alpha-1}\right\rceil$, which
    contradicts $(\U,\astore,\aheap) \models \len{U} <
    \left\lceil\frac{1-\beta}{\alpha-1}\right\rceil$, by Proposition
    \ref{prop:test-formulae}. Otherwise, we have $\card{\U} = n$, with
    $1 \leq n \leq
    \left\lfloor\frac{-\beta}{\alpha-1}\right\rfloor$. In this case
    $(\U,\astore,\aheap) \models \len{h} \geq \alpha \cdot n + \beta$,
    which implies $\card{\aheap} \geq \alpha \cdot \card{\U} + \beta$,
    by Proposition \ref{prop:test-formulae}. Conversely, assume that
    $\card{\aheap} \geq \alpha \cdot \card{\U} + \beta$.  Since
    necessarily $\card{\U} \geq \card{\aheap}$, we obtain $\card{\U}
    \geq \alpha \cdot \card{\U} + \beta$, i.e., $\card{\U} > \alpha
    \cdot \card{\U} + \beta - 1$ and thus $\card{\U} <
    \left\lceil\frac{1-\beta}{\alpha-1}\right\rceil$ hence
    $(\U,\astore,\aheap) \models \len{U} <
    \left\lceil\frac{1-\beta}{\alpha-1}\right\rceil$.  Moreover, if $n
    = \card{U}$ then $(\U,\astore,\aheap) \models \len{h} \geq
    \alpha\cdot n+\beta$ follows by Proposition
    \ref{prop:test-formulae}.
  \item If $\alpha < 0$ and $\beta \geq 0$, assume first that
    $(\U,\astore,\aheap) \models \len{h} \geq \alpha \cdot \len{U} +
    \beta$. If, moreover, $\card{\U} \geq \frac{-\beta}{\alpha}$, then
    $\alpha\cdot\card{\U}+\beta \leq 0$, thus $\card{\aheap} \geq 0
    \geq \alpha\cdot\card{\U}+\beta$ holds. Otherwise, $1 \leq
    \card{\U} < \left\lfloor\frac{-\beta}{\alpha}\right\rfloor$ and if
    $(\U,\astore,\aheap) \models \len{U} \teq n$, for some $1 \leq n <
    \left\lfloor\frac{-\beta}{\alpha}\right\rfloor$, we have
    $(\U,\astore,\aheap) \models \len{h} \geq \alpha \cdot n + \beta$,
    thus $\card{\aheap} \geq \alpha \cdot \card{\U} + \beta$, by
    Proposition \ref{prop:test-formulae}. Conversely, assume that
    $\card{\aheap} \geq \alpha \cdot \card{\U} + \beta$ and
    $(\U,\astore,\aheap) \models \len{U} \teq n$, for some integer $1
    \leq n < \left\lfloor\frac{-\beta}{\alpha}\right\rfloor$. By
    Proposition \ref{prop:test-formulae}, we have $\card{\U} = n$ and
    $\card{\aheap} \geq \alpha \cdot n + \beta$, thus
    $(\U,\astore,\aheap) \models \len{h} \geq \alpha \cdot \len{U} +
    \beta$. \qed
  \end{compactitem}}

\begin{definition}\label{def:anvar}
A variable $x$ is \emph{allocated} in an $\seplog$-structure $\I$ iff
$\I \models \alloc(x)$. For a set of variables $X \subseteq \vars$,
let $\alloc(X) \defequal \bigwedge_{x \in X} \alloc(x)$ and
$\nalloc(X) \defequal \bigwedge_{x \in X} \neg\alloc(x)$. For a set
$T$ of literals, we define:
\[\begin{array}{l}
\avar{T} \defequal \\ 
\quad \big\{x \in \vars \mid x \teq_T x',~ T \cap \{\alloc(x'), x'\pto \vec{y} \mid \vec{y} \in \vars^k\} \neq \emptyset\big\} \\
\nvar{T} \defequal \\ 
\quad \{x \in \vars \mid x \teq_T x',~ \neg\alloc(x') \in T\} \\
\footprint{T}{X} \defequal \\ 
\quad T \cap \{\alloc(x), \neg\alloc(x), x \pto \vec{y}, \neg x \pto \vec{y} \mid x \in X, \vec{y} \in \vars^k\}
\end{array}\]
We let $\allocno{T} \defequal \len{\avar{T}}_T$ be the number of
equivalence classes of $\teq_T$ containing variables allocated in
every model of $T$
% by a formula in $T$
and $\nallocno{X}{T} \defequal \len{X \cap \nvar{T}}_T$ be the number
of equivalence classes of $\teq_T$ containing variables from $X$ that
are not allocated in any model of $T$. We also let $\afootprint{T}
\defequal \footprint{T}{\avar{T}}$.
\end{definition}
Intuitively, $\avar{T}$ [$\nvar{T}$] is the set of variables that must
be [are never] allocated in every [any] model of $T$ and
$\footprint{T}{X}$ is the \emph{footprint} of $T$ relative to the set
$X \subseteq \vars$, i.e.\ the set of formulae describing allocation
and points-to relations over variables from $X$.
For example, if $T = \{x \teq z, \alloc(x), \neg\alloc(y), \neg z \pto
\vec{y}\}$, then $\avar{T} = \set{x,z}$, $\nvar{T} = \set{y}$,
$\afootprint{T} = \set{\alloc(x),\neg z \pto \vec{y}}$ and
$\footprint{T}{\nvar{T}} = \set{\neg\alloc(y)}$. 

\prop{prop:footprint-extension}{
  Given a set $T$ of test formulae and a structure
  $(\U,\astore,\aheap)$, if $(\U,\astore,\aheap) \models
  \afootprint{T}$, we have $(\U,\astore,\aheap') \models
  \afootprint{T}$ for any extension $\aheap'$ of $\aheap$.
}
\proof{By a case split on the form of the atom in $\afootprint{T}$,
  namely $\alloc(x), x \pto \vec{y}, \neg x \pto \vec{y}$, with $x \in
  \avar{T}$. For the case $\neg\alloc(x)$, since $x \in \avar{T}$ then
  $\alloc(x') \in T$ for some variable $x'$ such that $x \teq_T x'$,
  thus $\afootprint{T}$ is unsatisfiable, contradicting the assumption
  that $(\U,\astore,\aheap) \models \afootprint{T}$. \qed}

%Next, we denote by $\allocno{T} \defequal \len{\avar{T}}_T$ the number
%of equivalence classes of $\teq_T$ containing variables allocated by
%$T$, and by $\nallocno{X}{T} \defequal \len{X \cap \nvar{T}}_T$ the
%number of equivalence classes of $\teq_T$ containing variables from
%$X$ that are not allocated in any model of $T$.

\subsection{From Test Formulae to $\fol$}
\label{sec:tf-fol}

The introduction of test formulae (Definition \ref{def:test-formulae})
is motivated by the reduction of the (in)finite satisfiability problem
for quantified boolean combinations thereof to the same problem for
$\fol$. Given a quantified boolean combination of test formulae
$\phi$, the $\fol$ formula $\foltrans{\phi}$ is defined by induction
on the structure of $\phi$:
\[\begin{array}{l}
\begin{array}{rclcrcl}
\foltrans{\len{h} \geq n} & \defequal & \aconst_n &~& \foltrans{\len{U} \geq n} & \defequal & \bconst_n \\
\foltrans{\len{h} \geq \len{U} - n} & \defequal & \neg \cconst_{n+1} && \foltrans{\neg\phi_1} & \defequal & \neg\foltrans{\phi_1} \\
%% \foltrans{\finite} & \defequal & \fconst 
\end{array}
\\
\begin{array}{rcl}
\foltrans{x \pto \vec{y}} & \defequal & \pfunc(x,y_1,\ldots,y_k) \\
\foltrans{\alloc(x)} & \defequal & \exists y_1 \ldots \exists y_k ~.~ \pfunc(x,y_1,\ldots,y_k) 
\end{array} 
\\
\begin{array}{rcl}
\foltrans{\phi_1 \wedge \phi_2} & \defequal & \foltrans{\phi_1} \wedge \foltrans{\phi_2} \\
\foltrans{\exists x ~.~ \phi_1} & \defequal & \exists x ~.~ \foltrans{\phi_1}
\end{array}
\end{array}\]
where $\pfunc$ is a $(k+1)$-ary function symbol of sort $\Bool$ and
$\aconst_n,\bconst_n$ and $\cconst_n$ are constants of sort $\Bool$,
for all $n \in \nat$. These function symbols are related by the
following axioms, where $\uconst_n, \vconst_n$ and $\wconst_n$ are
constants of sort $U$, for all $n > 0$:
\begin{align}
\tag{$P$}\label{ax:p} & 
\forall x \forall \vec{y} \forall \vec{y'} ~.~ 
\pfunc(x,\vec{y}) \wedge \pfunc(x,\vec{y'}) \rightarrow \bigwedge_{i=1}^k y_i \teq y'_i
\\
\tag{$A_0$}\label{ax:a0} & \aconst_0 \\[-4pt]
\tag{$A_n$}\label{ax:an} & 
\left\{\begin{array}{l}
\exists \vec{y} ~.~ \aconst_n \rightarrow \aconst_{n-1} \wedge \pfunc(\uconst_n, \vec{y}) \wedge \bigwedge_{i=1}^{n-1} \neg \uconst_i \teq \uconst_n \\
\wedge~ \forall x \forall \vec{y} ~.~ \neg \aconst_n \wedge \pfunc(x, \vec{y}) \rightarrow \bigvee_{i=1}^{n-1} x \teq \uconst_i 
\end{array}\right\}
\\
\tag{$B_0$}\label{ax:b0} & \bconst_0 \\[-4pt]
\tag{$B_n$}\label{ax:bn} & 
\left\{\begin{array}{l}
\bconst_n \rightarrow \bconst_{n-1} \wedge \bigwedge_{i=1}^{n-1} \neg \vconst_i \teq \vconst_n \\
\wedge~ \forall x ~.~ \neg \bconst_n \rightarrow \bigvee_{i=1}^{n-1} x \teq \vconst_i 
\end{array}\right\}\\
\tag{$C_0$}\label{ax:c0} & \cconst_0 \\[-4pt]
\tag{$C_n$}\label{ax:cn} &
\begin{array}{l}
\forall \vec{y} ~.~ \cconst_n \rightarrow \cconst_{n-1} \wedge \neg \pfunc(\wconst_n,\vec{y}) \wedge \bigwedge_{i=1}^{n-1} \neg \wconst_n \teq \wconst_i 
\end{array}
\end{align}
Intuitively, $\aconst_n$ or $\bconst_n$ are true iff there are at
least $n$ locations in the domain of the heap and in the universe,
respectively $\uconst_1, \ldots, \uconst_n$ and $\vconst_1, \ldots,
\vconst_n$. However, if $\cconst_n$ is true, then there are at least
$n$ locations $\wconst_1, \ldots, \wconst_n$ outside of the domain of
the heap (free), but the converse does not hold (remark
\ref{rem:test-fol-card}). The following remarks motivate some of the
restrictions that define decidable fragments of $\bsr(\seplogk{k})$,
by reduction to $\bsr(\fol)$ (\S\ref{sec:bsr-sl-dec}).

\begin{remark}\label{rem:test-fol-alloc}
The translation of $\alloc(x)$ introduces existential quantifiers
depending on $x$. For instance, the $\bsr(\seplogk{k})$ formula $\phi
= \forall x ~.~ \alloc(x)$ is translated as $\foltrans{\phi} = \forall
x \exists y_1 \ldots \exists y_k ~.~ \pfunc(x,y_1,\ldots,y_k)$, which
lies outside of the $\bsr(\fol)$ fragment. Because upcoming results
(Thm. \ref{thm:pspaceBSR}) require that $\tau(\phi)$ be in
$\bsr(\fol)$, we consider quantified boolean combinations of test
formulae $\phi$ in which the $\alloc(x)$ formulae either occur at a
negative polarity, or $x$ is not universally quantified. In both such
cases, $\foltrans{\phi}$ is in $\bsr(\fol)$ (Lemma
\ref{lem:bsrfoltrans}).  \hfill$\blacksquare$
\end{remark}

\begin{remark}\label{rem:test-fol-card}
The $C_n$ axioms do not state the equivalence of $\cconst_n$ with the
existence of at least $n$ free locations. Such an equivalence seems to
be hard, if not impossible, to express in $\bsr(\fol)$\footnote{The
  converse of $C_n$: \(\forall x ~.~ (\neg \cconst_n \wedge \forall
  \vec{y}~.~ \neg \pfunc(x, \vec{y})) \rightarrow \bigvee_{i=1}^{n-1}
  x \teq \wconst_i\) is not in $\bsr(\fol)$.}. Note that if the domain
is infinite then this problem does not arise since the formulae
$\len{h} \geq \len{U} - n$ are always false.  \hfill$\blacksquare$
\end{remark}

\begin{definition}\label{def:maxn-axioms}
Given a quantified boolean combination of test formulae $\phi$, let
$\maxn{\phi}$ be the maximum integer parameter $n$ occurring in a test
formula $\theta \in \set{\len{h}\geq n, \len{U}\geq n,\len{h}\geq
  \len{U}-n \mid n \in \nat}$ from $\phi$ and let \(\axioms{\phi}
\defequal \set{P} \cup \set{A_i}_{i=0}^{\maxn{\phi}} \cup
\set{B_i}_{i=0}^{\maxn{\phi}} \cup \set{C_i}_{i=0}^{\maxn{\phi}+1}\)
be the set of axioms related to $\phi$.
%% For a universe $\U$, let $\fin{\U} \defequal \fconst$ if
%% $\card{\U}\in\nat$ and $\fin{\U} \defequal \neg\fconst$ if $\card{\U}
%% = \infty$.
\end{definition}

The relationship between %the satisfiability of 
a boolean combination of
test formulae $\phi$ and %that of 
its translation into $\fol$ is
%formally 
stated below. %Note that, if $\phi$ does not contain
%$\finite$, then $\foltrans{\phi}$ does not contain $\fconst$, thus the
%axiom $\fin{\U}$ is useless and can be omitted.
 
\begin{lemma}\label{lemma:test-fo-sat}
  Let $\phi$ be a quantified boolean combination of test formulae.
  The following hold, for any universe $\U$ and any store
  $\astore$: \begin{compactenum}
    \item\label{it:test-fo-sat1} if $(\U,\astore,\aheap) \models
      \phi$, for a heap $\aheap$, then $(\U,\astore,\afunc) \models
      \foltrans{\phi} \wedge \axioms{\phi}$, for an interpretation
      $\afunc$, and
    \item\label{it:test-fo-sat2} if each test formula $\len{h} \geq
      \len{U} - n$ in $\phi$ occurs at negative polarity and
      $(\U,\astore,\afunc) \models \foltrans{\phi} \wedge
      \axioms{\phi}$, for an interpretation $\afunc$, such that
      $\card{\pfunc^\afunc} \in \nat$, then $(\U,\astore,\aheap)
      \models \phi$, for a heap $\aheap$.
  \end{compactenum}
\end{lemma}
\proof{ (\ref{it:test-fo-sat1}) Let $(\U,\astore,\aheap)$ be a model
  of $\phi$. Considering an arbitrary interpretation $\bot^\afunc$ and
  $\top^\afunc$ for the boolean constants true and false, we extend
  $\afunc$ to the $(k+1)$-ary function symbol $\pfunc$, the constants
  $\aconst_i, \bconst_i, \cconst_j$ of sort $\Bool$ and $\uconst_i,
  \vconst_i, \wconst_i$ of sort $U$, for all $i \in [0,\maxn{\phi}]$
  and all $j \in [0,\maxn{\phi}+1]$, as follows. For all $\ell_0,
  \ldots, \ell_{k} \in \U$ we set
  $\pfunc^\afunc(\ell_0,\ldots,\ell_{k}) = \top^\afunc$ if
  $\aheap(\ell_0) = (\ell_1,\ldots,\ell_k)$ and
  $\pfunc^\afunc(\ell_0,\ldots,\ell_{k}) = \bot^\afunc$,
  otherwise. The interpretation of the boolean constants is defined
  below:
  \[
  \begin{array}{rcl}
    \aconst_i^\afunc & \defequal & \left\{\begin{array}{ll}
    \top^\afunc & \text{ if $0 \leq i \leq \min(\card{\aheap},\maxn{\phi})$} \\
    \bot^\afunc & \text{ if $i > \min(\card{\aheap},\maxn{\phi})$}       
    \end{array}\right.
    \\
    \bconst_i^\afunc & \defequal & \left\{\begin{array}{ll}
    \top^\afunc & \text{ if $0 \leq i \leq \min(\card{\U},\maxn{\phi})$} \\
    \bot^\afunc & \text{ if $i > \min(\card{\U},\maxn{\phi})$}       
    \end{array}\right.
    \\
    \cconst_i^\afunc & \defequal & \left\{\begin{array}{ll}
    \top^\afunc & \text{ if $0 \leq i \leq \min(\card{\U}-\card{\aheap},\maxn{\phi}+1)$} \\
    \bot^\afunc & \text{ if $i > \min(\card{\U}-\card{\aheap},\maxn{\phi}+1)$}       
    \end{array}\right.
    %% \\
    %% \fconst^\afunc & \defequal & \left\{\begin{array}{ll}
    %% \top^\afunc & \text{ if $\card{\U} \in \nat$} \\
    %% \bot^\afunc & \text{ if $\card{\U} = \infty$}
    %% \end{array}\right.
  \end{array}
  \]
  Finally, the constants of sort $U$ are interpreted as
  locations, as follows: \begin{compactitem}
  \item $\uconst_1^\afunc, \ldots,
    \uconst_{\min(\card{\aheap},\maxn{\phi})}^\afunc$ are the first
    $\min(\card{\aheap},\maxn{\phi})$ locations in $\dom(\aheap)$ and
    the rest are arbitrary.
  \item $\vconst_1^\afunc, \ldots,
    \vconst_{\min(\card{\U},\maxn{\phi})}^\afunc$ are the first
    $\min(\card{\U},\maxn{\phi})$ locations in $\U$ and the rest are
    arbitrary.
    \item $\wconst_1^\afunc, \ldots,
      \wconst_{\min(\card{\U}-\card{\aheap},\maxn{\phi}+1)}^\afunc$
      are the first $\min(\card{\U}-\card{\aheap},\maxn{\phi}+1)$
      locations in $\U \setminus \dom(\aheap)$ and the rest are
      arbitrary.
  \end{compactitem}
  Clearly $(\U,\astore,\afunc) \models
  \axioms{\phi}$. We prove $(\U,\astore,\aheap) \models \phi \iff
  (\U,\astore,\afunc) \models \foltrans{\phi}$ by induction on the
  structure of $\phi$: \begin{compactitem}
  \item $\phi = \len{h} \geq n$: $(\U,\astore,\aheap) \models
      \phi$ iff $\card{\aheap} \geq n$, by Proposition
      \ref{prop:test-formulae}. Since $n \leq \maxn{\phi}$, we
      have $\card{\aheap} \geq n \iff n \leq
      \min(\card{\aheap},\maxn{\phi}) \iff \aconst_n^\afunc =
      \top^\afunc \iff (\U,\astore,\afunc) \models
      \foltrans{\phi}$.
    \item $\phi = \len{U} \geq n$: $(\U,\astore,\aheap) \models
      \phi$ iff $\card{\U} \geq n$, by Proposition
      \ref{prop:test-formulae}. Since $n \leq \maxn{\phi}$, we have
      $\card{\U} \geq n \iff n \leq \min(\card{\U},\maxn{\phi})
      \iff \bconst_n^\afunc = \top^\afunc \iff
      (\U,\astore,\afunc) \models \foltrans{\phi}$.
    \item $\phi = \len{h} \geq \len{U} - n$: $(\U,\astore,\aheap)
      \models \phi$ iff $\card{\aheap} \geq \card{\U} - n$, by
      Proposition \ref{prop:test-formulae}. Equivalently,
      $(\U,\astore,\aheap) \models \neg\phi$ iff $n+1 \leq \card{\U} -
      \card{\aheap}$. Since $n+1 \leq \maxn{\phi}+1$, we have
      $(\U,\astore,\aheap) \models \neg\phi \iff n+1 \leq
      \min(\card{\U} - \card{\aheap}, \maxn{\phi}+1) \iff
      \cconst_{n+1}^\afunc = \top^\afunc$, thus $(\U,\astore,\aheap)
      \models \phi \iff (\U,\astore,\afunc) \models \neg c_{n+1} \iff
      (\U,\astore,\afunc) \models \foltrans{\phi}$.
      %
%    \item $\phi = \finite$: $(\U,\astore,\aheap) \models \phi$ iff
%      $\card{\U} \in \nat$, by Proposition
%      \ref{prop:test-formulae}. By definition of $\afunc$, we have
%      $\card{\U} \in \nat \iff \fconst^\afunc = \top^\afunc \iff
%      (\U,\astore,\aheap) \models \foltrans{\phi}$.
      %
    \item $\phi = x \pto (y_1,\ldots,y_k)$: $(\U,\astore,\aheap)
      \models \phi$ iff $\aheap(\astore(x)) = (\astore(y_1), \ldots,
      \astore(y_k))$ iff $\pfunc^\afunc(\astore(x), \astore(y_1),
      \ldots, \astore(y_k)) = \top^\afunc$ iff $(\U,\astore,\afunc)
      \models \pfunc(x,y_1,\ldots,y_k)$.
    \item $\phi = \alloc(x)$: $(\U,\astore,\aheap) \models \phi$ iff
      $\astore(x) \in \dom(\aheap)$ iff $\aheap(\astore(x)) =
      (\ell_1,\ldots,\ell_k)$, for some $\ell_1,\ldots,\ell_k \in \U$
      iff $\pfunc^\afunc(\astore(x),\ell_1,\ldots,\ell_k) =
      \top^\afunc$, for some $\ell_1,\ldots,\ell_k \in \U$ iff
      $(\U,\astore[y_1 \leftarrow \ell_1] \ldots [y_k \leftarrow
        \ell_k],\afunc) \models \pfunc(x,y_1,\ldots,y_k)$, for some
      $\ell_1,\ldots,\ell_k \in \U$ iff $(\U,\astore,\afunc) \models
      \exists y_1 \ldots \exists y_k ~.~ \pfunc(x,y_1,\ldots,y_k)$.
    \item the cases $\phi = \phi_1 \wedge \phi_2$, $\phi = \neg\phi_1$
      and $\phi = \exists x ~.~ \phi_1$ are by the inductive
      hypothesis, since $(\U,\astore,\aheap) \models \phi_i \iff
      (\U,\astore,\afunc) \models \foltrans{\phi_i}$, for all $i =
      1,2$.
    \end{compactitem}         

    \noindent(\ref{it:test-fo-sat2}) Let $(\U,\astore,\afunc)$ be a
    model of $\foltrans{\phi} \wedge \axioms{\phi}$,
    such that $\card{\pfunc^\afunc} \in \nat$.  We define a heap
    $\aheap$ as follows: for each $(k+1)$-tuple of locations $\ell_0,
    \ldots, \ell_k \in \U$ such that
    $\pfunc^\afunc(\ell_0,\ldots,\ell_k) = \top^\afunc$, we set
    $\aheap(\ell_0) = (\ell_1,\ldots,\ell_k)$. Since
    $(\U,\astore,\afunc) \models P$ and $\card{\pfunc^\afunc} \in
    \nat$, we obtain that $\aheap$ is a finite partial function. We
    prove that $(\U,\astore,\afunc) \models \foltrans{\phi}
    \Rightarrow (\U,\astore,\aheap) \models \phi$ by induction on the
    structure of the negation normal form of
    $\phi$: \begin{compactitem}
    \item $\phi = \len{h} \geq n$: $\foltrans{\phi} = \aconst_n$ and
      $(\U,\astore,\afunc) \models \aconst_n \Rightarrow
      \aconst_n^\afunc = \top^\afunc$. Since $n \leq \maxn{\phi}$ and
      $(\U,\astore,\afunc) \models \bigwedge_{i=0}^{\maxn{\phi}} A_j$,
      we have $\aconst_j^\afunc = \top^\afunc$ and $\uconst_j^\afunc
      \in \dom(\aheap)$, for all $j \in [1,n]$. Because
      $\uconst_j^\afunc$ are pairwise disjoint, for all $j \in [1,n]$,
      we obtain that $\card{\aheap} \geq n$, and $(\U,\astore,\aheap)
      \models \phi$ follows, by Proposition \ref{prop:test-formulae}.
    \item $\phi = \len{h} < n$: $\foltrans{\phi} = \neg\aconst_n$ and
      $(\U,\astore,\afunc) \models \neg\aconst_n \Rightarrow
      \aconst_n^\afunc = \bot^\afunc$. Since $n \leq \maxn{\phi}$ and
      $(\U,\astore,\afunc) \models \bigwedge_{i=0}^{\maxn{\phi}} A_j$,
      we have that each location $\ell\in\dom(\aheap)$ must be one of
      $\uconst_1^\afunc, \ldots, \uconst_{n-1}^\afunc$, thus
      $\card{\dom(\aheap)} \leq n-1$ and $(\U,\astore,\aheap) \models
      \len{h} < n$ follows, by Proposition \ref{prop:test-formulae}.
    \item $\phi = \len{U} \geq n$: $\foltrans{\phi} = \bconst_n$ and
      $(\U,\astore,\afunc) \models \bconst_n \Rightarrow
      \bconst_n^\afunc = \top^\afunc$. Since $n \leq \maxn{\phi}$ and
      $(\U,\astore,\afunc) \models \bigwedge_{i=0}^{\maxn{\phi}} B_j$,
      we have $\bconst_j^\afunc = \top^\afunc$, for all $j \in
      [1,n]$. Because $\vconst_j^\afunc$ are pairwise disjoint, for
      all $j \in [1,n]$, we obtain that $\card{\U} \geq n$, and
      $(\U,\astore,\aheap) \models \phi$ follows, by Proposition
      \ref{prop:test-formulae}.
    \item $\phi = \len{U} < n$: $\foltrans{\phi} = \neg\bconst_n$ and
      $(\U,\astore,\afunc) \models \neg\bconst_n \Rightarrow
      \bconst_n^\afunc = \bot^\afunc$. Since $n \leq \maxn{\phi}$ and
      $(\U,\astore,\afunc) \models \bigwedge_{i=0}^{\maxn{\phi}} B_j$,
      we have that each location $\ell\in\U$ must be one of
      $\vconst_1^\afunc, \ldots, \vconst_{n-1}^\afunc$, thus
      $\card{\U} \leq n-1$ and $(\U,\astore,\aheap) \models \phi$
      follows, by Proposition \ref{prop:test-formulae}.
    \item $\phi = \len{h} \geq \len{U} - n$: this case is impossible
      because $\len{h} \geq \len{U} - n$ must occur at negative
      polarity in $\phi$.
    \item $\phi = \len{h} < \len{U} - n$: $\foltrans{\phi} =
      \cconst_{n+1}$ and $(\U,\astore,\afunc) \models \cconst_{n+1}
      \Rightarrow \cconst_{n+1} = \top^\afunc$.  Since $n+1 \leq
      \maxn{\phi}+1$ and $(\U,\astore,\afunc) \models
      \bigwedge_{i=0}^{\maxn{\phi}+1} C_j$, we obtain that
      $\wconst_j^\afunc \in \U \setminus \dom(\aheap)$, 
      for all $j \in [1,n+1]$. Since $\wconst_j^\afunc$ are pairwise
      disjoint, we obtain $\card{U} - \card{\aheap} \geq n+1$
      thus $(\U, \astore, \aheap) \models \phi$ follows, by
      Proposition \ref{prop:test-formulae}.
      %
%    \item $\phi = \finite$: $\foltrans{\phi} = \fconst$ and
%      $(\U,\astore,\afunc) \models \fconst \Rightarrow \fconst^\afunc
%      = \top^\afunc$. Since $(\U,\astore,\afunc) \models \fin{\U}$ it
%      must be the case that $\fin{\U}=\fconst=\top$, thus $\card{U}
%      \in \nat$ and $(\U,\astore,\aheap) \models \phi$ follows, by
%      Proposition \ref{prop:test-formulae}.
      %
%    \item $\phi = \neg\finite$: $\foltrans{\phi} = \neg\fconst$ and
%      $(\U,\astore,\afunc) \models \neg\fconst \Rightarrow
%      \fconst^\afunc = \bot^\afunc$. Since $(\U,\astore,\afunc)
%      \models \fin{\U}$ it must be the case that
%      $\fin{\U}=\neg\fconst=\top$, thus $\card{U} = \infty$ and
%      $(\U,\astore,\aheap) \models \phi$ follows, by Proposition
%      \ref{prop:test-formulae}.
      %
    \item $\phi \in \set{x \pto (y_1,\ldots,y_k), \neg x \pto
      (y_1,\ldots,y_k), \alloc(x), \neg\alloc(x)}$:
      $(\U,\astore,\aheap) \models x \pto (y_1,\ldots,y_k) \iff
      (\U,\astore,\afunc) \models \pfunc(x,y_1,\ldots,y_k)$ and
      $(\U,\astore,\aheap) \models \alloc(x) \iff (\U,\astore,\afunc)
      \models \exists y_1 \ldots y_k ~.~ \pfunc(x,y_1,\ldots,y_k)$ are
      proved in the same way as for point (\ref{it:test-fo-sat1}).
    \item the cases $\phi = \phi_1 \wedge \phi_2$, $\phi = \phi_1
      \vee \phi_2, \exists x ~.~ \phi_1$ are by inductive
      hypothesis. \qed
\end{compactitem}}

\prop{prop:existsoutside}{
Let $\forall x.\phi[\exists y.\psi]$ be a $\fol$ formula, such that
$\exists y.\psi$ occurs at a positive polarity in $\phi$. If $x \not\in
\fv{\psi}$ and $y \not\in \fv{\phi}$ then \(\forall x.  \phi[\exists
  y.\psi] \equiv \exists y \forall x . \phi[\psi/\exists y.\psi]\).
}
\proof{ $\boxed{\forall x. \phi[\exists y.\psi] \models \exists y
    \forall x . \phi[\psi/\exists y.\psi]}$ Let $(\U,\astore,\afunc)$
  be a model of $\forall x. \phi[\exists y.\psi]$ and let $t \in \{
  \bot, \top \}$ be the boolean constant such that
  $(\U,\astore,\afunc) \models \exists y.\psi \leftrightarrow
  t$. Since $x \not\in \fv{\psi}$, we have $(\U,\astore,\afunc)
  \models \forall x.\phi[t/\exists y.\psi]$ and moreover
  $(\U,\astore[y \leftarrow \ell],\afunc) \models \psi \leftrightarrow
  t$, for some location $\ell \in \U$. Since $y \not\in \fv{\forall x
    . \phi[t/\exists y.\psi]}$, we also have $(\U,\astore[y \leftarrow
    \ell],\afunc) \models \forall x . \phi[\psi/\exists y.\psi]$, thus
  $(\U,\astore,\afunc) \models \exists y \forall x . \phi[\psi/\exists
    y.\psi]$.

 \noindent
  $\boxed{\exists y \forall x . \phi[\psi/\exists y.\psi] \models \forall
   x. \phi[\exists y.\psi]}$ Let $(\U,\astore,\afunc)$ be a model of
 $\exists y \forall x . \phi[\psi/\exists y.\psi]$. Then $(\U,\astore[y \leftarrow
   \ell],\afunc) \models \forall x . \phi[\psi/\exists y.\psi]$, for some location
 $\ell \in \U$. Let $t \in \set{\top,\bot}$ be the boolean constant
 such that $(\U,\astore[y \leftarrow \ell],\afunc) \models t
 \leftrightarrow \psi$. We distinguish the following cases: \begin{compactitem}
  \item $t = \bot$: since $t$ occurs positively in $\phi[t/\exists
    y.\psi]$ by hypothesis, we have $(\U,\astore[y \leftarrow \ell],\afunc) \models
    \forall x . \phi[\exists y . \psi]$;
  \item $t = \top$: $(\U,\astore[y \leftarrow \ell],\afunc) \models
    \exists y . \psi \leftrightarrow \top$ hence $(\U,\astore[y
      \leftarrow \ell],\afunc) \models \forall x . \phi [\exists y
      . \psi]$.
  \end{compactitem}
In both cases $(\U,\astore,\afunc) \models \forall x . \phi[\exists y
  . \psi]$ follows because $y\not\in\fv{\phi}$. \qed}

We end this section by delimiting a fragment of $\seplogk{k}$ whose
translation into $\fol$ falls into $\bsr(\fol)$.

\begin{lemma}\label{lem:bsrfoltrans}
Given an $\seplogk{k}$ formula $\varphi = \forall y_1 \ldots \forall
y_m ~.~ \phi$, where $\phi$ is a boolean combination of test formulae
containing no positive occurrence of $\alloc(y_i)$, for any $i \in
[1,m]$, $\foltrans{\varphi}$ is equivalent to a $\bsr(\fol)$ formula
with the same constants and free variables as
$\foltrans{\varphi}$.
\end{lemma}
\proof{ By definition of $\foltrans{.}$, we have $\foltrans{\varphi} =
  \forall y_1 \ldots \forall y_m ~.~ \foltrans{\phi}$ and the only
  existential quantifiers in $\foltrans{\varphi}$ are those in the
  formulae $\exists z_1 \ldots \exists z_k~.~ \pfunc(x,z_1,\dots,z_k)$
  introduced by translating formulae of the form $\alloc(x)$ occurring
  positively in $\phi$.  Since $\phi$ contains no positive occurrence
  of $\alloc(y_i)$ by hypothesis, $x \not \in
  \set{y_1,\ldots,y_m}$. Consequently, for all $i\in[1,m]$, $y_i$ does
  not occur as an argument of $\pfunc$ and by Proposition
  \ref{prop:existsoutside}, the quantifiers $\exists z_1 \ldots
  \exists z_k$ may be shifted to the root of the formula
  $\foltrans{\varphi}$ without affecting equivalence. By repeating
  this operation for each formula $\alloc(x)$ in $\phi$, we eventually
  get a formula in $\bsr(\fol)$. \qed}

\section{From Quantifier-free $\seplogk{k}$ to Test Formulae}
\label{sec:spatial-elim}

This section gives the expressive completeness result of the paper,
namely that any quantifier-free $\seplogk{k}$ formula is equivalent,
on both finite and infinite models, to a quantifier-free boolean
combination of test formulae. Starting from a quantifier-free
$\seplogk{k}$ formula $\varphi$, we define, inductively on the
structure of $\varphi$, a set $\mu(\varphi)$ of conjunctions of test
formulae and their negations, called \emph{minterms}, such that
$\varphi \equiv \bigvee_{M \in \mu(\varphi)} M$. Although the number
of minterms in $\mu(\varphi)$ is, in general, exponential in the size
of $\varphi$, checking the membership of a given minterm $M$ in
$\mu(\varphi)$ can be done in \pspace. Together with the translation
of minterms into $\fol$ (\S\ref{sec:tf-fol}), this fact is used to
prove \pspace\ membership of the two decidable fragments of
$\bsr(\seplogk{k})$, defined next (\S\ref{sec:bsr-sl-dec}).

In the rest of this section we view a conjunction $T$ of literals as a
set\footnote{The empty set is thus considered to be true.}, thus we
use the same symbol to denote both a set and the formula obtained by
conjoining the elements of the set. We define $\fv{T} \defequal
\bigcup_{\ell \in T} \fv{\ell}$. The equivalence relation $x \teq_T y$
is defined as $T \models x \teq y$ and we write $x \not\teq_T y$ for
$T \models \neg x \teq y$. Observe that $x \not\teq_T y$ is not the
complement of $x \teq_T y$. For a set $X$ of variables, $\len{X}_T$ is
the number of equivalence classes of $\teq_T$ in $X$.

\subsection{Minterms}

A \emph{minterm} $M$ is a set (conjunction) of literals containing: 
\begin{compactitem}
\item exactly one literal $\len{h} \geq \minheap_M$ and one literal
  $\len{h} < \maxheap_M$, where $\minheap_M \in \nat \cup \set{\len{U}
  - n \mid n \in \nat}$ and $\maxheap_M \in \nat_{\infty} \cup
  \set{\len{U} - n \mid n \in \nat}$, and
\item at most\footnote{This condition is not restrictive: if $M$
  contains two literals $\len{U} \geq n_1$ and $\len{U} \geq n_2$ with
  $n_1 < n_2$ then $\len{U} \geq n_1$ is redundant and can be
  removed.} one literal of the form $\len{U} \geq n$, respectively
  $\len{U} < n$.
\end{compactitem}
For an $\seplog$-structure $\I = (\U,\astore,\aheap)$, let
$\minheap_M^\I, \maxheap_M^\I \in \nat_{\infty}$ be the values
obtained by replacing $\len{U}$ with $\card{\U}$ in $\minheap_M$ and
$\maxheap_M$, respectively.

\begin{definition}\label{def:minterm-sets}
Given a minterm $M$, we define the sets:
\[\begin{array}{rcl}
%M^f & \defequal & M \cap \set{\finite,\neg\finite} \\
M^e & \defequal & M \cap \set{x \teq y, \neg x \teq y \mid x,y \in \vars} \\ 
M^a & \defequal & M \cap \set{\alloc(x),\neg\alloc(x) \mid x \in \vars} \\
M^u & \defequal & M \cap \set{\len{U} \geq n, \len{U} < n \mid n \in \nat} \\
M^p & \defequal & M \cap \{x \pto \vec{y}, \neg x \pto
  \vec{y} \mid x \in \vars, \vec{y} \in \vars^k\}
\end{array}\]
\end{definition}
%Thus, $M = M^f \cup M^e \cup M^u \cup M^a \cup M^p \cup
Thus, $M = M^e \cup M^u \cup M^a \cup M^p \cup
\set{\len{h} \geq \minheap_M, \len{h} < \maxheap_M}$, for each minterm
$M$.

\prop{prop:heap-indep}{
  Given a minterm $M$, for all structures $\I=(\U,\astore,\aheap)$ and
%  $\I'=(\U,\astore,\aheap')$ we have $\I \models M^f \wedge M^e \wedge
  $\I'=(\U,\astore,\aheap')$ we have $\I \models M^e \wedge
%  M^u \iff \I' \models M^f \wedge M^e \wedge M^u$.
  M^u \iff \I' \models M^e \wedge M^u$.
}
%\proof{The test formulae in $M^f \cup M^e \cup M^u$ do not depend on  the heap. \qed}
\proof{The test formulae in $M^e \cup M^u$ do not depend on  the heap. \qed}

Given a set of variables $X \subseteq \vars$, a minterm $M$
is \begin{inparaenum}[(1)]
\item \emph{E-complete} for $X$ iff for all $x,y \in X$ exactly one of
  $x \teq y \in M$, $\neg x \teq y \in M$ holds, and
\item \emph{A-complete} for $X$ iff for each $x \in X$ exactly one of
  $\alloc(x) \in M$, $\neg\alloc(x) \in M$ holds.
\end{inparaenum}

\prop{prop:ecomp-eqclass}{
  If $M$ is E-complete for $\fv{M}$, $(\U,\astore,\aheap) \models M$
  and $X\subseteq \fv{M}$, then $\len{X}_M = \card{\astore(\fv{M})}$.
}
\proof{
  This is an immediate consequence of the fact that if $x,x'\in
  \fv{M}$, then $\astore(x) = \astore(x')$ if and only if $M \models
  x\teq x'$. \qed}

For a literal $\ell$, we denote by $\overline{\ell}$ its complement,
i.e.\ $\overline{\theta} \defequal \neg \theta$ and $\overline{\neg
  \theta} \defequal \theta$, where $\theta$ is a test formula. Let
$\overline{M}$ be the minterm obtained from $M$ by replacing each
literal with its complement. The \emph{complement closure} of $M$ is
$\cclose{M} \defequal M \cup \overline{M}$.

Two tuples $\vec{y}, \vec{y'} \in \vars^k$ are \emph{$M$-distinct} if
$y_i \not\teq_M y'_i$, for some $i \in [1,k]$.  Given a minterm $M$
that is E-complete for $\fv{M}$, its \emph{points-to closure} is
\(\pclose{M} \defequal \bot\) if there exist literals $x \pto \vec{y},
x' \pto \vec{y'} \in M$ such that $x \teq_M x'$ and $\vec{y}$,
$\vec{y'}$ are $M$-distinct, and $\pclose{M} = M$, otherwise.
Intuitively, $\pclose{M}$ is $\bot$ iff $M$ contradicts the fact that
the heap is a partial function. Note that we do not assert the
equality $\vec{y} \teq \vec{y'}$, instead we only check that it is not
falsified. This is sufficient for our purpose because in the following
we always assume that the considered minterms are E-complete.

The \emph{domain closure} of $M$ is $\uclose{M} \defequal \bot$ if either
$\minheap_M = n_1$ and $\maxheap_M = n_2$ for some $n_1,n_2 \in \zed$
such that $n_1 \geq n_2$, or $\minheap_M = \len{U} - n_1$ and $\maxheap_M
= \len{U} - n_2$, where $n_2 \geq n_1$; and otherwise:
\[\begin{array}{l}
\uclose{M} \defequal M \cup  \set{\len{U} \geq \left\lceil\sqrt[k]{\maxheap_{x \in \avar{M}}(\datano{x}{M}+1)}\right\rceil}  \\ 
\cup \set{\len{U} \geq n_1+n_2+1 \mid \minheap_M = n_1, \maxheap_M = \len{U}-n_2, n_1, n_2 \in \nat} \\
\cup \set{\len{U} < n_1+n_2 \mid \minheap_M = \len{U}-n_1, \maxheap_M = n_2, n_1, n_2 \in \nat} 
\end{array}\]
where $\datano{x}{M}$ is the number of pairwise $M$-distinct tuples
$\vec{y}$ for which there exists $\neg x' \pto \vec{y} \in M$ such
that $x \teq_M x'$. Intuitively, $\uclose{M}$ asserts that $\minheap_M
< \maxheap_M$ and that the domain contains enough elements to allocate
all cells. Moreover, for every allocated variable $x$,
there must exist at least $\datano{x}{M}+1$ distinct $k$-vectors of
elements of the domain: the $\datano{x}{M}$ that $x$ cannot point to,
plus the image of $x$.  For instance, if $M = \{ \neg x \pto y_i,
\alloc(x), y_i \not \teq y_j \mid i,j \in [1,n], i \not = j \}$, then
it is clear that $M$ is unsatisfiable if there are less than $n$
locations, since $x$ cannot be allocated in this case.

\prop{prop:closure-equiv}{ For any minterm $M$, we have $M \equiv
  \pclose{M} \equiv \uclose{M}$.  } \proof{ It is manifest that
  $\pclose{M} \models M$ and $\uclose{M} \models M$. Let $\I =
  (\U,\astore,\aheap)$ be a model of $M$. Then for each two variables
  $x,x' \in \fv{M}$ such that $x \pto (y_1,\ldots,y_k), x' \pto
  (z_1,\ldots,z_k) \in M$ and $x \teq_M x'$, we have
  $\aheap(\astore(x)) = (\astore(y_1), \ldots, \astore(y_k))$,
  $\aheap(\astore(x')) = (\astore(y_1), \ldots, \astore(y_k))$ and
  $\aheap(\astore(x)) = \aheap(\astore(x'))$, thus $\astore(y_i) =
  \astore(z_i)$, for all $i \in [1,k]$, thus $\I \models
  \pclose{M}$. For a variable $x \in \avar{M}$, let $x_1 \not\pto
  \vec{y}_1, \ldots, x_n \not\pto \vec{y}_n \in M$ be all literals
  such that $x_1 \teq_M \ldots \teq_M x_n \teq_M x$ and $\vec{y}_i
  \not\teq_M \vec{y}_j$ for all $i \neq j$. Then $\aheap(\astore(x))
  \in \U^k \setminus
  \set{\astore(\vec{y}_1),\ldots,\astore(\vec{y}_n)}$, thus
  $\card{\U}^k \geq n+1 = \datano{x}{M}+1$. Since this holds for each
  $x \in \avar{M}$, we have $\I \models \len{U} \geq
  \left\lceil\sqrt[k]{\max_{x \in \avar{M}}
    (\datano{x}{M}+1)}\right\rceil$. Further, if $\I \models M$ and
  $\len{h}\geq n_1, \len{h} < \len{U}-n_2 \in M$ then $\card{\U} - n_2
  > \card{\aheap} \geq n_1$, thus $\card{U} \geq n_1+n_2+1$ and $\I
  \models \len{U} \geq n_1+n_2+1$. Analogously, we obtain $\I \models
  \len{U} < n_1+n_2$ in the case $\len{h} < n_1, \len{h} \geq
  \len{U}-n_2 \in M$. \qed}

\prop{prop:domain-closure}{
 Given a minterm $M$, $\minheap_M^\I < \maxheap_M^\I$ for any model
 $\I$ of $\uclose{M}^u$.
}
\proof{ Let $\I = (\U,\astore,\aheap)$ and $n_1, n_2 \in
  \nat_\infty$. We distinguish the following cases:\begin{compactitem}
    \item if $\minheap_M = n_1$ and $\maxheap_M = n_2$ then $n_1 \geq
      n_2$ must be the case, or else $\uclose{M} \equiv \bot$, in
      contradiction with $\I \models \uclose{M}^u$.
    \item if $\minheap_M = n_1$ and $\maxheap_M = \len{U} - n_2$ then
      $\len{U} \geq n_1+n_2+1 \in \uclose{M}$ and since $\I \models
      \uclose{M}^u$, we obtain $n_1 < \card{\U} - n_2$.
    \item if $\minheap_M = \len{U}-n_1$ and $\maxheap_M = n_2$ then
      $\len{U} < n_1+n_2 \in \uclose{M}$ and since $\I \models
      \uclose{M^u}$, we obtain $\card{\U} - n_1 < n_2$.
    \item if $\minheap_M = \len{U}-n_1$ and $\maxheap_M = \len{U}-n_2$
      then $n_2 < n_1$ must be the case, or else $\uclose{M} \equiv
      \bot$, in contradiction with $\I \models \uclose{M}^u$. \qed
  \end{compactitem}
}

\begin{definition}
A minterm $M$ is \emph{footprint-consistent} if for all $x,x' \in
\vars$ and $\vec{y}, \vec{y'} \in \vars^k$, such that $x \teq_M x'$
and $y_i \teq_M y'_i$ for all $i \in [1,k]$, we have \begin{inparaenum}[(1)]
\item if $\alloc(x) \in M$ then $\neg\alloc(x') \not\in M$, and
\item if $x \pto \vec{y} \in M$ then $\neg\alloc(x'), \neg x' \pto
  \vec{y'} \not\in M$.
\end{inparaenum}
\end{definition}
Note that footprint-consistency is a necessary, yet not sufficient,
condition for satisfiability of minterms. For example, the minterm
$M=\set{x \pto y,x'\pto y',\neg y\teq y', \len{h} < 2}$ is at the same
time footprint-consistent and unsatisfiable.

\prop{prop:footprint-consistent}{
 If $M$ is a footprint-consistent minterm, then $\nvar{M} \cap
 \avar{M} = \emptyset$. If, moreover, $M$ is E-complete for $\fv{M}$,
 then $\astore(X) \cap \astore(\avar{M}) = \emptyset$ for each set $X$
 disjoint from $\avar{M}$ and each model $(\U,\astore,\aheap)$ of $M$.
}
\proof{ Suppose first that $x \in \nvar{M} \cap \avar{M}$. Then there
  exists literals $\neg\alloc(x')$ and $\alloc(x'')$ in $M$ such that
  $x \teq_M x'$ and $x \teq_M x''$, which contradicts the footprint
  consistency of $M$. For the second point, suppose that $\ell \in
  \astore(X) \cap \astore(\avar{M})$. Then there exists variables $x
  \in X$ and $x' \in \avar{M}$ such that $\astore(x) = \astore(x') =
  \ell$. If $M$ is E-complete, either $x \teq x' \in M$ or $\neg x
  \teq x' \in M$. The first case contradicts $x \not\in \avar{M}$ and
  the second case contradicts $(\U,\astore,\aheap) \models M$. \qed}

We are now ready to define a boolean combination of test formulae that
is equivalent to $M_1 * M_2$, where $M_1$ and $M_2$ are minterms
satisfying a number of additional conditions. Let \(\negpto{M_1}{M_2}
\defequal (M_1 \cap M_2) \cap \{\neg x \pto \vec{y} \mid x \not\in
\avar{M_1 \cup M_2}, \vec{y} \in \vars^k\}\) be the set of negative
points-to literals common to $M_1$ and $M_2$, involving left-hand side
variables not allocated in either $M_1$ or $M_2$.

\begin{lemma}\label{lemma:star-elim}
  Let $M_1$ and $M_2$ be two minterms that are footprint-consistent
  and E-complete for $\fv{M_1 \cup M_2}$, with $\cclose{M_1^p}
  = \cclose{M_2^p}$. Then $M_1 * M_2 \equiv \elim_*(M_1,M_2)$, where
  $\elim_*(M_1,M_2)$ is:
  \begin{eqnarray}
    && %M_1^f \wedge M_2^f \wedge 
    M_1^e \wedge M_2^e \wedge \uclose{M_1}^u \wedge \uclose{M_2}^u \wedge \label{eq:star-elim-pure} \\
    && \bigwedge_{
        \scriptstyle{x \in \avar{M_1}},~ 
        \scriptstyle{y \in \avar{M_2}}
    } \!\!\!\!\!\!\!\!\!\!\!\!\neg x \teq y \wedge \afootprint{M_1} \wedge \afootprint{M_2} \wedge 
    \label{eq:star-elim-footprint} \\
    && \nalloc(\nvar{M_1} \cap \nvar{M_2}) \wedge \negpto{M_1}{M_2}
    \wedge \label{eq:star-elim-negfp} \\
    && \len{h} \geq \minheap_{M_1} + \minheap_{M_2} ~\wedge~ \len{h} < \maxheap_{M_1} + \maxheap_{M_2} - 1 \label{eq:star-elim-heapsize} \\
    && \wedge~ \eta_{12} \wedge \eta_{21} \label{eq:star-elim-eta}
  \end{eqnarray}
  where $\eta_{ij} \defequal$ \[\bigwedge_{Y \subseteq \nvar{M_j} \setminus \avar{M_i}} 
  \!\!\!\!\!\!\!\! \alloc(Y) \rightarrow 
    \left(\begin{array}{l} \len{h} \geq \allocno{M_i} + \len{Y}_{M_i} + \minheap_{M_j} \\
    \wedge~ \allocno{M_i} + \len{Y}_{M_i} < \maxheap_{M_i}\end{array}\right)\]
\end{lemma}
Intuitively, if $M_1$ and $M_2$ hold separately, then all
heap-indepen\-dent literals from $M_1 \cup M_2$ must be satisfied
(\ref{eq:star-elim-pure}), the variables allocated in $M_1$ and $M_2$
must be pairwise distinct and their footprints, relative to the
allocated variables, jointly asserted
(\ref{eq:star-elim-footprint}). Moreover, unallocated variables on
both sides must not be allocated and common negative points-to
literals must be asserted (\ref{eq:star-elim-negfp}). Since the heap
satisfying $\elim_*(M_1,M_2)$ is the disjoint union of the heaps for
$M_1$ and $M_2$, its bounds are the sum of the bounds on both sides
(\ref{eq:star-elim-heapsize}) and, moreover, the variables that $M_2$
never allocates [$\nvar{M_2}$] may occur allocated in the heap of
$M_1$ and viceversa, thus the constraints $\eta_{12}$ and $\eta_{21}$,
respectively (\ref{eq:star-elim-eta}).

\proof{ 
  Suppose first that $M_1^e \neq M_2^e$. Since $M_1$ and $M_2$ are
  E-complete for $\fv{M_1 \cup M_2}$, there must exist a literal $x
  \teq y \in M_1^e$ such that $\neg x \teq y \in M_2^e$, or
  viceversa. In both cases however $M_1 * M_2 \equiv \elim_*(M_1,M_2)
  \equiv \bot$. Thus we consider from now on that $M_1^e = M_2^e$.
    
  \noindent
  $\boxed{M_1 * M_2 \models \elim_*(M_1,M_2)}$ Let $\I = (\U, \astore,
  \aheap)$ be a model of $M_1 * M_2$. Then there exists disjoint heaps
  $\aheap_1$ and $\aheap_2$ such that $\aheap = \aheap_1 \uplus
  \aheap_2$ and $(\U, \astore, \aheap_i) \models M_i$, for all $i =
  1,2$. Below we show that $\I$ is a model of the formulae
  (\ref{eq:star-elim-pure}), (\ref{eq:star-elim-footprint}),
  (\ref{eq:star-elim-negfp}), (\ref{eq:star-elim-heapsize}) and
  (\ref{eq:star-elim-eta}).

 (\ref{eq:star-elim-pure}) Since $(\U, \astore, \aheap_i) \models
%  M_i^f \wedge 
  M_i^e \wedge M_i^u$, by Proposition
  \ref{prop:heap-indep}, we also have $(\U, \astore, \aheap) \models
%  M_i^f \wedge 
  M_i^e \wedge M_i^u$, for $i = 1,2$.  By Proposition
  \ref{prop:closure-equiv}, we obtain further that $(\U, \astore,
  \aheap) \models \uclose{M_i}^u$, for $i = 1,2$.

(\ref{eq:star-elim-footprint}) Since $\dom(\aheap_1) \cap \dom(\aheap_2)
  = \emptyset$, for every $x \in \avar{M_1}$ and $y \in
  \avar{M_2}$, we must have $\astore(x) \neq \astore(y)$, hence $\I
  \models \neg x \teq y$. Further, we have $(\U,\astore,\aheap_i)
  \models M_i$, thus $(\U,\astore,\aheap_i) \models \afootprint{M_i}$
  and, by Proposition \ref{prop:footprint-extension},
  $(\U,\astore,\aheap) \models \afootprint{M_i}$, for all $i=1,2$.

  (\ref{eq:star-elim-negfp}) Let $x \in \nvar{M_1} \cap \nvar{M_2}$ be
  a variable. Then there exists variables $x_1$ and $x_2$ such that
  $\neg\alloc(x_1) \in M_1$, $x \teq_{M_1} x_1$, $\neg\alloc(x_2) \in
  M_2$ and $x \teq_{M_2} x_2$. Hence $\astore(x) = \astore(x_1)
  \not\in \dom(\aheap_1)$ and $\astore(x) = \astore(x_2) \not\in
  \dom(\aheap_2)$, thus $\astore(x) \not\in \dom(\aheap)$ and
  $(\U,\astore,\aheap) \models \neg\alloc(x)$. Since $x$ was chosen
  arbitrarily, we have $(\U,\astore,\aheap) \models \nalloc(\nvar{M_1}
  \cap \nvar{M_2})$. Secondly, let $\neg x \pto \vec{y} \in M_1 \cap
  M_2$, for some $x \not\in \avar{M_1 \cup M_2}$. Since
  $\dom(\aheap_1) \cap \dom(\aheap_2) = \emptyset$, only the following
  are possible: \begin{compactenum}
  \item $\astore(x) \in \dom(\aheap_1)$. Since $(\U,\astore,\aheap_1)
    \models M_1$, we must have $\aheap_1(x) \neq
    \astore(\vec{y})$. Then $\aheap(x) \neq \astore(\vec{y})$ thus
    $(\U,\astore,\aheap) \models \neg x \pto \vec{y}$.
  \item $\astore(x) \in \dom(\aheap_2)$ and $\aheap_2(x) \neq
    \astore(\vec{y})$ is symmetrical.
  \item $\astore(x) \not\in \dom(\aheap_1) \cup \dom(\aheap_2)$, then
    $\astore(x) \not\in \dom(\aheap)$ and $(\U,\astore,\aheap) \models
    \neg x \pto \vec{y}$.
  \end{compactenum}
  Since $\neg x \pto \vec{y} \in \negpto{M_1}{M_2}$ was chosen
  arbitrarily, $(\U,\astore,\aheap) \models \negpto{M_1}{M_2}$.

  (\ref{eq:star-elim-heapsize}) Since $\aheap = \aheap_1 \uplus
  \aheap_2$, we have $\card{\aheap} = \card{\aheap_1} + \card{\aheap_2}$,
  thus the first two constraints are obtained by summing up the
  constraints $\min^\I_{M_i} \leq \card{\aheap_i} < \max^\I_{M_i}$, for
  $i=1,2$. 

  (\ref{eq:star-elim-eta}) We prove $\I \models \eta_{12}$, the proof
  for $\I \models \eta_{21}$ being symmetrical. Consider a set $Y
  \subseteq \nvar{M_2} \setminus \avar{M_1}$ and suppose that
  $(\U,\astore,\aheap) \models \alloc(Y)$. For each $y \in Y$ we must
  have $\astore(y) \in \dom(\aheap_1)$, because $\astore(y) \not\in
  \dom(\aheap_2)$ and $\astore(y) \in \dom(\aheap)$.  Moreover,
  $\astore(Y) \cap \astore(\avar{M_1}) = \emptyset$ because $Y \cap
  \avar{M_1} = \emptyset$ and $M_1$ is E-complete for $\fv{M_1 \cup
    M_2}$, by Proposition \ref{prop:footprint-consistent}. Thus
  $\allocno{M_1} + \len{Y}_{M_1} \leq \card{\aheap_1} <
  \maxheap^\I_{M_1}$ and $\card{\aheap} = \card{\aheap_1} +
  \card{\aheap_2} \geq \allocno{M_1} + \len{Y}_{M_1} +
  \minheap^\I_{M_2}$, as required.
 
  \noindent
  $\boxed{\elim_*(M_1,M_2) \models M_1 * M_2}$ Let $\I =
  (\U,\astore,\aheap)$ be a model of $\elim_*(M_1,M_2)$. We shall find
  $\aheap_1$ and $\aheap_2$ such that $\aheap = \aheap_1 \uplus
  \aheap_2$ and $(\U,\astore,\aheap_i) \models M_i$, for all $i=1,2$.
  Since $\I \models \minheap_{M_1} + \minheap_{M_2} \leq \len{h}
  \wedge \len{h} < \maxheap_{M_1} + \maxheap_{M_2} - 1$ by
  (\ref{eq:star-elim-heapsize}), we have, by Proposition
  \ref{prop:test-formulae}:
  \begin{equation}
    \minheap^\I_{M_1} + \minheap^\I_{M_2} \leq \card{\aheap} < \maxheap^\I_{M_1} + \maxheap^\I_{M_2} - 1
    \label{eq:heap-min-max}
  \end{equation}
  Let us now define the following sets, for $i=1,2$: 
  \[\begin{array}{rcl}
  L_i & = & \set{\astore(x) \in \dom(\aheap) \mid x \in \nvar{M_{3-i}} \setminus \avar{M_i}} \\
  Y_i & = & \set{x \in \vars \mid \astore(x) \in L_i} \\
  A_i & = & \set{\astore(x) \mid x \in \avar{M_i}}
  \end{array}\]
  First, we prove that $L_1 \cap L_2 = \emptyset$. By contradiction,
  suppose that there exists $\ell \in L_1 \cap L_2$. Then $\ell =
  \astore(y_1) = \astore(y_2)$ for some $y_1 \in \nvar{M_1}$ and $y_2
  \in \nvar{M_2}$. Because $M_1$ is E-complete for $\fv{M_1 \cup
    M_2}$, exactly one of $y_1 \teq y_2$, $\neg y_1 \teq y_2$ belongs
  to $M_1$. But $\neg y_1 \teq y_2 \in M_1$ contradicts with
  $\astore(y_1) = \astore(y_2)$ and $y_1 \teq y_2 \in M_1$ leads to
  $y_2 \in \nvar{M_1}$. Symmetrically, $y_1 \in \nvar{M_2}$, thus
  $y_1,y_2 \in \nalloc(\nvar{M_1} \cap \nvar{M_2})$. Since
  $(\U,\astore,\aheap) \models \nalloc(\nvar{M_1} \cap \nvar{M_2})$ by
  (\ref{eq:star-elim-negfp}), we have $\ell \not\in \dom(\aheap)$,
  which contradicts with the fact that $L_1 \cup L_2 \subseteq
  \dom(\aheap)$, according to the definition of $L_1$ and $L_2$.

  Next, we show that $L_i \cap (A_1 \cup A_2) = \emptyset$, for $i =
  1,2$. First, $L_i \cap A_i = \emptyset$ because $M_i$ are E-complete
  for $\fv{M_1 \cup M_2}$, by Proposition
  \ref{prop:footprint-consistent}. Second, $L_i \cap A_{3-i} =
  \emptyset$ because $M_i$ are E-complete for $\fv{M_1 \cup M_2}$
  and $\nvar{M_{3-i}} \cap \avar{M_{3-i}} = \emptyset$, by Proposition
  \ref{prop:footprint-consistent}. 

  Moreover $\I \models \alloc(Y_1) \wedge \alloc(Y_2)$ because $L_1
  \cup L_2 \subseteq \dom(\aheap)$ by definition and, because
  $(\U,\astore,\aheap) \models \eta_{12} \wedge \eta_{21}$, the
  following hold, for $i=1,2$:
  \begin{center}
    \begin{minipage}{6cm}
      \begin{equation}
        \card{\aheap} \geq \card{A_i} + \card{L_i} + \minheap^\I_{M_{3-i}} \label{eq:lambda-min}
      \end{equation}
    \end{minipage}
    \begin{minipage}{6cm}
      \begin{equation}
        \card{A_i} + \card{L_i} < \maxheap^I_i \label{eq:lambda-max}
      \end{equation}
    \end{minipage}
  \end{center}
  We prove the following relation by distinguishing the cases below:
  \begin{equation}\label{eq:max-max}
    \max(\minheap^\I_{M_1},\card{A_1} + \card{L_1}) +
    \max(\minheap^\I_{M_2},\card{A_2} + \card{L_2}) \leq \card{\aheap}
  \end{equation}
  \begin{compactenum}
  \item if $\minheap^\I_{M_1} \geq \card{A_1} + \card{L_1}$ then we
    have \(\minheap^\I_{M_1} + \max(\minheap^\I_{M_2},\card{A_2} +
    \card{L_2}) \leq \card{\aheap}\) by (\ref{eq:star-elim-heapsize})
    and (\ref{eq:lambda-min}). The case $\minheap^\I_{M_2} \geq
    \card{A_2} + \card{L_2}$ is symmetric, and
  \item otherwise, if $\minheap^\I_{M_1} < \card{A_1} + \card{L_1}$
    and $\minheap^\I_{M_2} < \card{A_2} + \card{L_2}$, because $\I
    \models \bigwedge_{x \in \avar{M_1},~ y \in \avar{M_2}} \neg
    x \teq y$, the sets of locations $L_1$, $L_2$, $A_1$ and $A_2$ are
    pairwise disjoint and, since $L_1 \cup L_2 \cup A_1 \cup A_2
    \subseteq \dom(\aheap)$, it must be the case that
    $\card{\aheap} \geq \card{A_1} + \card{L_1} + \card{A_2} +
    \card{L_2}$.
  \end{compactenum}
  \begin{figure}[thb]
    \vspace*{-\baselineskip}
    \centerline{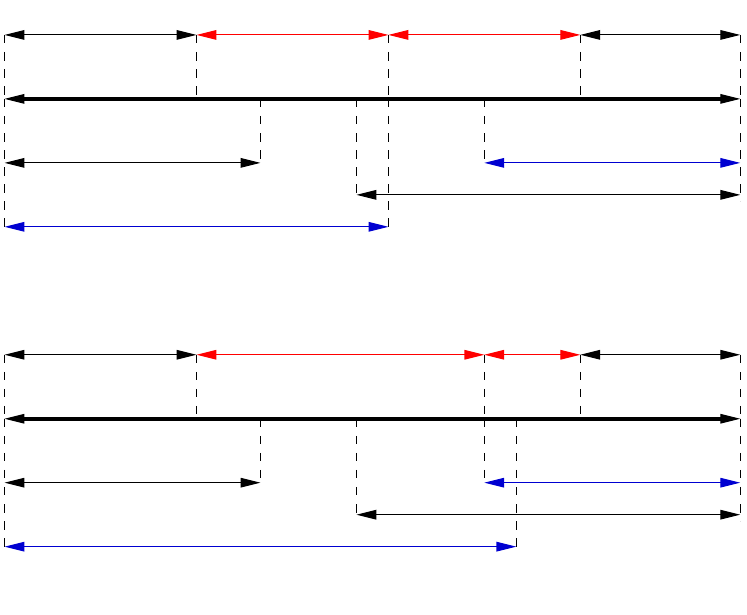}
    \caption{}\label{fig:heap-split}
    \vspace*{-\baselineskip}
  \end{figure}  
  Furthermore, we have $\card{\aheap} < \maxheap^\I_{M_1} +
  \maxheap^\I_{M_2} - 1$ by (\ref{eq:heap-min-max}) and one of the
  following cases occurs (see Fig. \ref{fig:heap-split}): \begin{compactenum}
    \item if $\maxheap^\I_{M_1} - 1 \leq \card{\aheap} -
      \max(\card{A_2} + \card{L_2}, \minheap^\I_{M_2})$ then let $n_1
      = \maxheap^\I_{M_1} - \card{A_1} - \card{L_1} - 1$ and $n_2 =
      \card{\aheap} - \maxheap^\I_{M_1} - \card{A_2} - \card{L_2} + 1$
      (Fig. \ref{fig:heap-split} (a)). We have that $n_1 \geq 0$ by
      (\ref{eq:lambda-max}) and $n_2 \geq 0$ by the hypothesis
      $\maxheap^\I_{M_1} - 1 \leq \card{\aheap} - \max(\card{A_2} +
      \card{L_2}, \minheap^\I_{M_2})$.
    \item otherwise, let $n_1 = \card{\aheap} - \card{A_1} - \card{L_1}
      - \maxheap^\I(\card{A_2} + \card{L_2},\minheap^\I_{M_2})$ and
      $n_2 = \maxheap^\I(\card{A_2} + \card{L_2},\minheap^\I_{M_2}) -
      \card{A_2} - \card{L_2}$ (Fig. \ref{fig:heap-split} (b)). We
      have $n_1 \geq 0$ by (\ref{eq:max-max}) and $n_2 \geq 0$ is
      immediate.
  \end{compactenum}
  In both cases, the following holds, for $i = 1,2$:
  \begin{equation}\label{eq:heap-part-size}
    \minheap^\I_{M_i} \leq \card{A_i} + \card{L_i} + n_i < \maxheap^\I_{M_i}
  \end{equation}
  We have used the fact that $\minheap^\I_{M_i} < \maxheap^\I_{M_i}$,
  for $i=1,2$, which is a consequence of the fact that $\I \models
  \uclose{M_i}^u$, by (\ref{eq:star-elim-pure}) and Proposition
  \ref{prop:domain-closure}. 

  Further, we have that $\card{\aheap} = \sum_{i=1,2} \card{A_i} +
  \card{L_i} + n_i$. Moreover, there are exactly $n_1 + n_2$ locations
  in $\dom(\aheap) \setminus (A_1 \cup L_1 \cup A_2 \cup L_2)$, thus
  we can partition this set into $N_1$ and $N_2$ such that $\card{N_i}
  = n_i$ and define $\aheap_i$ to be the restriction of $\aheap$ to
  $A_i \cup L_i \cup N_i$, for $i = 1,2$. It remains to be shown that
  $(\U,\astore,\aheap_i) \models M_i$, for $i=1,2$. Below we do the
  proof for $i=1$, the case $i=2$ being symmetric.

  Clearly, $(\U,\astore,\aheap_1) \models %M_1^f \wedge M_1^e \wedge
  M_1^u$, because $(\U,\astore,\aheap) \models %M_1^f \wedge M_1^e
  \wedge \uclose{M_1}^u$, by Proposition
  \ref{prop:heap-indep}. Further, by (\ref{eq:heap-part-size}) and
  Proposition \ref{prop:test-formulae}, we have $(\U,\astore,\aheap_1)
  \models \len{h} \geq \minheap_{M_1} \wedge \len{h} <
  \maxheap_{M_1}$. It remains to show that $(\U,\astore,\aheap_1)
  \models M_1^a \wedge M_1^p$.

  ($M_1^a$) Let $\alloc(x) \in M_1^a$ be a literal. Then $x \in
  \avar{M_1}$, thus $\astore(x) \in A_1$ and $(\U,\astore,\aheap_1)
  \models \alloc(x)$ follows, by the definition of $\aheap_1$. Dually,
  let $\neg\alloc(x) \in M_1^a$ be a literal. Then, we have $x \in
  \nvar{M_1}$. We distinguish the cases: \begin{compactitem}
  \item if $x \in \avar{M_2}$ then $\astore(x) \in A_2$ and since
    $\dom(\aheap_1) \cap A_2 = \emptyset$, we have $\astore(x) \not\in
    \dom(\aheap_1)$, thus $(\U,\astore,\aheap_1) \models
    \neg\alloc(x)$.
  \item otherwise, $x \in \nvar{M_1} \setminus
    \avar{M_2}$. Again, we distinguish the cases: \begin{compactitem}
    \item if $x \in Y_2$ then $\astore(x) \in L_2$ and because
      $\dom(\aheap_1) \cap L_2 = \emptyset$, we obtain $\astore(x)
      \not\in \dom(\aheap_1)$, thus $(\U,\astore,\aheap_1) \models
      \neg\alloc(x)$.
    \item otherwise, $x \not\in Y_2$, thus $\astore(x) \not\in
      L_2$. But since $x \in \nvar{M_1} \setminus \avar{M_2}$,
      by the definition of $L_2$, it must be the case that $\astore(x)
      \not\in \dom(\aheap)$, thus $(\U,\astore,\aheap) \models
      \neg\alloc(x)$ and $(\U,\astore,\aheap_1) \models \neg\alloc(x)$
      follows.
    \end{compactitem}
  \end{compactitem} 

  ($M_1^p$) Let $x \pto \vec{y} \in M_1^p$ be a literal. Then $x \in
  \avar{M_1}$ and $\astore(x) \in A_1$. Moreover, we have $x \pto
  \vec{y} \in \afootprint{M_1}$, thus $(\U,\astore,\aheap) \models x
  \pto \vec{y}$, by (\ref{eq:star-elim-footprint}). Since $\aheap$ and
  $\aheap_1$ agree on $A_1$, we also have $(\U,\astore,\aheap_1)
  \models x \pto \vec{y}$. Dually, let $\neg x \pto \vec{y} \in
  M_1^p$. If $x \in \avar{M_1}$ then $\neg x \pto \vec{y} \in
  \afootprint{M_1}$, thus $(\U,\astore,\aheap_1) \models \neg x \pto
  \vec{y}$, since $\aheap$ and $\aheap_1$ agree on $A_1$. Otherwise,
  if $x \not\in \avar{M_1}$, we distinguish the
  cases: \begin{compactitem}
  \item if $x \in \avar{M_2}$ then $\astore(x) \in A_2$, and since
    $\dom(\aheap_1) \cap A_2 = \emptyset$, we have $\astore(x) \not\in
    \dom(\aheap_1)$, thus $(\U,\astore,\aheap_1) \models \neg x \pto
    \vec{y}$. 
  \item otherwise, $x \not\in \avar{M_2}$, and since $\cclose{M_1^p} =
    \cclose{M_2^p}$, we have $\set{x \pto \vec{y}, \neg x \pto
      \vec{y}} \cap M_2 \neq \emptyset$. Since $x \not\in
    \avar{M_2}$, the only possibility is $\neg x \pto \vec{y} \in M_2$,
    thus $\neg x \pto \vec{y} \in \negpto{M_1}{M_2}$ and
    $(\U,\astore,\aheap) \models \neg x \pto \vec{y}$, by
    (\ref{eq:star-elim-negfp}). Since $\aheap$ is an extension of
    $\aheap_1$, we obtain that $(\U,\astore,\aheap_1) \models \neg x
    \pto \vec{y}$ as well. \qed
  \end{compactitem}}

\begin{remark}\label{rem:star-elim-alloc}
Note that $\elim_*(M_1,M_2)$ introduces literals of the form
$\neg\alloc(x)$ that do not occur in $M_1 \cup M_2$. For example, if
$M_1 = \set{\neg\alloc(x), x \teq y, \len{h} \geq 0, \len{h} <
  \infty}$ and $M_2 = \set{\len{h} \geq 0, \len{h} < \infty}$, then $y
\in \nvar{M_1}$ and $\alloc(y)$ occurs at negative polarity in
$\elim_*(M_1,M_2)$. This is problematic because upcoming results
depend on the fact that the polarity of $\alloc(x)$ formulae is
preserved (Lemma \ref{lemma:polarity}). Moreover, if $\neg \alloc(x)
\in \elim_*(M_1,M_2)$, then there exists a literal $\neg\alloc(x') \in
M_1 \cup M_2$, such that $\elim_*(M_1,M_2) \models x \teq x'$, making
$\neg \alloc(x)$ actually redundant. Consequently, equivalence is
preserved when only the literals $\neg\alloc(x) \in M_1 \cup M_2$ are
included in $\elim_*(M_1,M_2)$. This refined version of
$\elim_*(M_1,M_2)$ is used in the proof of Lemma
\ref{lemma:polarity}. However, taking this observation into account at
this point would clutter the definition of $\elim_*(M_1,M_2)$.
\hfill$\blacksquare$
\end{remark}

Next, we prove a similar result for the separating implication. For
technical convenience, we translate the septraction $M_1 \septraction
M_2$, instead of $M_1 \wand M_2$, as an equivalent boolean combination
of test formulae. This is without loss of generality, because $M_1
\wand M_2 \equiv \neg(M_1 \septraction \neg M_2)$. Unlike with the
case of the separating conjuction (Lemma \ref{lemma:star-elim}), here
the definition of the boolean combination of test formulae depends on
whether the universe is finite or infinite. 

If the complement of some literal $\ell \in \afootprint{M_1}$ belongs
to $M_2$ then no extension by a heap that satisfies $\ell$ may satisfy
$\neg\ell$. Therefore, as an additional simplifying assumption, we
suppose that \(\afootprint{M_1} \cap \overline{M_2} = \emptyset\),
so that $M_1 \septraction M_2$ is not trivially unsatisfiable.

\begin{lemma}\label{lemma:septraction-elim}
  Let $M_1$ and $M_2$ be two footprint-consistent minterms that are
  E-complete for $\fv{M_1 \cup M_2}$, such
  that \begin{inparaenum}[(a)]
  \item $M_1$ is A-complete for $\fv{M_1 \cup M_2}$,
  \item $M_2^a \cup M_2^p \subseteq \cclose{M_1^a \cup M_1^p}$, and 
  \item $\afootprint{M_1} \cap \overline{M_2} = \emptyset$\label{eq:footprint-const}.
  \end{inparaenum}
  Then, we have:
  \[\begin{array}{rcl}
  M_1 \septraction M_2 & \equivfin & \elim^\fincard_\septraction(M_1,M_2) \\ 
  M_1 \septraction M_2 & \equivinf & \elim^\infcard_\septraction(M_1,M_2)
  \end{array}\]
  where $\elim^\gencard_\septraction(M_1,M_2)$, for $\gencard \in
  \set{\fincard,\infcard}$ is defined as:
  \begin{eqnarray}
    && %M_1^f \wedge M_2^f \wedge 
    {\pclose{M_1}}^e \wedge M_2^e \wedge {\uclose{M_1}}^u \wedge {\uclose{M_2}}^u \wedge 
    \label{eq:septraction-elim-pure} \\
    && \nalloc(\avar{M_1}) \wedge \footprint{M_2}{\nvar{M_1}} \wedge 
    \label{eq:septraction-elim-footprint} \\
    && \len{h} \geq \minheap_{M_2} - \maxheap_{M_1} + 1 \wedge 
    \len{h} < \maxheap_{M_2} - \minheap_{M_1} \label{eq:septraction-elim-heapsize} \\
    && \wedge~ \lambda^\gencard \label{eq:septraction-elim-lambda}
    \end{eqnarray}
  where \(\lambda^\infcard \defequal \top\)  and \(\lambda^\fincard \defequal\)
  \[\bigwedge_{\scriptstyle{Y \subseteq \fv{M_1 \cup M_2}}} \!\!\!\! \!\!\!\!
  \nalloc(Y) \rightarrow \left(\begin{array}{l}
  \len{h} < \len{U} - \minheap_{M_1} - \nallocno{Y}{M_1} + 1 \\
  \wedge~ \len{U} \geq \minheap_{M_2} + \nallocno{Y}{M_1}\end{array}\right)\]
\end{lemma}
Observe that a heap satisfies $M_1 \septraction M_2$ iff it has an
extension, by a disjoint heap satisfying $M_1$, that satisfies $M_2$.
Thus, $\elim^\gencard_\septraction(M_1,M_2)$ must entail the
heap-independent literals of both $M_1$ and $M_2$
(\ref{eq:septraction-elim-pure}). Next, no variable allocated by $M_1$
must be allocated by $\elim^\gencard_\septraction(M_1,M_2)$, otherwise
no extension with a heap satisfying $M_1$ is possible and, moreover,
the footprint of $M_2$ relative to the unallocated variables of $M_1$
must be asserted (\ref{eq:septraction-elim-footprint}). The heap's
cardinality constraints depend on the bounds of $M_1$ and $M_2$
(\ref{eq:septraction-elim-heapsize}) and, if $Y$ is a set of variables
not allocated in the heap, these variables might occur allocated in
the extension (\ref{eq:septraction-elim-lambda}).

Actually, this is where the finite universe assumption first comes
into play. If the universe is infinite, then there are enough many
locations outside the heap to be assigned to $Y$. However, if the
universe is finite, then it is necessary to ensure that there are at
least $\nallocno{Y}{M_1}$ free locations to be assigned to $Y$
(\ref{eq:septraction-elim-lambda}).  Note that
$\finelim_\septraction(M_1,M_2)$ introduces positive $\alloc(x)$ test
formulae (remark \ref{rem:test-fol-alloc}), in the definition of
$\lambda^\fincard$ (\ref{eq:septraction-elim-lambda}). These formulae
do not match the polarity of the separating implication $M_1 \wand M_2
= \neg(M_1 \septraction \neg M_2)$.

\proof{ 
  If $\pclose{M_1} = \bot$ then $M_1 \septraction M_2 \equiv
  \elim_{\septraction}(M_1,M_2) \equiv \bot$. Also, since $M_1$ and
  $M_2$ are E-complete for $\fv{M_1 \cup M_2}$, if we suppose that
  $M_1^e \neq M_2^e$ then $M_1 \septraction M_2 \equiv
  \elim_{\septraction}(M_1,M_2) \equiv \bot$. From now on, we shall
  assume that $\pclose{M_1} = M_1$ and $M_1^e = M_2^e$.

  \noindent
  $\boxed{M_1 \septraction M_2 \models \elim_{\septraction}(M_1,M_2)}$
  Let $\I = (\U,\astore,\aheap)$ be a structure such that $\I \models
  M_1 \septraction M_2$. Then there exists a heap $\aheap'$ disjoint
  from $\aheap$ such that $(\U,\astore,\aheap') \models M_1$ and
  $(\U,\astore,\aheap \uplus \aheap') \models M_2$. Below we prove
  that $\I$ is also a model of the formulae
  (\ref{eq:septraction-elim-pure}),
  (\ref{eq:septraction-elim-footprint}),
  (\ref{eq:septraction-elim-heapsize}) and
  (\ref{eq:septraction-elim-lambda}), respectively.

  (\ref{eq:septraction-elim-pure}) We have $(\U,\astore,\aheap')
  \models %M_1^f \wedge 
  M_1^e \wedge M_1^u$, thus $(\U,\astore,\aheap)
  \models %M_1^f \wedge 
  M_1^e \wedge M_1^u$ by Proposition
  \ref{prop:heap-indep}, and by Proposition \ref{prop:closure-equiv},
  we deduce that $(\U,\astore,\aheap) \models %M_1^f \wedge
  \pclose{M_1}^e \wedge \uclose{M_1}^u$. Analogously,
  $(\U,\astore,\aheap) \models %M_2^f \wedge 
  M_2^e \wedge \uclose{M_2}^u$
  follows from $(\U,\astore,\aheap \uplus \aheap') \models M_2$ 
%  and
%  $(\U,\astore,\aheap) \models \uclose{M_2}^u$, 
  by Propositions
  \ref{prop:heap-indep} and \ref{prop:closure-equiv}.

  (\ref{eq:septraction-elim-footprint}) Since $(\U,\astore,\aheap')
  \models M_1$, also $(\U,\astore,\aheap') \models \alloc(\avar{M_1})$
  and since $\dom(\aheap') \cap \dom(\aheap) = \emptyset$, we have
  $(\U,\astore,\aheap) \models \nalloc(\avar{M_1})$. To prove that
  $(\U,\astore,\aheap) \models \footprint{M_2}{\nvar{M_1}}$, we
  consider four cases, depending on the form of the
  literal: \begin{compactitem}
    \item If $\alloc(x) \in M_2$ and $x \in \nvar{M_1}$, then
      $\astore(x) \in \dom(\aheap) \cup \dom(\aheap')$ and $\astore(x)
      \not\in \dom(\aheap')$, thus $\astore(x) \in \dom(\aheap)$ and
      $(\U,\astore,\aheap) \models \alloc(x)$, by Proposition
      \ref{prop:test-formulae}.
    \item The case $x \pto \vec{y} \in M_2$ and $x \in \nvar{M_1}$
      uses a similar argument.
    \item If $\neg \alloc(x) \in M_2$ and $x\in \nvar{M_1}$, then
      $\astore(x) \notin \dom(\aheap\cup \aheap')$, hence
      $\astore(x)\notin\dom(\aheap)$ and $(\U, \astore, \aheap)
      \models \neg\alloc(x)$, by Proposition
      \ref{prop:test-formulae}.
    \item If $\neg x \pto \vec{y} \in M_2$ and $x \in \nvar{M_1}$ then
      $\astore(x) \not\in \dom(\aheap')$ and
      either: \begin{compactitem}
      \item $\astore(x) \not\in \dom(\aheap)$ and $(\U,\astore,\aheap)
        \models \neg x \pto \vec{y}$, by Proposition
        \ref{prop:test-formulae}, or
      \item $\astore(x) \in \dom(\aheap)$ in which case $\aheap' \uplus \aheap$
        and $\aheap$ agree on $\astore(x)$ and $(\U,\astore,\aheap)
        \models \neg x \pto \vec{y}$.
    \end{compactitem}
  \end{compactitem}

  (\ref{eq:septraction-elim-heapsize}) We have $\card{\aheap \uplus
    \aheap'} = \card{\aheap} + \card{\aheap'}$ and since
  $(\U,\astore,\aheap \uplus \aheap') \models M_2$, we obtain
  $\minheap^\I_{M_2} \leq \card{\aheap} + \card{\aheap'} <
  \maxheap^\I_{M_2}$.  Since $(\U,\astore,\aheap') \models M_1$ we
  also have $\minheap^\I_{M_1} \leq \card{\aheap'} <
  \maxheap^\I_{M_1}$, thus $\minheap^\I_{M_1}
  \leq \card{\aheap'} \leq \maxheap^\I_{M_1}-1$, i.e.,
  $-\maxheap^\I_{M_1}+1 \leq -\card{\aheap'} \leq -\minheap^\I_{M_1}$
  so that $\minheap^\I_{M_2} - \maxheap^\I_{M_1} +1 \leq \card{\aheap}
  < \maxheap^\I_{M_2} - \minheap^\I_{M_1}$. 

  (\ref{eq:septraction-elim-lambda}) Assume that $(\U,\astore,\aheap)
  \models \nalloc(Y)$ for a set $Y \subseteq \fv{M_1 \cup M_2}$, which
  implies that $\dom(\aheap) \cap \astore(Y) = \emptyset$. Since
  $(\U,\astore,\aheap') \models M_1$, we also have $\dom(\aheap') \cap
  \astore(\nvar{M_1}) = \emptyset$. Thus $\card{\U} \geq \card{\aheap}
  + \card{\aheap'} + \card{\astore(Y \cap \nvar{M_1})} \geq
  \card{\aheap} + \minheap^\I_{M_1} + \nallocno{Y}{M_1}$, because
  $\card{\aheap'} \geq \minheap^\I_{M_1}$ and $\card{\astore(Y \cap
    \nvar{M_1})} = \len{Y \cap \nvar{M_1}}_{M_1} = \nallocno{Y}{M_1}$,
  by Proposition \ref{prop:ecomp-eqclass}, since $M_1$ is
  E-complete. Therefore, $\card{\aheap} \leq \card{\U} -
  \minheap^\I_{M_1} - \nallocno{Y}{M_1}$.  Moreover, since
  $(\U,\astore,\aheap\uplus\aheap') \models M_2$, we obtain $\len{U}
  \geq \card{\aheap \uplus \aheap'} + \nallocno{Y}{M_1} \geq
  \minheap^\I_{M_2} + \nallocno{Y}{M_1}$.

  \noindent
  $\boxed{\elim_{\septraction}(M_1,M_2) \models M_1 \septraction M_2}$
  Let $\I = (\U,\astore,\aheap)$ be a structure such that $\I \models
  \elim_{\septraction}(M_1,M_2)$. We shall build a heap $\aheap'$ such
  that $\dom(\aheap) \cap \dom(\aheap') = \emptyset$,
  $(\U,\astore,\aheap') \models M_1$ and
  $(\U,\astore,\aheap\uplus\aheap') \models M_2$. First, for each
  variable $x \in \avar{M_1}$ such that $x' \pto \vec{y} \in M_1^p$
  for some variable $x'$ with $x \teq_{M_1} x'$, we add the tuple
  $(\astore(x),\astore(\vec{y}))$ to $\aheap'$. Since
  $(\U,\astore,\aheap) \models \pclose{M_1}^e$, for any pair of
  variables $x \teq_{M_1} x'$ if $x \pto \vec{y}, x' \pto \vec{y'} \in
  M_1$ then $y_i \teq_{M_1} y'_i$, and the result is a functional
  relation. We define:
  \[\begin{array}{rcl}
  A & = & \{x \in \avar{M_1} \mid \forall x' \forall \vec{y} ~.~ 
  x \teq_{M_1} x' \Rightarrow x' \pto \vec{y} \not\in M_1^p\} \\
  V_x & = & \{(\astore(y_1),\ldots,\astore(y_k)) \in \U^k \mid 
  x \teq_{M_1} x',~ \neg x' \pto \vec{y} \in M_1^p\}, \text{ for $x \in \avar{M_1}$} \\
  N & = & \{x \in \fv{M_1 \cup M_2} \mid \astore(x) \not\in \dom(\aheap)\}
  \end{array}\]
  Intuitively, $A$ denotes the set of variables that must be allocated
  but with no constraint on their image; this set is independent of
  the interpretation under consideration. The set $V_x$ denotes the
  set of images the allocated variable $x$ cannot point to, and $N$
  denotes the set of variables that are not allocated in $\aheap$.

  Then for each $x \in A$ we choose a tuple $(\ell_1,\ldots,\ell_k) \in
  \U^k \setminus V_x$ and let $\aheap'(\astore(x)) =
  (\ell_1,\ldots,\ell_k)$. Since $M_1$ is E-complete, we have $\card{V_x}
  \leq \datano{x}{M_1}$ for each $x \in A$, and such a choice is possible
  because $(\U,\astore,\aheap) \models \uclose{M_1}^u$, thus
  $\card{\U^k} \geq \datano{x}{M_1}+1$.

  Since $(\U,\astore,\aheap) \models \nalloc(N)$, if $\U$ is finite, by
  (\ref{eq:septraction-elim-heapsize}) it must be the case that:
  \begin{eqnarray}
    \card{\aheap} < \card{\U} - \minheap^{\I}_{M_1} - \nallocno{N}{M_1} + 1 \label{eq:lambda-heap} \\
    \card{\U} \geq \minheap^{\I}_{M_2} + \nallocno{N}{M_1} \label{eq:lambda-univ}
  \end{eqnarray}
  Finally, let $L \subseteq \U \setminus (\dom(\aheap) \cup
  \astore(\avar{M_1}) \cup \astore(\nvar{M_1}))$ be a finite set of
  locations of cardinality $\card{L} =
  \max(\minheap^\I_{M_1},\minheap^\I_{M_2} - \card{\aheap}) -
  \allocno{M_1}$. Choosing such a set $L$ is possible, because either $\U$
  is infinite, or $\U$ is finite, in which case:
  \[\begin{array}{rcl}
  \card{\U} & \geq & \max(\minheap^\I_{M_1} + \card{\aheap}, \minheap^\I_{M_2}) + \nallocno{N}{M_1} \text{, by (\ref{eq:lambda-heap}) and (\ref{eq:lambda-univ})} \\
  & \geq & \max(\minheap^\I_{M_1},\minheap^\I_{M_2} - \card{\aheap}) - \allocno{M_1} + \card{\aheap} + \allocno{M_1} + \nallocno{N}{M_1} \\ 
  & = & \card{L} +\card{\aheap} + \allocno{M_1} + \nallocno{N}{M_1}\\
  & \geq & \card{L} + \card{\dom(h) \cup \astore(\avar{M_1})\cup \astore(\nvar{M_1})}
  \end{array}\]
  where the last inequality is a consequence of Proposition
  \ref{prop:ecomp-eqclass}. We choose an arbitrary tuple
  $(\ell_1,\ldots,\ell_k) \in \U^k$ and let $\aheap'(\ell) =
  (\ell_1,\ldots,\ell_k)$ for all $\ell \in L$. Because $\U$ is
  non-empty, such a tuple exists. Consequently, we have $\dom(\aheap')
  = \astore(\avar{M_1}) \cup L$ and $\dom(\aheap') \cap \dom(\aheap) =
  \emptyset$ because $\astore(\avar{M_1}) \cap \dom(\aheap) =
  \emptyset$ by (\ref{eq:septraction-elim-footprint}) and $L \cap
  \dom(\aheap) = \emptyset$ by construction. We now prove:
  
  \noindent
  $\underline{(\U,\astore,\aheap') \models M_1}$. Clearly
  $(\U,\astore,\aheap) \models %M_1^f \wedge 
  M_1^e \wedge M_1^u$ by
  (\ref{eq:septraction-elim-pure}) and Proposition
  \ref{prop:closure-equiv}. To show $(\U,\astore,\aheap') \models
  M_1^a$, observe that $\astore(x) \in \dom(\aheap')$ for each $x \in
  \avar{M_1}$, hence for each literal $\alloc(x) \in M_1$ we have
  $(\U,\astore,\aheap') \models \alloc(x)$. Moreover, we have
  $\dom(\aheap') \cap \astore(\nvar{M_1}) = (\astore(\avar{M_1}) \cup
  L) \cap \astore(\nvar{M_1}) = \emptyset$, because $M_1$ is footprint
  consistent and E-complete for $\fv{M_1 \cup M_2}$, by Proposition
  \ref{prop:footprint-consistent}. Thus $(\U,\astore,\aheap') \models
  \neg\alloc(x)$ for each literal $\neg\alloc(x) \in M_1^a$. For each
  literal $x \pto \vec{y} \in M_1^p$ we have $\aheap'(\astore(x)) =
  (\astore(y_1),\ldots,\astore(y_k))$ by construction, thus
  $(\U,\astore,\aheap') \models x \pto \vec{y}$. For each literal
  $\neg x \pto \vec{y} \in M_1^p$, we distinguish two
  cases. \begin{compactitem}
    \item If $x \in \avar{M_1}$, then
      $(\astore(y_1),\ldots,\astore(y_k)) \in V_x$ hence
      $\aheap(\astore(x)) \neq (\astore(y_1),\ldots,\astore(y_k))$ by construction.
    \item If $x \not\in \avar{M_1}$, then since $M_1$ is A-complete for
      $\fv{M_1 \cup M_2}$, we have $x \in \nvar{M_1}$, thus
      $\astore(x) \not\in \dom(\aheap') = \astore(\avar{M_1}) \cup L$.
  \end{compactitem}
  
  We finally prove that $(\U,\astore,\aheap') \models \len{h} \geq
  \minheap_{M_1} \wedge \len{h} < \maxheap_{M_1}$.  Since
  $\dom(\aheap') = \astore(\avar{M_1}) \cup L$ and
  $\astore(\avar{M_1}) \cap L = \emptyset$, we have $\card{\aheap'} =
  \card{\astore(\avar{M_1})} + \card{L} = \max(\minheap^\I_{M_1},
  \minheap^\I_{M_2} - \card{\aheap})$.  If $\card{\aheap'} =
  \minheap^I_{M_1}$ then $\card{\aheap'} < \maxheap^I_{M_1}$ because
  $\I \models \uclose{M_1}^u$, which implies that $\minheap^\I_{M_1} <
  \maxheap^\I_{M_1}$, by Proposition
  \ref{prop:domain-closure}. Otherwise $\card{\aheap'} =
  \minheap^\I_{M_2} - \card{\aheap} \geq \minheap^I_{M_1}$
  and we have by
  (\ref{eq:septraction-elim-heapsize}) $\card{\aheap} \geq
  \minheap^I_{M_2} - \maxheap^I_{M_1} + 1$, thus $\card{\aheap} >
  \minheap^I_{M_2} - \maxheap^I_{M_1}$, and therefore $\card{\aheap'}
  <\maxheap^I_{M_1}$.

  \noindent
  $\underline{(\U,\astore,\aheap' \uplus \aheap) \models M_2}$. We
  have $(\U,\astore,\aheap'\uplus\aheap) \models %M_2^f \wedge 
  M_2^e
  \wedge M_2^u$ because $(\U,\astore,\aheap) \models %M_2^f \wedge
  M_2^e \wedge M_2^u$ and these formulae do not depend on the
  heap. Next, for a given variable $x$, let $\alpha_x \in \{\alloc(x),
  \neg\alloc(x), x \pto \vec{y}, \neg x \pto \vec{y} \mid \vec{y} \in
  \vars^k\} \cap M_2$ be a literal and let $\overline{\alpha}_x$
  denote its complement. If $x \in \nvar{M_1}$ then $\alpha_x \in
  \footprint{M_2}{\nvar{M_1}}$ and $(\U,\astore,\aheap) \models
  \alpha_x$ by (\ref{eq:septraction-elim-footprint}). Moreover,
  because $\aheap$ and $\aheap \uplus \aheap'$ agree on
  $\astore(\nvar{M_1})$, we obtain $(\U,\astore,\aheap\uplus\aheap')
  \models \alpha_x$. Otherwise $x \not\in \nvar{M_1}$ hence $x \in
  \avar{M_1}$ because $M_1$ is A-complete for $\fv{M_1 \cup M_2}$, and
  since $\alpha_x \in M_2^a \cup M_2^p \subseteq \cclose{M_1^a \cup
    M_1^p}$, we have $\alpha_x \in \afootprint{M_1}$, because the case
  $\overline{\alpha}_x \in \afootprint{M_1}$ is in contradiction with
  $\afootprint{M_1}\cap \overline{M_2} = \emptyset$ (condition
  (\ref{eq:footprint-const}) of the Lemma).  But then
  $(\U,\astore,\aheap') \models \alpha_x$ and
  $(\U,\astore,\aheap\uplus\aheap') \models \alpha_x$ follows, by
  Proposition \ref{prop:footprint-extension}. We have thus proved that
  $(\U,\astore,\aheap\uplus\aheap') \models M_2^a \cup M_2^p$. We are
  left with proving that $\minheap^\I_{M_2} \leq \card{\aheap} +
  \card{\aheap'} = \max(\minheap^I_{M_1} + \card{\aheap},
  \minheap^\I_{M_2}) < \maxheap^\I_{M_2}$. If $\minheap^\I_{M_1} +
  \card{\aheap} \leq \minheap^\I_{M_2}$ the result follows from the
  fact that $\I \models \uclose{M_2}^u$, which implies
  $\minheap_{M_2}^\I < \maxheap_{M_2}^\I$, by Proposition
  \ref{prop:domain-closure}. Otherwise, $\card{\aheap} +
  \card{\aheap'} = \minheap^\I_{M_1} + \card{\aheap} >
  \minheap^\I_{M_2}$ and $\card{\aheap} + \card{\aheap'} <
  \maxheap^\I_{M_2}$ follows from
  (\ref{eq:septraction-elim-heapsize}). \qed}

\subsection{Translating Quantifier-free $\seplogk{k}$ into Minterms}

We prove next that each quantifier-free $\seplogk{k}$ formula is
equivalent to a finite disjunction of minterms. Given minterms $M_1$
and $M_2$, we define:
\[\begin{array}{c}
\minconj{M_1}{M_2} \defequal \\ 
\left\{\begin{array}{l}
\set{\len{h} \geq \max(\minheap_{M_1},\minheap_{M_2})} \\
\hspace*{5mm} \text{if $\minheap_{M_1}, \minheap_{M_2} \in \nat$} \\
\left\{\begin{array}{l}
\len{h} \geq \minheap_{M_i} \wedge \len{U} < \minheap_{M_i}+m+1, \\
\len{h} \geq \minheap_{M_{3-i}} \wedge \len{U} \geq \minheap_{M_i}+m+1 
\end{array}\right\} \\ 
\hspace*{5mm} \text{if $\minheap_{M_i} \in \nat$, $\minheap_{M_{3-i}} = \len{U}-m$, $i=1,2$} \\
\{ \len{h} \geq \len{U} - \min(m_1,m_2) \} \\ 
\hspace*{5mm}\text{if $\minheap_{M_i} = \len{U}-m_i$, $i=1,2$}
\end{array}\right. 
\\\\
\maxconj{M_1}{M_2} \defequal \\ 
\left\{\begin{array}{ll}
\set{\len{h} < \min(\maxheap_{M_1},\maxheap_{M_2})} \\
\hspace*{5mm}\text{if $\maxheap_{M_1}, \maxheap_{M_2} \in \nat_{\infty}$} \\
\set{\len{h} < \maxheap_{M_i}} \\
\hspace*{5mm}\text{if $\maxheap_{M_{3-i}} = \infty, \maxheap_{M_i} = \len{U}-m$, $i=1,2$} \\
\left\{\begin{array}{c}
\len{h} < \maxheap_{M_i} \wedge \len{U} \geq \maxheap_{M_i} + m, \\
\len{h} < \len{U}-m \wedge \len{U} < \maxheap_{M_i} + m
\end{array}\right\} \\ 
\hspace*{5mm}\text{if $\maxheap_{M_i} \in \nat$, $\maxheap_{M_{3-i}} = \len{U}-m$, $i=1,2$} \\
\{ \len{h} < \len{U}-\max(m_1,m_2) \} \\
\hspace*{5mm}\text{if $\maxheap_{M_i} = \len{U}-m_i$, $i=1,2$}
\end{array}\right.
\end{array}\]
Intuitively, we merge the cardinality constraints occurring in $M_1$
and $M_2$, by taking the conjunction and keeping only the most
restrictive bounds.  For instance, if $M_1 = \{ \len{h} \geq 2,
\len{h} < \len{U}-1 \}$ and $M_2 = \{ \len{h} \geq 3, \len{h} <
\len{U}-2 \}$, then $\minconj{M_1}{M_2} = \{ \len{h} \geq 3 \}$ and
$\maxconj{M_1}{M_2} = \{ \len{h} < \len{U}-2 \}$.  Heterogeneous
constraints are merged by performing a case split on the value of
$\len{U}$, which explains why $\minconj{M_1}{M_2}$ is a set of
conjunctions, and not a single conjunction.  For example, if $M_1' =
\{ \len{h} \geq \len{U}-4 \}$ and $M_2' = \{ \len{h} \geq 1 \}$, then
the first condition prevails if $\len{U} \geq 5$ yielding:
$\minconj{M_1'}{M_2'} = \{ \len{h} \geq 1 \wedge \len{U}<5, \len{h}
\geq \len{U}-4 \wedge \len{U}\geq 5\}$.  The disjunction of minterms
equivalent to a conjunction of two minterms is then defined as:
$\conj{M_1,M_2} \defequal$
\[\begin{array}{rcl}
\big\{\bigwedge_{i=1,2} M_i^e \wedge M_i^a \wedge M_i^p \wedge M_i^u \wedge \mu \wedge \nu 
& \mid & \mu \in \minconj{M_1}{M_2}, \\ 
&& \nu \in \maxconj{M_1}{M_2}\big\}
\end{array}\]
We extend this notation recursively to any set of minterms of size
$n>2$, as $\conj{M_1,M_2,\ldots,M_n} \defequal \bigcup_{M \in
  \conj{M_1, \ldots, M_{n-1}}} \conj{M, M_n}$.

\prop{prop:conjequiv}{
Given minterms $M_1, \ldots, M_n$, we have $\bigwedge_{i=1}^n M_i
\equiv \bigvee_{M\in \conj{M_1,\ldots, M_n}} M$.
}
\proof{ We prove the result for $n=2$, the general result follows by
  induction. For $n=2$, this is a consequence of the fact that
  $\len{h} \geq \minheap_{M_1} \wedge \len{h} \geq \minheap_{M_2}
  \equiv \bigvee_{\mu \in \minconj{M_1}{M_2}}\mu$, and $\len{h} <
  \maxheap_{M_1} \wedge \len{h} < \maxheap_{M_2} \equiv \bigvee_{\nu
    \in \maxconj{M_1}{M_2}}\nu$. We prove the first fact in the case
  where $\minheap_{M_1} = m_1$ and $\minheap_{M_2} = \len{U} - m_2$,
  the other cases are similar. Consider a structure $\I = (\U, \aheap,
  \astore)$ such that $\I \models \len{h} \geq m_1 \wedge \len{h} \geq
  \len{U} - m_2$. Then $\card{\aheap} \geq m_1$ and $\card{\aheap}
  \geq \card{\U} - m_2$, and we distinguish two cases.
  \begin{compactitem}
  \item if $m_1\geq \card{U} - m_2$, then necessarily $\card{\U} < m_1
    + m_2 + 1$, so that $\I \models \len{h} \geq m_1 \wedge \len{U} <
    m_1 + m_2 +1$.
  \item otherwise, we have $\card{U} \geq m_1 + m_2 +1$, so that $\I
    \models \len{h} \geq \len{U} - m_2 \wedge \len{U} \geq m_1 + m_2
    +1$.
  \end{compactitem}
  Conversely, if $\I$ is a structure such that either $\I \models
  \len{h} \geq m_1 \wedge \len{U} < m_1 + m_2 +1$ or $\I \models
  \len{h} \geq \len{U} - m_2 \wedge \len{U} \geq m_1 + m_2 +1$, then
  it is straightforward to verify that $\I \models \len{h} \geq m_1
  \wedge \len{h} \geq \len{U} - m_2$.  \qed}

\prop{prop:minterm-conj-bound}{
  Given minterms $M_1, \ldots, M_n$ and $M \in
  \conj{M_1,\ldots,M_n}$, if $\ell \in M$ is a literal then either
  $\ell \in M_i$, for some $i = 1,\dots,n$, or $\ell \in \{ \len{U}
  \geq m_1+m_2, \len{U} < m_1+m_2, \len{U} \geq m_1+m_2+1, \len{U} <
   m_1+m_2+1\}$, where $M_1 \cup \dots \cup M_n$ contains two literals
   $\ell_i \in \set{\len{h} \geq m_i, \len{h} < m_i, \len{h} \geq
     \len{U} - m_i, \len{h} < \len{U} - m_i}$, for $i=1,2$. 
}
\proof{Assume that $n=2$. If $\ell \not \in M_1\cup M_2$ then by
  definition of $\conj{M_1,M_2}$, necessarily $\ell$ occurs in
  $\minconj{M_1}{M_2} \cup \maxconj{M_1}{M_2}$ and the proof is
  immediate, by definition of these sets. The proof for $n > 2$
  goes by induction on $n$. \qed}

Given a set $L$ of literals and a subset $B \subseteq L$, let $L^{B}
\defequal B \cup \{\overline\ell \mid \ell \in L \setminus B\}$. For a
set $K$ of literals, let $\completion{K}{L} \defequal \{K \cup L^{B}
\mid B \subseteq L \}$ be the set of completions of $K$ using literals
from $L$ and their complements, so that $K \subseteq
\completion{K}{L}$ and $\completion{K}{L}$ contains either $\ell$ or
$\overline\ell$, for every $\ell \in L$.

\prop{prop:completion}{
  If $K$ and $L$ are sets of literals, then $K \equiv \bigvee_{\psi
    \in \completion{K}{L}} \psi$. If further $K$ is a minterm and $L$
  contains no literals of the form $\len{h}\geq t$ or $\len{h} < t$,
  then every set $P\in \completion{K}{L}$ is a minterm such that
  $\fv{P} = \fv{K}\cup \fv{L}$, $\minheap_P = \minheap_K$ and
  $\maxheap_P = \maxheap_K$.
}
\proof{ Immediate, by the definition of $\completion{K}{L}$. \qed}

For a literal $\ell$, let $\minterm{\ell}$ be an equivalent minterm
obtained from $\ell$ by adding the missing lower/upper bounds on the
cardinality of the heap, namely $\len{h}\geq0$ if $\ell \not\in
\{\len{h} \geq n, \len{h} \geq \len{U}-n \mid n \in \zed\}\}$ and
$\len{h} < \infty$ if $\ell \not\in \{\len{h} < n, \len{h} < \len{U}-n
\mid n \in \zed\}$. We extend this notation to sets of literals as
$\minterm{\ell_1,\dots,\ell_n} \defequal
\conj{\minterm{\ell_1},\dots,\minterm{\ell_n}}$. We have $\ell \equiv
\minterm{\ell}$ for any literal $\ell$ and $L \equiv
\bigvee_{M\in\minterm{L}} M$, for any set $L$ of literals.  For a
boolean combination of literals $\phi$, we denote by $\dnf{\phi}$ its
disjunctive normal form. Given a formula $\phi$ in disjunctive normal
form $\phi = \bigvee_{i=1}^n L_i$, where each conjunctive clause
$L_i$ is identified with the set of its elements, we define
$\minterm{\phi} \defequal \bigcup_{i=1}^n \minterm{L_i}$. We have
$\minterm{\phi} \equiv \bigvee_{M\in \minterm{\phi}} M$. Further, let
$\eqs{L} \defequal \set{x \teq y \mid x,y \in \fv{L}}$ and $\allocs{L}
\defequal \set{\alloc(x) \mid x \in \fv{L}}$, for a set $L$ of
literals.

For each $\gencard \in \set{\fincard, \infcard}$, we define the set of
minterms $\genmt{\phi}$ recursively on the structure of $\phi$, as
follows:
\[\begin{array}{rcl}
  \genmt{\emp} & \defequal & \set{\len{h} \teq 0} \\ 
  \genmt{x \mapsto \vec{y}} & \defequal & \set{x \pto \vec{y} \wedge \len{h} \teq 1} \\
  \genmt{x \teq y} & \defequal & \set{x \teq y \wedge \len{h} \geq 0 \wedge \len{h} < \infty} \\
  \genmt{\phi_1 \wedge \phi_2} & \defequal &  \bigcup_{\begin{array}{l}
      \scriptstyle{M_i \in \genmt{\phi_i}} \\[-2mm] 
      \scriptstyle{i=1,2}
  \end{array}} \conj{M_1,M_2} \\
  \genmt{\neg\phi_1} & \defequal & \bigcup \Big\{ \minterm{\overline{\ell_1},\ldots,\overline{\ell_n}} \Big\rvert 
  \ell_i \in M_i,~ i\in[1,n] \Big\} \\
  && \text{where } \genmt{\phi_1} = \set{M_1,\ldots,M_n}
\end{array}\]
\[\begin{array}{ll}
\genmt{\phi_1 * \phi_2} \defequal \\
\bigcup_{\begin{array}{l}
    \scriptstyle{M_i \in \genmt{\phi_i}} \\[-2mm]
    \scriptstyle{i=1,2}
\end{array}}
\ifLongVersion\else\!\!\!\!\fi
\Big\{\minterm{\dnf{\elim_*(P_1,P_2)}} \Big\rvert & 
\ifLongVersion\else\!\!\!\!\fi N_j \in \completion{M_j}{\eqs{M_1\cup M_2}}, \\[-4mm]
& \ifLongVersion\else\!\!\!\!\fi P_j \in \completion{N_j}{N_{3-j}^p},j=1,2 \Big\}
\end{array}\]
\[\begin{array}{ll}
\genmt{\phi_1 \septraction \phi_2} \defequal \\
  \bigcup_{\begin{array}{l}
      \scriptstyle{M_i \in \genmt{\phi_i}} \\[-2mm]
      \scriptstyle{i=1,2}
  \end{array}} 
  \ifLongVersion\else\!\!\!\!\fi
  \Big\{\minterm{\dnf{\genelim_{\septraction}(Q_1,N_2)}} \Big\rvert & 
 \ifLongVersion\else \!\!\!\!\fi N_j \in \completion{M_j}{\eqs{M_1 \cup M_2}}, \\[-3mm]
  & \ifLongVersion\else\!\!\!\!\fi P_1 \in \completion{N_1}{\allocs{M_1 \cup M_2}}, \\[-1mm]
  & \ifLongVersion\else\!\!\!\!\fi Q_1 \in \completion{P_1}{M_2^a \cup M_2^p}, j=1,2 \Big\}
\end{array}\]  
Intuitively, $\genmt{\phi_1 * \phi_2}$ and $\genmt{\phi_1 \septraction
  \phi_2}$ are computed by first computing recursively
$\genmt{\phi_1}$ and $\genmt{\phi_2}$, then extending the obtained
minterms in such a way that the hypotheses of Lemmas
\ref{lemma:star-elim} or \ref{lemma:septraction-elim} are satisfied,
and finally applying $\genelim_*$ and $\genelim_{\septraction}$,
respectively.
\begin{example}
For instance, consider the formula $x \mapsto x \septraction y \mapsto
y$.  It is easy to check that $\genmt{x \pto x} = \{ M_1 \}$, where
$M_1 = x \pto x \wedge \len{h} \geq 1 \wedge \len{h} < 2$ and
$\genmt{y \mapsto y} = \{ M_2 \}$, where $M_2 = y \pto y \wedge
\len{h} \geq 1 \wedge \len{h} < 2$.  To apply Lemma
\ref{lemma:septraction-elim}, we need to ensure that $M_1$ and $M_2$
are E-complete, which may be done by adding either $x \teq y$ or $x
\not \teq y$ to each minterm.  We also have to ensure that $M_1$ is
A-complete, thus we add either $\alloc(z)$ or $\neg \alloc(z)$ to
$M_1$, for $z\in \set{x,y}$. Finally, we must have $M_2^a \cup M_2^p
\subseteq \cclose{M_1^a \cup M_1^p}$, thus we add either $x \pto x$ or
$\neg x \pto x$ into $M_1$. After removing redundancies, we get (among
others) the minterms: $M_1' = x \pto x \wedge \len{h} \geq 1 \wedge
\len{h} < 2 \wedge x \teq y$ and $M'_2= y \pto y \wedge \len{h} \geq 1
\wedge \len{h} < 2 \wedge x \teq y$.  Afterwards we compute
$\finelim_{\septraction}(M_1',M_2') = x \teq y \wedge \neg \alloc(x)
\wedge \len{h} \geq 0 \wedge \len{h} < 1$. \hfill$\blacksquare$
\end{example}

\begin{lemma}\label{lemma:sl-mnf}
  Given a quantifier-free $\seplogk{k}$ formula $\phi$, the following
  equivalences hold:
  \begin{inparaenum}[(1)]
  \item\label{it:sl-mnf-fin} $\phi \equivfin 
    \bigvee_{M \in \finmt{\phi}} M$, and
  \item\label{it:sl-mnf-inf} $\phi \equivinf 
     \bigvee_{M \in \infmt{\phi}} M$.
  \end{inparaenum}
\end{lemma}
\proof{
(\ref{it:sl-mnf-fin}) We show that $\phi \equivfin
\bigvee_{M \in \finmt{\phi}} M$ by induction on the
structure of $\phi$. The base cases are immediate and the inductive
cases are dealt with below: 
\begin{compactitem}
\item if $\phi = \phi_1 \wedge \phi_2$ and $\phi_i \equivfin
  \bigvee_{M_i \in \finmt{\phi_i}} M_i$ for $i=1,2$ by the inductive
  hypothesis and Proposition \ref{prop:conjequiv}, we have:
  \[\begin{array}{rcl}
  \phi & \equivfin & \bigvee_{M_i \in \finmt{\phi_i},~ i=1,2} M_1 \wedge M_2 \\
  & \equivfin & \bigvee_{M_i \in \finmt{\phi_i},~ i=1,2}
  \bigvee_{M \in \conj{M_1,M_2}} M
  \end{array}\]
\item if $\phi = \neg\phi_1$, $\finmt{\phi_1} = \set{M_1,\ldots,M_n}$,
  $M_i = \set{\ell_{i1},\ldots,\ell_{in_i}}$ for all $i\in[1,n]$, then
  since $\phi_1 \equivfin \bigvee_{i=1}^n \bigwedge_{j=1}^{n_i}
  \ell_{ij}$ by the inductive hypothesis, we have:
  \[\begin{array}{rcl}
  \neg\phi_1 & \equivfin & \bigwedge_{i=1}^n \bigvee_{j=1}^{n_i} \overline{\ell_{ij}} \\ 
  & \equivfin & \bigwedge_{i=1}^n \bigvee_{j=1}^{n_i} \minterm{\overline{\ell_{ij}}} \\
  & \equivfin & \bigvee \setof{\minterm{\overline{\ell_1}}\wedge\ldots \wedge\minterm{\overline{\ell_n}}}{\ell_i \in M_i,~ i \in [1,n]}\\
  & \equivfin & \bigvee\setof{\minterm{\overline{\ell_1},\ldots,\overline{\ell_n}}}{\ell_i \in M_i,~ i \in [1,n]}
  \end{array}\]
\item if $\phi = \phi_1 * \phi_2$ and $\phi_i \equivfin \bigvee_{M \in
  \finmt{\phi_i}} M$ for $i=1,2$ by the induction hypothesis, we
  compute successively\footnote{See Definition \ref{def:minterm-sets}
    for the definition of $N^p$.}:
  \[\begin{array}{rl}
  & (\phi_1 * \phi_2) \text{ [distributivity of $*$ with $\vee$]} \\
  \equivfin & \bigvee_{M_i \in \finmt{\phi_i},~ i=1,2} M_1 * M_2 \\
  & \left[\text{because } M_i \equiv \bigvee_{N_i \in \completion{M_i}{\eqs{M_1 \cup M_2}}} N_i\right] \\
  \equivfin & \bigvee_{M_i \in \finmt{\phi_i},~ i=1,2} \bigvee_{N_i \in \completion{M_i}{\eqs{M_1 \cup M_2}}} N_1 * N_2 \\
  & \left[\text{because } N_i \equiv \bigvee_{P_i \in \completion{N_i}{N_{3-i}^p}} P_i\right] \\
  \equivfin & \bigvee_{M_i \in \finmt{\phi_i},~ i=1,2} \bigvee_{N_i \in \completion{M_i}{\eqs{M_1 \cup M_2}}} \\ 
  & \bigvee_{P_i \in \completion{N_i}{N_{3-i}^p}} P_1 * P_2
  \end{array}\]
  At this point, observe that $N_i$, and thus $P_i$, are E-complete
  for $\fv{M_1 \cup M_2}$, for $i=1,2$. Moreover, $\cclose{P_1^p} =
  \cclose{P_2^p}$, because $P_i \in \completion{N_i}{N_{3-i}^p}$, for
  $i=1,2$. We can thus apply Lemma \ref{lemma:star-elim} and infer
  that:
  \[\begin{array}{rcl}
  P_1 * P_2 & \equiv & \elim_*(P_1,P_2) \\ 
  & \equiv & \dnf{\elim_*(P_1,P_2)} \\ 
  & \equiv & \bigvee_{M \in
    \minterm{\dnf{\elim_*(P_1,P_2)}}} M
  \end{array}\]
\item if $\phi = \phi_1 \septraction \phi_2$ and $\phi_i \equivfin
  \bigvee_{M \in \finmt{\phi_i}} M$, $i=1,2$, by the induction
  hypothesis, we compute, successively:
  \[\begin{array}{rl}
  & (\phi_1 \septraction \phi_2) \left[\text{distributivity of } \septraction \text{ with }\vee\right] \\
  \equivfin & \bigvee_{M_i \in \finmt{\phi_i},~ i=1,2} M_1 \septraction M_2 \\
  & \left[\text{because } M_i \equiv \bigvee_{N_i \in \completion{M_i}{\eqs{M_1 \cup M_2}}} N_i\right] \\
  \equivfin & \bigvee_{M_i \in \finmt{\phi_i},~ i=1,2} \bigvee_{N_i \in \completion{M_i}{\eqs{M_1 \cup M_2}}} N_1 \septraction N_2 \\
  & \left[\text{because } N_1 \equiv \bigvee_{P_1 \in \completion{N_1}{\allocs{M_1 \cup M_2}}} P_1\right] \\
  \equivfin & \bigvee_{M_i \in \finmt{\phi_i},~ i=1,2} \bigvee_{N_i \in \completion{M_i}{\eqs{M_1 \cup M_2}}} \\
  & \bigvee_{P_1 \in \completion{N_1}{\allocs{M_1 \cup M_2}}} P_1 \septraction N_2 \\
  & \left[\text{because } P_1 \equiv \bigvee_{Q_1 \in \completion{P_1}{N_2^a \cup N_2^p}} Q_1\right] \\
  \equivfin & \bigvee_{M_i \in \finmt{\phi_i},~ i=1,2} \bigvee_{N_i \in \completion{M_i}{\eqs{M_1 \cup M_2}}} \\
  & \bigvee_{P_1 \in \completion{N_1}{\allocs{M_1 \cup M_2}}} \bigvee_{Q_1 \in \completion{P_1}{N_2^a \cup N_2^p}} Q_1 \septraction N_2
  \end{array}\]
  Observe that $N_i$ and thus $P_i$ are E-complete for $\fv{M_1 \cup
    M_2}$, for $i=1,2$. Moreover, $P_1$ is A-complete for $\fv{M_1
    \cup M_2}$, because $P_1 \in \completion{N_1}{\allocs{M_1 \cup
      M_2}}$ and $N_2^a \cup N_2^p \subseteq \cclose{Q_1^a \cup
    Q_1^p}$, because $Q_1 \in \completion{P_1}{N_2^a \cup
    N_2^p}$. Then we can apply Lemma \ref{lemma:septraction-elim} and
  infer that: \[\begin{array}{rcl}
  Q_1 \septraction N_2 & \equivfin & \finelim_\septraction(Q_1,N_2) \\ 
  & \equiv & \dnf{\finelim_\septraction(Q_1,N_2)} \\
  & \equiv & \bigvee_{M \in \minterm{\dnf{\finelim_\septraction(Q_1,N_2)}}} M
  \end{array}\]
  \end{compactitem}

\vspace*{\baselineskip}\noindent (\ref{it:sl-mnf-inf}) This point uses
a similar argument. \qed
}

The following lemma relates the polarity of a test formula $\len{h}
\geq \len{U} - n$ or $\alloc(x)$ that occur in some minterm $M \in
\finmt{\phi} \cup \infmt{\phi}$ with that of a separating implication
within $\phi$ (Remarks \ref{rem:test-fol-alloc} and
\ref{rem:test-fol-card}).

\begin{lemma}\label{lemma:polarity}
For any quantifier-free $\seplogk{k}$ formula $\phi$, we have:\begin{compactenum}
\item\label{it:polarity0} For all $M \in \infmt{\phi}$, we have \(M
  \cap \{\len{h} \geq \len{U}-n, \len{h} < \len{U}-n \mid n \in
    \nat\} = \emptyset\).
\item\label{it:polarity1} If $\len{h} \geq \len{U}-n \in M$ [$\len{h}
  < \len{U}-n \in M$] for some minterm $M \in \finmt{\phi}$, then a
  formula $\psi_1 \wand \psi_2$ occurs at a positive [negative] polarity
  in $\phi$.
\item\label{it:polarity2} If $\alloc(x) \in M$ [$\neg\alloc(x) \in M$]
  for some minterm $M \in \infmt{\phi}$, then a formula $\psi_1 \wand
  \psi_2$, such that $x \in \fv{\psi_1} \cup \fv{\psi_2}$, occurs at a
  positive [negative] polarity in $\phi$.
\item\label{it:polarity3} If $M \cap \set{\alloc(x),\neg\alloc(x) \mid
  x \in \vars} \neq \emptyset$ for some minterm $M \in \finmt{\phi}$,
  then a formula $\psi_1 \wand \psi_2$, such that $x \in \fv{\psi_1}
  \cup \fv{\psi_2}$, occurs in $\phi$ at some polarity $p \in
  \set{-1,1}$. Moreover, $\alloc(x)$ occurs at a polarity $-p$, only if
  $\alloc(x)$ is in the scope of a $\lambda^\fincard$ subformula
  (\ref{eq:septraction-elim-lambda}) of a formula
  $\finelim_{\septraction}(M_1,M_2)$ used to compute
  $\bigvee_{M\in\finmt{\phi}} M$.  
\end{compactenum}
\end{lemma}
\proof{ 
  \noindent(\ref{it:polarity0}) By induction on the structure of
  $\phi$, one shows that no literal from $\set{\len{h} \geq \len{U}-n,
    \len{h} < \len{U}-n \mid n \in \nat}$ is introduced during the
  construction of $\infmt{\phi}$.

  \noindent(\ref{it:polarity1}) Let $\ell \in M \cap \set{\len{h} \geq
    \len{U}-n, \len{h} < \len{U}-n \mid n \in \nat}$ be a literal. The
  proof is by induction on the structure of
  $\phi$: \begin{compactitem}
  \item the cases $\phi=\emp$, $\phi=x \pto \vec{y}$ and $\phi= x \teq
    y$ are trivial, because $\ell \not\in \finmt{\phi}$.
  \item $\phi=\phi_1\wedge\phi_2$: we have $M \in [M_1,M_2]$, for some
    minterms $M_i \in \finmt{\phi_i}$, for $i=1,2$. By Proposition
    \ref{prop:minterm-conj-bound}, since $\ell \not\in
    \set{\len{U}\geq n, \len{U}<n \mid n \in \nat}$, we deduce that
    $\ell \in M_1 \cup M_2$ and the proof follows from the induction
    hypothesis, since any formula occurring in $\phi_i$, $i=1,2$,
    occurs at the same polarity in $\phi$.
  \item $\phi=\neg\phi_1$: assuming $\finmt{\phi_1} =
    \set{M_1,\ldots,M_m}$, we have $M \in \minterm{\overline{\ell_1},
      \ldots, \overline{\ell_m}}$, for some literals $\ell_i \in M_i$,
    $i \in [1,m]$. By Proposition \ref{prop:minterm-conj-bound}, we
    deduce that $\ell = \overline{\ell_i}$ for some $i=1,\ldots,n$,
    because $\ell \not\in \set{\len{U}\geq n, \len{U}<n \mid n \in
      \nat}$. By the induction hypothesis, there exists a formula
    $\psi_1 \wand \psi_2$ occurring at polarity $p \in \set{1,-1}$ in
    $\phi_1$, where $p = 1$ if $\ell_i = \len{h} \geq \len{U}-n$ and
    $p=-1$ if $\ell_i = \len{h} < \len{U}-n$. Then $\ell$ occurs at
    polarity $-p$ in $M$ and $\psi_1 \wand \psi_2$ occurs at polarity
    $-p$ in $\phi$.
  \item $\phi = \phi_1 * \phi_2$: for $i=1,2$, there exist minterms
    $M_i \in \finmt{\phi_i}$, $N_i \in \completion{M_i}{\eqs{M_1\cup
      M_2}}$ and $P_i \in \completion{N_i}{N_{3-i}^p}$, such that $M
    \in \minterm{\dnf{\elim_*(P_1,P_2)}}$.  Since by hypothesis $\ell
    \in \set{\len{h} \geq \len{U}-n, \len{h} < \len{U}-n \mid n \in
      \nat}$, by Proposition \ref{prop:minterm-conj-bound}, this
    literal is necessarily introduced by $\elim_*(P_1,P_2)$ and, by
    inspection of $\elim_*(P_1,P_2)$, one of the following must
    hold: \begin{compactitem}
    \item $\ell = \len{h} \geq \minheap_{M_1} + \minheap_{M_2}$, where
      $\minheap_{M_1}$ and/or $\minheap_{M_2}$ is of the form
      $\len{U}-n$. By the induction hypothesis $\phi_i$ contains a
      formula $\psi_1 \wand \psi_2$ at polarity $1$, for some $i=1,2$,
      and the proof is completed.
    \item $\ell = \len{h} < \maxheap_{M_1} + \maxheap_{M_2} -1$, where
      $\maxheap_{M_1}$ and/or $\maxheap_{M_2}$ is of the form
      $\len{U}-n$. The proof is similar, with polarity $-1$.
    \item $\ell = \len{h} \geq \allocno{M_i} + \len{Y}_{M_i} +
      \minheap_{M_j}$, where $\minheap_{M_j}$ is of the form
      $\len{U}-n$. The proof is similar.
    \end{compactitem}
  \item $\phi = \phi_1\septraction\phi_2 =
    \neg(\phi_1\wand\neg\phi_2)$: there exist minterms $M_i \in
    \finmt{\phi_i}$, $N_i \in \completion{M_i}{\eqs{M_1 \cup M_2}}$, for
    $i=1,2$, $P_1 \in \completion{N_1}{\allocs{M_1 \cup M_2}}$ and
    $Q_1 \in \completion{P_1}{M_2^a \cup M_2^p}$, such that $M \in
    \minterm{\dnf{\finelim_\septraction(Q_1,N_2)}}$. By inspection of
    $\finelim_\septraction(Q_1,N_2)$, one of the following cases must
    occur:\begin{compactitem}
    \item $\ell=\len{h} \geq \minheap_{M_2} - \maxheap_{M_1} - 1$,
      where $\minheap_{M_2}$ is of the form $\len{U}-n_2$. By the
      induction hypothesis, $\phi_2$ contains a formula $\psi_1 \wand
      \psi_2$ at polarity $1$, and this formula also occurs at
      polarity $1$ in $\phi$, thus the proof is completed. Note that
      if $\maxheap_{M_1} = \len{U}-n_1$ then either $\minheap_{M_2} =
      \len{U}-n_2$ and $\len{h} \geq \minheap_{M_2} - \maxheap_{M_1} -
      1 = \len{h} \geq n_1-n_2$, or $\minheap_{M_2} = n_2 \in \nat$
      and $\len{h} \geq \minheap_{M_2} - \maxheap_{M_1} - 1 = \len{h}
      \geq -\len{U} + (n_1+n_2) = \bigwedge_{\scriptscriptstyle{1 \leq
          n < n_1+n_2}} \len{U} \teq n \rightarrow \len{h} \geq
      n_1+n_2-n$ by Definition \ref{def:lin-comb}, thus $\len{h} \geq \minheap_{M_2} - \maxheap_{M_1} -
      1$ contains no literal of the above form.
    \item $\ell = \len{h} < \maxheap_{M_2} - \minheap_{M_1}$. The
      proof is similar.
    \item $\ell = \len{h} < \len{U} - \minheap_{M_1} -
      \nallocno{Y}{M_1} + 1$. In this case since $(\phi_1 \wand \neg
      \phi_2)$ occurs at polarity $-1$ in $\phi$, the proof is
      completed.
    \end{compactitem}
  \end{compactitem}

  \noindent(\ref{it:polarity2}) Let $\ell \in M \cap \set{\alloc(x),
    \neg\alloc(x) \mid x \in \vars}$ be a literal occurring in some
  minterm $M \in \infmt{\phi}$. The proof is by induction on the
  structure of $\phi$: \begin{compactitem}
  \item the cases $\phi=\emp$, $\phi=x \pto \vec{y}$ and $\phi= x \teq
    y$ are trivial, because $\ell \not\in \infmt{\phi}$.
  \item the cases $\phi=\phi_1\wedge\phi_2$ and $\phi=\neg\phi_1$ are
    similar to point (\ref{it:polarity1}) of the Lemma.
  \item $\phi = \phi_1 * \phi_2$: there exist minterms $M_i \in
    \infmt{\phi_i}$, $N_i \in \completion{M_i}{\eqs{M_1\cup M_2}}$ and
    $P_i \in \completion{N_i}{N_{3-i}^p}$, such that $M \in
    \minterm{\dnf{\elim_*(P_1,P_2)}}$, for all $i=1,2$. By inspection
    of $\elim_*(P_1,P_2)$, one of the following cases must
    occur:\begin{compactitem}
    \item $\ell = \neg \alloc(x)$ with $x \in \nvar{M_1} \cap
      \nvar{M_2}$. Assuming that the definition of $\elim_*(P_1,P_2)$
      is changed according to Remark \ref{rem:star-elim-alloc}, it
      must be the case that $\neg\alloc(x)$ occurs at a positive polarity in $M_1$
      or $M_2$. Then, by the induction hypothesis $\phi_i$ contains a
      subformula $\psi_1 \wand \psi_2$ at polarity $-1$ with $x \in
      \fv{\psi_1} \cup \fv{\psi_2}$. But then $\psi_1 \wand \psi_2$
      also occurs at polarity $-1$ in $\phi$ and the proof is
      completed.
    \item $\ell = \neg \alloc(x)$ with $x \in Y \subseteq
      \nvar{M_j}$. Similar to the previous case.
    \end{compactitem}
  \item $\phi = \phi_1\septraction\phi_2 =
    \neg(\phi_1\wand\neg\phi_2)$: there exist minterms $M_i \in
    \infmt{\phi_i}$, $N_i \in \completion{M_i}{\eqs{M_1 \cup M_2}}$,
    for $i=1,2$, $P_1 \in \completion{N_1}{\allocs{M_1 \cup M_2}}$ and
    $Q_1 \in \completion{P_1}{M_2^a \cup M_2^p}$, such that $M \in
    \minterm{\dnf{\infelim_\septraction(Q_1,N_2)}}$. By inspection of
    $\infelim_\septraction(Q_1,N_2)$, the only case possible is $\ell
    = \neg \alloc(x)$ (\ref{eq:septraction-elim-footprint}) with $x
    \in \avar{M_1}$, thus $x \in \fv{\phi_1}\cup \fv{\phi_2}$ and
    $(\phi_1 \wand \neg \phi_2)$ occurs at polarity $-1$ in $\phi$,
    which completes the proof. 
  \end{compactitem}
  
  \noindent(\ref{it:polarity3}) The proof is similar to point
  (\ref{it:polarity2}). The only difference is that $\alloc(x)$ may
  occur in the $\lambda^\fincard$ subformula
  (\ref{eq:septraction-elim-lambda}) of the
  $\finelim_\septraction(Q_1,N_2)$, in which case its polarity may be
  different from that of $\phi_1\wand\phi_2$. \qed}

\subsection{A Decision Problem}
\label{sec:pspacemt}

Given a quantifier-free $\seplogk{k}$ formula $\phi$, the number of
minterms occurring in $\finmt{\phi}$ [$\infmt{\phi}$] is
exponential in the size of $\phi$, in the worst case. Therefore, an
optimal decision procedure cannot generate and store these sets
explicitly, but rather must enumerate minterms lazily. We show
that \begin{inparaenum}[(i)]
\item the size of the minterms in $\finmt{\phi} \cup \infmt{\phi}$ is
  bounded by a polynomial in the size of $\phi$ (Corollary
  \ref{cor:sl-mnf-size}), and that
\item the problem ``\emph{given a minterm $M$, does $M \in
  \finmt{\phi}$ [$\infmt{\phi}$]} ?'' is in \pspace\ (Lemma
  \ref{lemma:pspacemt}). \end{inparaenum}

To start with, we define a measure on a quantifier-free formula
$\phi$, which bounds the size of the minterms in the sets
$\finmt{\phi}$ and $\infmt{\phi}$, inductively on the structure of the
formulae, as follows:
\[\begin{array}{rclrcl}
\bound{x \teq y} & \defequal & 0 &
\bound{\bot} & \defequal & 0 \\
\bound{\emp} & \defequal & 1 &
\bound{x \mapsto \vec{y}} & \defequal & 2 \\
\bound{\neg\phi_1} & \defequal & \bound{\phi_1} &
\bound{\phi_1 \wedge \phi_2} & \defequal & \max(\bound{\phi_1},\bound{\phi_2}) \\
\end{array}\]
\[\begin{array}{rcl}
\bound{\phi_1 * \phi_2} & \defequal & \sum_{i=1}^2 (\bound{\phi_i} + \card{\fv{\phi_i}}) \\
\bound{\phi_1 \wand \phi_2} & \defequal & \sum_{i=1}^2 (\bound{\phi_i} + \card{\fv{\phi_i}})
\end{array}\]

\prop{prop:test-formula-bound}{
  For any $n \in \nat$, we have: 
  \[\begin{array}{l}
  \bound{\len{h} \geq n} = \bound{\len{U} \geq n} = n \\ 
  \bound{\len{h} \geq \len{U}-n} = n+1
  \end{array}\]
}
\proof{By induction on $n\geq0$. \qed}

\prop{prop:boundsize}{ For any formula $\phi$, $\bound{\phi} =
  \bigO(\size{\phi}^2)$.  } \proof{By induction on $\phi$. The most
  interesting cases are $\phi_1 * \phi_2$ and $\phi_1
  \wand \phi_2$: 
  \[\begin{array}{rcl}
  \bound{\phi_1 * \phi_2} & = & \sum_{i=1}^2 (\bound{\phi_i} + \card{\fv{\phi_i}}) \\
  & \leq & \sum_{i=1}^2 (\bound{\phi_i} + \size{\phi_i}) \\
  & = & \bigO(\sum_{i=1}^2 (\size{\phi_i}^2 + \size{\phi_i}) \\
  & = & \bigO((\size{\phi_1} + \size{\phi_2})^2)
  \end{array}\]
The case $\phi_1 \wand \phi_2$ is identical. \qed}

\begin{definition}\label{def:bounded}
A minterm $M$ is \emph{\bounded}\ by a formula $\phi$, if for each
literal $\ell \in M$, the following hold: \begin{inparaenum}[(i)]
 \item $\bound{\ell} \leq \bound{\phi}$ if $\ell \in \set{\len{h}
   \geq \min_{M_i}, \len{h} < \max_{M_i}}$, and
 \item $\bound{\ell} \leq 2 \bound{\phi} + 1$, if $\ell \in
   \set{\len{U} \geq n, \len{U} < n \mid n \in \nat}$.
 \end{inparaenum}
\end{definition} 

 \prop{prop:heap-univ-bound}{
   Given minterms $M_1, \ldots, M_n$ all \bounded\ by $\phi$, each
   minterm $M \in \conj{M_1,\ldots,M_n}$ is also \bounded\ by $\phi$. 
}
\proof{An immediate corollary of Proposition
  \ref{prop:minterm-conj-bound}. \qed}

\begin{lemma}\label{lemma:sl-mnf-bounded}
  Given a quantifier-free $\seplogk{k}$ formula $\phi$, each minterm $M
  \in \finmt{\phi} \cup \infmt{\phi}$ is \bounded\ by $\phi$.
\end{lemma}
\proof{ We prove that each $M \in \finmt{\phi}$ is \bounded\ by
  $\phi$. The proof for $M \in \infmt{\phi}$ follows from the
  observation that, because of the definition of
  $\elim_{\septraction}^\infcard$, for each $M \in \infmt{\phi}$ there
  exists $M' \in \finmt{\phi}$ such that $\bound{M} \leq
  \bound{M'}$. By induction on the structure of $\phi$:
  \begin{compactitem}
  \item If $\phi=\emp$ then $\finmt{\phi} = \set{\len{h} \geq 0 \wedge
    \len{h} < 1}$, $\bound{\len{h} \geq 0} = 0$, $\bound{\len{h} < 1}
    = \bound{\len{h} \geq 1} = 1$ and $\bound{\emp} = 1$, by
    definition.
  \item If $\phi=x \mapsto \vec{y}$ then $\finmt{\phi} = \set{x \pto
    \vec{y} \wedge \len{h} \geq 1 \wedge \len{h} < 2}$,
    $\bound{\len{h} \geq 1} = 1$, $\bound{\len{h} < 2} = 2$ and
    $\bound{x \mapsto \vec{y}} = 2$, by definition.
  \item If $\phi=x\teq y$ then $\finmt{\phi} = \set{x\teq y \wedge
    \len{h} \geq 0 \wedge \len{h} < \infty}$ and $\bound{\len{h} \geq
    0} = \bound{\len{h} < \infty} = 0$, by definition.
  \item If $\phi=\phi_1\wedge\phi_2$, let $\ell \in M$ be a literal,
    where $M \in \finmt{\phi_1\wedge\phi_2}$ is a minterm. Then $M \in
    [M_1,M_2]$, for some minterms $M_i \in \finmt{\phi_i}$, $i=1,2$
    and the proof follows from Proposition \ref{prop:heap-univ-bound},
    because $M_i$ is \bounded\ by $\phi_i$ and $\bound{\phi_i} \leq
    \bound{\phi}$, thus $M_i$ is \bounded\ by $\phi$, for $i=1,2$.
  \item If $\phi=\neg\phi_1$ assume that $\finmt{\phi_1} =
    \set{M_1,\ldots,M_m}$. Let $\ell \in M$ be a literal, where $M \in
    \finmt{\neg\phi_1}$ is a minterm.  Then $M \in
    \conj{\minterm{\overline{\ell_1}}, \ldots,
      \minterm{\overline{\ell_n}}}$, for some literals $\ell_i \in
    M_i$, $i \in [1,m]$. By the induction hypothesis,
    $\overline{\ell_i}$ is \bounded\ by $\phi$, for every $i\in
    1,\dots,n$, thus the same holds for $\ell_i$. Since
    $\bound{\len{h} \geq 0} = \bound{\len{h} < \infty} = 0$, we deduce
    that $\minterm{\overline{\ell_i}}$ is \bounded\ by $\phi$, and the
    proof follows from Proposition \ref{prop:heap-univ-bound}.
  \item If $\phi=\phi_1*\phi_2$, let $\ell \in M$ be a literal, where
    $M \in \finmt{\phi_1*\phi_2}$. Then there exist minterms $M_i \in
    \finmt{\phi_i}$, $N_i \in \completion{M_i}{\eqs{M_1\cup M_2}}$ and
    $P_i \in \completion{N_i}{N_{3-i}^p}$, such that $M \in
    \minterm{\dnf{\elim_*(P_1,P_2)}}$, for $i=1,2$. First assume that $\ell$
    is of the form $\len{h} \geq t$ or $\len{h} < t$.  We only
    consider the case where $\ell$ occurs in $\elim_*(P_1,P_2)$, the
    rest of the cases follow from Proposition
    \ref{prop:heap-univ-bound}. We distinguish the following
    cases: \begin{compactitem}
      \item $\ell$ is a subformula of $\len{h} \geq \minheap_{P_1} +
        \minheap_{P_2} = \len{h} \geq \minheap_{M_1} +
        \minheap_{M_2}$, because $\minheap_{P_i} = \minheap_{M_i}$,
        for $i=1,2$, by Proposition \ref{prop:completion}. By the
        inductive hypothesis we have $\bound{\len{h} \geq
          \minheap_{M_i}} \leq \bound{\phi_i}$, for $i=1,2$. If
        $\minheap_{M_i} \in \nat$ for $i=1,2$ then $\ell = \len{h}
        \geq \minheap_{M_1} + \minheap_{M_2}$ and we have: 
        \[\begin{array}{rcl}
        \bound{\ell} = \bound{\len{h} \geq \minheap_{M_1} + \minheap_{M_2}} 
        & = & \bound{\len{h} \geq \minheap_{M_1}} + \bound{\len{h} \geq \minheap_{M_2}} \\
        & \leq & \bound{\phi_1} + \bound{\phi_2} \leq \bound{\phi}.
        \end{array}\] 
        If $\minheap_{M_i} = \len{U} - n_i$ and $n_i,
        \minheap_{M_{3-i}} \in \nat$, then $\ell = \len{h} \geq
        \minheap_{M_1} + \minheap_{M_2}$ and we obtain:
        \[\begin{array}{rcl}
        \bound{\ell} = \bound{\len{h} \geq \minheap_{M_1} + \minheap_{M_2}} 
        & = & \bound{\len{h} \geq \len{U} - (n_i - \minheap_{M_{3-i}})} \\
        & \leq & \bound{\len{h} \geq \len{U} - n_i} \\
        & \leq & \bound{\phi_i} \leq \bound{\phi}.
        \end{array}\] 
        Otherwise, $\minheap_{M_i} = \len{U} - n_i$, for $i=1,2$,
        where $n_1,n_2 \in \nat$, thus by Definition \ref{def:lin-comb}:
        \[\begin{array}{ll}
        \len{h} \geq \minheap_{M_1} + \minheap_{M_2} & = \\ 
        \len{h} \geq 2\cdot\len{U} - n_1 - n_2  & = \\
        \len{U} < 1 + n_1 + n_2 ~\wedge \\ 
        \bigwedge_{1 \leq n \leq n_1 + n_2} 
        \len{U} \teq n \rightarrow \len{h} \geq 2n - n_1 - n_2
        \end{array}\]
        and either:\begin{compactitem}
        \item $\ell \in \set{\len{U} \geq n, \len{U} < n+1}$ for some
          $n \in [1,n_1+n_2]$: we have $\bound{\ell} \leq n+1 \leq
          n_1+n_2+1 \leq 2(\bound{\phi_1} + \bound{\phi_2}) + 1 =
          2\bound{\phi}+1$, or
        \item $\ell = \len{h} \geq 2n - n_1 - n_2$ for some $n \in
          [1,n_1+n_2]$: we have $\bound{\ell} = 2n-n_1-n_2 \leq
          n_1+n_2 = \bound{\phi_1} + \bound{\phi_2} = \bound{\phi}$. 
        \end{compactitem}
      \item The proof in the case where $\ell$ is a subformula of $\len{h} <
        \maxheap_{M_1} + \maxheap_{M_2} - 1$ is analogous.
      \item $\ell = \len{h} \geq \allocno{P_i} + \len{Y}_{P_i} +
        \minheap_{P_{3-i}}$, where $Y \subseteq \nvar{P_{3-i}}
        \setminus \avar{P_i}$, for some $i=1,2$. Because $Y \cap
        \avar{P_i} = \emptyset$, we have (Definition \ref{def:anvar}
        and Proposition \ref{prop:completion}): $\allocno{P_i} +
        \len{Y}_{P_i} \leq \card{\fv{P_i}} + \card{\fv{P_{3-i}}} \leq
        \card{\fv{\phi_1}} + \card{\fv{\phi_2}}$ and thus
        $\bound{\ell} \leq \bound{\len{h} \geq \minheap_{P_{3-i}}} +
        \card{\fv{\phi_1}} + \card{\fv{\phi_2}} \leq \bound{\phi}$.
    \end{compactitem}
    Now assume $\ell \in \set{\len{U} \geq m, \len{U} < m \mid m \in \nat}$.
    Then one of the following holds:
    \begin{compactitem}
    \item $\ell \in \uclose{P_i}^u$, for some $i=1,2$, and we have two cases:
      \begin{compactitem}
      \item $\ell \in \set{\len{U} \geq n_1 + n_2 + 1,\len{U} <
        n_1+n_2}$, where $\minheap_{P_i}=\minheap_{M_i}=n_1$ and
        $\maxheap_{P_i}=\maxheap_{M_i}=\len{U}-n_2$. By the induction
        hypothesis, we have $n_1, n_2 \leq \bound{\phi_i}$, thus
        $\bound{\ell} \leq 2\bound{\phi_i}+1 \leq 2\bound{\phi}+1$.
         \item $\ell = \len{U} \geq \left\lceil\sqrt[k]{\max_{x \in
             \avar{M}}(\datano{x}{P_i}+1)}\right\rceil$, in which case
           either $\fv{M_1} \cup \fv{M_2} = \emptyset$ so that
           $\left\lceil\sqrt[k]{\max_{x \in
               \avar{M}}(\datano{x}{P_i}+1)}\right\rceil = 0$ and the
           proof is immediate, or we have $\bound{\ell} \leq
           \sqrt[k]{\card{\fv{M_i}}^k+1} \leq \len{\fv{M_i}}+1 \leq
           2\bound{\phi} +1$.
      \end{compactitem}
    \item $\ell = \len{U} > n_i + \allocno{P_i} + \len{Y}_{M_i}$,
      where $Y \subseteq \nvar{M_{3-i}} \setminus \avar{M_i}$ and
      $\maxheap_{M_i} = \len{U} - n_i$, for some $i=1,2$. Because $Y
      \cap \avar{P_i} = \emptyset$, we have $\allocno{P_i} +
      \len{Y}_{P_i} \leq \card{\fv{P_i}} + \card{\fv{P_{3-i}}} \leq
      \card{\fv{\phi_1}} + \card{\fv{\phi_2}}$ and thus $\bound{\ell}
      \leq \bound{\phi_i} + \card{\fv{\phi_1}} + \card{\fv{\phi_2}}
      \leq 2\bound{\phi}+1$.
    \end{compactitem}
  \item If $\phi = \phi_1 \septraction \phi_2$, consider a literal
    $\ell \in M$, where $M \in \finmt{\phi_1 \septraction
    \phi_2}$. Then there exist minterms $M_i \in \finmt{\phi_i}$ and
    $N_i \in \completion{M_i}{\eqs{M_1 \cup M_2}}$, for $i=1,2$, and
    minterms $P_1 \in \completion{N_1}{\allocs{M_1 \cup M_2}}$ and
    $Q_1 \in \completion{P_1}{M_2^a \cup M_2^p}$, such that $M \in
    \minterm{\dnf{\finelim_\septraction(Q_1,N_2)}}$. We only consider
    the case where $\ell$ occurs in $\finelim_\septraction(Q_1,N_2)$,
    in the remaining cases, the result follows directly from
    Proposition \ref{prop:heap-univ-bound}. If $\ell$ is of the form
    $\len{h} \geq t$ or $\len{h} < t$ then either:
    \begin{compactitem}
      \item $\ell$ is a subformula of $\len{h} \geq \minheap_{N_2} -
        \maxheap_{Q_1} - 1 = \len{h} \geq \minheap_{M_2} -
        \maxheap_{M_1} - 1$, because $\minheap_{N_2} = \minheap_{M_2}$
        and $\maxheap_{Q_1} = \maxheap_{P_1} = \maxheap_{N_1} =
        \maxheap_{M_1}$ by Proposition \ref{prop:completion}. Then
        $\minheap_{M_2} \in \{ n_2, \len{U} - n_2 \}$ and
        $\maxheap_{M_1} \in \{ n_1, \len{U} - n_1 \}$ with $n_1,n_2
        \in \nat_\infty$, and by the induction hypothesis $n_i \leq
        \bound{\phi_i}$. If $\maxheap_{M_1} = n_1$ or $\minheap_{M_2}
        \not = n_2$, then by an inspection of the different cases and
        using Proposition \ref{prop:test-formula-bound}, we have $\ell
        = \len{h}\geq\minheap_{M_2}-\maxheap_{M_1}+1$, thus:
        \[\bound{\ell} = \bound{\len{h}\geq\minheap_{M_2}-\maxheap_{M_1}+1} \leq n_1 + n_2 
        \leq \bound{\phi_1} + \bound{\phi_2} \leq \bound{\phi}\]
        Otherwise, $\minheap_{M_2} = n_2$ and $\maxheap_{M_1} =
        \len{U}-n_1$ hence either: \begin{compactitem}
        \item $\ell \in \set{\len{U} \geq n, \len{U} < n+1}$, for some
          $n \in [1,n_1+n_2-1]$ and we have $\bound{\ell} \leq n+1 \leq
          n_1+n_2 \leq 2(\bound{\phi_1}+\bound{\phi_2})+1 =
          \bound{\phi}$, or 
        \item $\ell = \len{h} \geq n_1+n_2-n$, for some $n \in
          [1,n_1+n_2-1]$ and we have $\bound{\ell} = n_1+n_2-n \leq
          n_1+n_2-1 \leq \bound{\phi_1} + \bound{\phi_2} =
          \bound{\phi}$.
        \end{compactitem}
        %% $\ell \iff \len{h} \geq -\len{U} + (n_1+n_2)
        %% \iff \bigwedge_{\scriptscriptstyle{1 \leq n < n_1+n_2}}
        %% \len{U} \teq n \rightarrow \len{h} \geq n_1+n_2-n$, and $n_1 +
        %% n_2 \leq 2\bound{\phi}+1$.
        %
      \item The case $\ell = \len{h} < \maxheap_{N_2} -
        \minheap_{Q_1}$ is proved in a similar way. 
      \item $\ell = \len{h} < \len{U} - \minheap_{Q_1} -
        \nallocno{Y}{Q_1} + 1$, for some $Y \subseteq \fv{Q_1 \cup
          N_2}$. Because $\nvar{Q_1} \subseteq \nvar{P_1} \subseteq
        \fv{\phi_1} \cup \fv{\phi_2}$, we have $\nallocno{Y}{Q_1} \leq
        \card{\fv{\phi_1}} + \card{\fv{\phi_2}}$. Moreover,
        $\minheap_{Q_1} = \minheap_{M_1}$ by Proposition \ref{prop:completion}. We distinguish the following
        cases: \begin{compactitem}
        \item If $\minheap_{M_1} \in \nat$, we compute:
          \[\begin{array}{rcl}
          \bound{\ell} & = & \minheap_{M_1} + \nallocno{Y}{Q_1} - 1
          \text{, by Proposition \ref{prop:test-formula-bound}} \\
          & \leq & \bound{\phi_1} + \card{\fv{\phi_1}} + \card{\fv{\phi_2}} \leq \bound{\phi},
          \end{array}\]
          since $\bound{\len{h} \geq \minheap_{M_1}} \leq
          \bound{\phi_1}$, by the inductive hypothesis.
        \item Otherwise, $\minheap_{M_1} = \len{U} - n_1$, for some
          $n_1 \in \nat$, thus $\ell = \len{h} < n_1 -
          \nallocno{Y}{Q_1} + 1$. By Proposition
          \ref{prop:test-formula-bound}, we have $\bound{\len{h} \geq
            \minheap_{M_1}} = n_1+1$ and $\bound{\ell} = n_1 - \nallocno{Y}{Q_1} + 1$, therefore: 
          \[\begin{array}{rcl}
          \bound{\ell} & = & \bound{\len{h} \geq \minheap_{M_1}} - \nallocno{Y}{Q_1} \\ 
          & \leq & \bound{\phi_1} \leq \bound{\phi}
          \end{array}\]
        \end{compactitem}
    \end{compactitem}
    If $\ell$ is of the form $\len{U} \geq m$ or $\len{U} < m$, with
    $m \in \nat$, then either: \begin{compactitem}
      \item if $\ell \in \uclose{Q_1} \cup \uclose{N_2}$ the argument
        is similar to the previous case $\phi = \phi_1 * \phi_2$, 
      \item otherwise, $\ell = \len{U} \geq \minheap_{M_2} +
        \nallocno{Y}{M_1}$ and either $\minheap_{M_2} \in \nat$, in
        which case $\bound{\ell} = \minheap_{N_2} + \nallocno{Y}{Q_1}
        \leq \bound{\phi_2} + \card{\fv{\phi_1}} +
        \card{\fv{\phi_2}} \leq \bound{\phi}$ as in the previous, or
        $\minheap_{M_2} = \len{U} - n_2$, for some $n_2 \in \nat$, in
        which case $\ell \equiv n_2 \geq \nallocno{Y}{Q_1}$ and
        $\bound{\ell} = 0$. \qed
    \end{compactitem}
  \end{compactitem}}

\begin{corollary}\label{cor:sl-mnf-size}
  Given a quantifier-free $\seplogk{k}$ formula $\phi$ and a minterm
  $M \in \finmt{\phi} \cup \infmt{\phi}$, we have $\size{M} =
  \bigO(\size{\phi}^2)$.
\end{corollary}
\proof{ We give the proof for $M \in \finmt{\phi}$, the case $M \in
  \infmt{\phi}$ being similar. Let $\ell \in M$ be a literal. We
  distinguish the following cases, based on the form of
  $\ell$: \begin{compactitem}
  \item $\ell \in \set{\alloc(x),\neg\alloc(x) \mid x \in \vars}$:
    $\ell$ occurs in $\phi$ or has been introduced by $\finmt{.}$, in
    which case, at most $2\card{\fv{\phi}}$ such literals are
    introduced.
  \item $\ell \in \{x \pto \vec{y}, \neg x \pto \vec{y} \mid x \in
    \vars, \vec{y} \in \vars^k\}$: $\ell$ occurs in $\phi$, since
    $\finmt{.}$ does not introduce literals of this form.
  \item $\ell \in \{x \teq y, \neg x \teq y \mid x, y \in \vars\}$:
    $\ell$ occurs in $\phi$ or has been introduced by $\finmt{.}$, in
    which case at most $2\card{\fv{\phi}}^2$ such literals are
    introduced.
  \item $\ell \in \{\len{U} \geq n, \len{U} < n \mid n \in \nat\}$: by
    Lemma \ref{lemma:sl-mnf-bounded}, $\bound{\ell} \leq 2\bound{\phi}
    + 1$, thus $\size{\ell} = \bigO(\size{\phi}^2)$ for each such
    literal. Furthermore, $M$ contains at most two literals of this
    form (up to redundancy).
  \item $\ell \in \{\len{h} \geq \minheap_M, \len{h} < \maxheap_M\}$:
    by Lemma \ref{lemma:sl-mnf-bounded}, $\bound{\ell} \leq
    \bound{\phi}$ and consequently, $\size{\ell} =
    \bigO(\size{\phi}^2)$ for each such literal. Furthermore, $M$
    contains exactly two literals of this form by definition of
    minterms.
  \end{compactitem}
  Summing up, we obtain that $\size{M}=\bigO(\size{\phi}^2)$. \qed}

\prop{prop:pspacednf}{
Let $L$ be a set of literals and $\phi$ be a boolean combination of
literals. The problem whether $L \in \dnf{\phi}$ is in
$\bigonspace{\size{L}+\size{\phi}}$.
}
\proof{ W.l.o.g., we may assume that $\phi$ is in negation normal
  form. The algorithm is nondeterministic and proceeds recursively on
  the structure of $\phi$: \begin{compactitem}
  \item $\phi=\ell$ is a literal: then $\dnf{\phi} = \set{\ell}$ hence
    it suffices to check that $L = \set{\ell}$, using 
    $\bigO(\size{L}+\size{\phi})$ space. 
  \item $\phi = \phi_1 \vee \phi_2$: then $\dnf{\phi} = \dnf{\phi_1}
    \cup \dnf{\phi_2}$ and we check that one of $L \in \dnf{\phi_1}$
    and $L \in \dnf{\phi_2}$ holds. By the induction hypothesis,
    checking $L \in \dnf{\phi_i}$ can be done using
    $\bigO(\size{L}+\size{\phi_i})$ space. Since the working space
    used for $L \in \dnf{\phi_1}$ can be reused for $L \in
    \dnf{\phi_2}$, the entire check takes
    $\bigO(\size{L}+\size{\phi})$ space.
  \item $\phi = \phi_1 \wedge \phi_2$: then $L \in \dnf{\phi} \iff L =
    L_1 \cup L_2$, with $L_1 \in \dnf{\phi_1}$ and $L_2 \in
    \dnf{\phi_2}$, thus we guess two subsets $L_1$ and $L_2$ with $L_1
    \cup L_2 = M$ and check that $L_i \in \dnf{\phi_i}$, using
    $\bigO(\size{L_i}+\size{\phi_i})$ space, for $i = 1,2$. Since we
    must store $L_2$ during the check $L_1 \in \dnf{\phi_1}$ and the
    working space can be reused for $L_2 \in \dnf{\phi_2}$, the entire
    check takes $\bigO(\size{L}+\size{\phi})$ space. \qed
  \end{compactitem}}

\prop{prop:pspaceand}{
Let $L$ be a set of literals and let $M_1,M_2$ be minterms.  Checking
whether $L \in \dnf{(\elim_{*}(M_1,M_2)}$ is in
$\bigonspace{\size{L}+\size{M_1}+\size{M_2}}$.
}
\proof{ The algorithm proceeds by induction on the structure of
  $\dnf{\elim_{*}(M_1,M_2)}$ as in the proof of Proposition
  \ref{prop:pspacednf}. The only difference concerns the subformulae
  $\eta_{ij}$ (\ref{eq:star-elim-eta}) which cannot be
  constructed explicitly since they are of exponential size. However,
  $\eta_{ij}$ is of positive polarity, and to check that $L \in
  \dnf{\eta_{ij}}$, it suffices to guess a set of variables $Y
  \subseteq \nvar{M_j} \setminus \avar{M_i}$ and check whether: \[L
  \in \dnf{\alloc(Y) \rightarrow (\len{h} \geq \allocno{M_i} +
    \len{Y}_{M_i} + \minheap_{M_j} ~\wedge~ \allocno{M_i} +
    \len{Y}_{M_i} < \maxheap_{M_i})}\] The size of the above formula
  is of the order of $\bigO(\size{M_1}+\size{M_2})$, thus $L \in
  \dnf{(\elim_{*}(M_1,M_2)}$ can be checked in
  $\bigonspace{\size{L}+\size{M_1}+\size{M_2}}$, by Proposition
  \ref{prop:pspacednf}. \qed}

\prop{prop:pspacewand}{ Let $L$ be a set of literals and let $M_1,M_2$
  be minterms. The problems whether $L \in
  \dnf{(\finelim_{\septraction}(M_1,M_2)}$ and $L \in
  \dnf{(\infelim_{\septraction}(M_1,M_2)}$ are both in
  $\bigonspace{\size{L}+\size{M_1}+\size{M_2}}$.  } 
\proof{ The proof is similar to that of Proposition
  \ref{prop:pspaceand} (again, the formula $\lambda^\gencard$ is
  exponential, but we do not have to construct it explicitly). \qed}

\prop{prop:pspaceconj}{
Checking whether $M \in \conj{M_1,\dots,M_n}$, where $M,M_1,\dots,M_n$
are minterms, $n \geq 2$, is in
$\bigonspace{\size{M}+(\size{M_1} + \ldots + \size{M_n})^2}$.
}
\proof{ The proof is by induction on $n \geq 2$. If $n = 2$ then by
  definition of $\conj{M_1,M_2}$ it suffices to check that $M = M_1^f
  \wedge M_1^e \wedge M_1^a \wedge M_1^p \wedge M_1^u \wedge M_2^f
  \wedge M_2^e \wedge M_2^a \wedge M_2^p \wedge M_2^u \wedge \mu
  \wedge \nu$ for some $\mu \in \minconj{M_1}{M_2}$, $\nu \in
  \maxconj{M_1}{M_2}$. By definition, the size of each formula in
  $\minconj{M_1}{M_2} \cup \maxconj{M_1}{M_2}$ is of the order of
  $\bigO(\size{M_1}+\size{M_2})$, thus the algorithm requires
  $\bigO(\size{M}+\size{M_1}+\size{M_2})$ space.

  If $n > 2$, $M \in \conj{M_1,\dots,M_n} \iff M \in \conj{M',M_n}$
  for some $M' \in \conj{M_1,\dots,M_{n-1}}$.  By Proposition
  \ref{prop:minterm-conj-bound}, the literals in $M'$ are either
  literals from $M_1,\dots,M_{n-1}$ or occur in $\{ \len{U} \geq
  m_1+m_2, \len{U} < m_1+m_2, \len{U} \geq m_1+m_2+1, \len{U} <
  m_1+m_2+1\}$, where $M_1 \cup \dots \cup M_{n-1}$ contains two
  literals $\ell_1$ and $\ell_2$ and $\ell_i$ is of the form $\len{h}
  \geq m_i, \len{h} < m_i, \len{h} \geq \len{U} - m_i$ or $\len{h} <
  \len{U} - m_i$, for $i=1,2$. Thus $\size{M'} \leq \sum_{i=1}^{n-1}
  \size{M_i}$. The nondeterministic algorithm guesses and stores a
  minterm $M'_1$ of size at most $\sum_{i=1}^{n-1} \size{M_i}$ and
  checks that $M \in \conj{M'_1,M_n}$ and that $M'_1 \in
  \conj{M_1,\dots,M_{n-1}}$. According to the base case $n=2$, the
  first check takes up $\bigO(\size{M}+\size{M'_1}+\size{M_n}) =
  \bigO(\size{M} + \sum_{i=1}^n \size{M_i})$ space, and the second
  check takes space $\bigO(\size{M'_1}+(\sum_{i=1}^{n-1}
  \size{M_i})^2) = \bigO((\sum_{i=1}^n \size{M_i})^2)$, by the
  induction hypothesis. Because we only need to store $M'_1$ between
  the two checks, the algorithm takes $\bigO(\size{M}+(\sum_{i=1}^{n}
  \size{M_i})^2)$ space. \qed}

\prop{prop:pspacemt}{
Let $M$ be a minterm and let $L$ be a set of literals. The problem of
checking whether $M = \minterm{L}$ is in
$\bigonspace{\size{M}+(\sum_{\ell\in L} \size{\ell})^2}$.
}
\proof{ By definition, $\minterm{L} =
  [\minterm{\ell_1},\dots,\minterm{\ell_n}]$, with $L = \set{
    \ell_1,\dots,\ell_n}$, and each minterm $\minterm{\ell_i}$ is of
  size $\bigO(\size{\ell_i})$, thus the proof follows immediately from
  Proposition \ref{prop:pspaceconj}. \qed}

\begin{lemma}\label{lemma:pspacemt}
Given a minterm $M$ and an $\seplogk{k}$ formula $\phi$, the problems of
checking whether $M \in \finmt{\phi}$ and $M \in \infmt{\phi}$ are
in \pspace.
\end{lemma}
\proof{ We show the existence of a nondeterministic algorithm that
  decides $M \in \finmt{\phi}$ in space
  $\bigO(\size{M}+\size{\phi}^8)$. The \pspace\ upper bound is by an
  application of Savitch's Theorem \cite{Savitch70}. We only give the
  proof for $M \in \finmt{\phi}$, the proof for $M \in \infmt{\phi}$
  being similar and omitted. By induction on the structure of
  $\phi$, we distinguish the following cases: \begin{compactitem}
  \item $\phi=\emp$: we check $M = \len{h} \teq 0$ in space
    $\bigO(\size{M}+\size{\phi})$.
  \item $\phi=x \mapsto \vec{y}$: we check $M = \set{x \pto \vec{y}
    \wedge \len{h} \teq 1}$ in space $\bigO(\size{M}+\size{\phi})$.
  \item $\phi = \phi_1 \wedge \phi_2$: $M \in \finmt{\phi} \iff M\in
    \conj{M_1,M_2}$ with $M_i \in \finmt{\phi_i}$, for every $i =1,2$.
    Since, by Corollary \ref{cor:sl-mnf-size}, $\size{M_i} =
    \bigO(\size{\phi_i}^2) = \bigO(\size{\phi}^2)$, for $i=1,2$, it
    suffices to guess two such minterms $M_1$ and $M_2$, check that
    $M_i \in \finmt{\phi_i}$, $i=1,2$ and that $M \in
    \conj{M_1,M_2}$. By the induction hypothesis, checking $M_i \in
    \finmt{\phi_i}$ requires space
    $\bigO(\size{M_i}+\size{\phi_i}^8)$, for each $i=1,2$, and by the
    proof of Proposition \ref{prop:pspaceconj} in the case $n=2$,
    checking $M \in \conj{M_1,M_2}$ requires space
    $\bigO(\size{M}+\size{M_1}+\size{M_2})=\bigO(\size{M}+\size{\phi})$.
    Since we only need to store $M_1$ and $M_2$ between the checks,
    the entire procedure takes space $\bigO(\size{M}+\size{\phi}^8)$.
  \item $\phi = \neg \phi_1$: $M \in \finmt{\phi}$ if and only if $M
    \in \conj{\minterm{\overline{\ell_1}}, \ldots,
      \minterm{\overline{\ell_m}}}$, for some literals $\ell_i \in
    M_i$, $i \in [1,m]$, where $\finmt{\phi} = \set{M_1,\dots,M_m}$.
    For any $i \in [1,m]$, we distinguish the following
    cases: \begin{compactitem}
    \item if $\ell_i \in \set{x \pto \vec{y}, \neg x \pto \vec{y} \mid
      x \in \vars, \vec{y} \in \vars^k}$ then $\ell_i$ occurs in
      $\phi_1$, thus there are at most $\size{\phi_1}$ such literals,
    \item if $\ell_i \in \set{x \teq y, \neg x \teq y \mid x,y \in
      \vars}$ then there are at most $2\card{\fv{\phi}}^2$ such
      literals,
    \item if $\ell_i \in \set{\len{U} \geq n, \len{U} < n \mid n \in
      \nat}$, by Lemma \ref{lemma:sl-mnf-bounded}, $\bound{\ell_i}
      \leq 2\bound{\phi_1}+1$, thus there are at most
      $2\bound{\phi_1}+1 = \bigO(\size{\phi_1})^2$ such literals.
    \end{compactitem}
    Summing up, we obtain that $\card{\set{\ell_i \mid i \in [1,m]}} =
    \bigO(\size{\phi_1}^2)$. 
    Thus it suffices to guess a set
    $\set{\ell'_1, \ldots, \ell'_n}$ of literals and a set of minterms
    $\set{M'_1, \ldots, M'_n}$ such that $\ell'_i \in M'_i$, where $n
    = \bigO(\size{\phi_1}^2)$ and $\size{M'_i} =
    \bigO(\size{\phi_1}^2)$, for all $i \in [1,n]$. Then we can check
    that: \begin{compactitem}
    \item $M'_i \in \finmt{\phi_1}$, which can be done in space
      $\bigO(\size{M'_i}+\size{\phi_1}^8) = \bigO(\size{\phi_1}^2 +
      \size{\phi_1}^8) = \bigO(\size{\phi_1}^8)$, by the inductive
      hypothesis,
    \item 
    $M \in \conj{\minterm{\overline{\ell_1}},
      \ldots, \minterm{\overline{\ell_n}}}$,
    %$M \in \conj{\minterm{\finite},\minterm{\overline{\ell_1}},
     % \ldots, \minterm{\overline{\ell_n}}}$,
      which can be done in
      space $\bigO(\size{M}+(n\cdot\size{\phi_1}^2)^2) =
      \bigO(\size{M}+\size{\phi_1}^8)$, by Proposition
      \ref{prop:pspaceconj}.
    \end{compactitem}
     To ensure that the set $\{ \ell_1,\dots,\ell_m \}$ contains no
     literal other than $\ell'_1,\dots,\ell'_n$, we also have to check
     that every minterm $M_j$, for $j \in [1,m]$ contains a literal
     $\ell'_i$, for some $i\in [1,n]$. To this aim, we use a non
     deterministic algorithm for the complement: we guess a minterm
     $M'$ \bounded\ by $\phi_1$, check that $M' \in \mu(\phi_1)$ and
     that it contains no literal $\ell_i$, for $i\in[1,n]$. By the
     inductive hypothesis, this is possible in space
     $\bigO(\size{M'}+\size{\phi_1}^8) = \bigO(\size{\phi_1}^2 +
     \size{\phi_1}^8) = \bigO(\phi_1^8)$. Then, checking that every
     minterm $M_j$, for $j\in[1,m]$ contains a literal $\ell'_i$, for
     some $i\in[1,n]$ can be done in the same amount of space, using a
     nondeterministic algorithm, see e.g. \cite[Corollary
       4.21]{AroraBarak09}.

    %% Note that, although $m$ may be exponential w.r.t.\ the size of $\phi_1$, the  set $\{ \ell_1,\dots,\ell_m \}$ must be of polynomial size, since each literal $\ell_i$ is \bounded\ by $\phi_1$ \todo{use lemma relating \bounded\ sets and size} (there may be repetitions among the $\ell_i$'s). Thus it suffices to guess a set $\{ \ell_1,\dots,\ell_n \}$ of test formulae \bounded\ by $\phi$ (with no repetition, hence we may have $n < m$),
    %% to guess  minterms $M_i$ \bounded\ by $\phi_1$ and containing $\ell_i$ for each $i = 1,\dots,n$, to call recursively the algorithm to check that $M_i \in \mu(\phi_1)$ and to verify that 
    %% $M \in  \conj{\minterm{\overline{\ell_1}}, \ldots,
    %%   \minterm{\overline{\ell_n}}}$ using Proposition \ref{prop:pspaceconj}.
    %%   To ensure that the set $\{ \ell_1,\dots,\ell_m \}$ contains no literal
    %%   other than $\ell_1,\dots,\ell_n$, we also have to check that every minterm $M_j$, for $j = 1,\dots,m$
    %%   contains a literal $\ell_i$, for some $i=1,\dots,n$. To this aim, we may use a non deterministic algorithm checking the negation: 
    %%   we guess a minterm $M'$ \bounded\ by $\phi_1$, check that $M' \in \mu(\phi_1)$ and that it contains no literal $\ell_i$, for $i=1,\dots,n$.

   \item $\phi=\phi_1*\phi_2$: $M \in \finmt{\phi}$ iff there exist
     minterms $M_i \in \mu(\phi_i)$, $N_i \in
     \completion{M_i}{\eqs{M_1\cup M_2}}$ and $P_i \in
     \completion{N_i}{N_{3-i}^p}$, such that $M \in
     \minterm{\dnf{\elim_*(P_1,P_2)}}$, for $i=1,2$. We first guess
     minterms $M_1,M_2$ of size $\bigO(\size{\phi_1}^2)$ and
     $\bigO(\size{\phi_2}^2)$, respectively, check that $M_i \in
     \finmt{\phi_i}$, then guess $N_i \in
     \completion{M_i}{\eqs{M_1\cup M_2}}$ and $P_i \in
     \completion{N_i}{N_{3-i}^p}$, for $i=1,2$. This is feasible since
     by definition each minterm in these sets is of size
     $\bigO(\size{M_1}+\size{M_2})$. Next, we guess minterms $M',
     M''$, of size $\bigO(\size{M_1}+\size{M_2})$ as well, and check
     that $M' \in \dnf{\elim_*(P_1,P_2)}$ in space
     $\bigO(\size{M'}+\size{P_1}+\size{P_2})$, by Proposition
     \ref{prop:pspaceconj} and $M'' \in \minterm{M'}$ in space
     $\bigO(\size{M''}+\size{M'}^2)$, by Proposition
     \ref{prop:pspacemt}.
   \item $\phi_1 \septraction \phi_2$: the proof is similar to the previous case. \qed
\end{compactitem}}

\begin{corollary}\label{cor:pspacemc}
Given a finite $\fol$ structure $\astruct = (\U,\astore,\afunc)$ and a
formula $\forall y_1 \ldots \forall y_m ~.~ \phi$, where $\phi$ is
quantifier-free, the problem $\astruct \models \foltrans{\forall y_1 \ldots
\forall y_m ~.~ \bigvee_{M \in \genmt{\phi}} M}$ is in \pspace, for each
$\gencard \in \set{\fincard,\infcard}$.
\end{corollary}
\proof{ We have the equivalences:
\[\begin{array}{l}
\neg \foltrans{\forall y_1 \ldots \forall y_m ~.~\bigvee_{M \in \genmt{\phi}} M} \equiv \\ 
\neg \forall y_1 \ldots \forall y_m ~.~\foltrans{\bigvee_{M \in \genmt{\phi}} M} \equiv \\ 
\exists y_1 \ldots \exists y_m ~.~ \neg \foltrans{\bigvee_{M \in \genmt{\phi}} M} \equiv \\ 
\exists y_1 \ldots \exists y_m ~.~ \foltrans{\neg \bigvee_{M \in \genmt{\phi}} M} 
\stackrel{\text{(Lemma \ref{lemma:sl-mnf})}}{\equiv} \\ 
\exists y_1 \ldots \exists y_m ~.~ \foltrans{\bigvee_{M \in \genmt{\neg \phi}} M} \equiv \\ 
\bigvee_{M \in \genmt{\neg \phi}} \exists y_1 \ldots \exists y_m ~.~ \foltrans{M}.
\end{array}\]
To check that $\astruct \not\models \foltrans{\forall y_1 \ldots
  \forall y_m ~.~ \bigvee_{M \in \genmt{\phi}} M}$, we guess locations
$\ell_1, \ldots, \ell_m \in \U$ and a minterm $M$ that is \bounded\ by
$\phi$, then check that $M \in \genmt{\neg\phi}$ and that
$(\U,\astore[y_1 \leftarrow \ell_1]\ldots[y_m \leftarrow
  \ell_m],\afunc) \models \foltrans{M}$. The first check is
in \pspace, by Lemma \ref{lemma:pspacemt}. The second check is also
in \pspace, because $\foltrans{M}$ is a $\bsr(\fol)$ formula of size
polynomially bounded by $\size{M}$, and the only quantifiers
introduced by $\foltrans{\ell}$, where $\ell\in M$ is a literal
are: \begin{compactitem}
\item $\exists y_1, \ldots \exists y_k ~.~ \pfunc(x,y_1,\ldots,y_k)$
  if $\ell=\alloc(x)$, and
\item $\forall y_1, \ldots \forall y_k ~.~
  \neg\pfunc(x,y_1,\ldots,y_k)$ if $\ell=\neg\alloc(x)$.
\end{compactitem}
In both cases, one can check whether $\astruct \models \ell$, by inspection of
$\pfunc^\afunc$, which requires no additional space. Finally,
\pspace\ is closed under complement (see, e.g., \cite[Corollary
  4.21]{AroraBarak09}), which concludes the proof. \qed}

\section{Bernays-Sch\"onfinkel-Ramsey $\seplogk{k}$}
\label{sec:bsr-sl}

This section gives the main results of the paper, concerning the
(un)decidability of the (in)finite satisfiability problems within the
$\bsr(\seplogk{k})$ fragment. From the satisfiability point of view,
we can assume w.l.o.g. that $\bsr(\seplogk{k})$ is the set of
sentences $\forall y_1 \ldots \forall y_m ~.~ \phi$, where $\phi$ is a
quantifier-free $\seplogk{k}$ formula, with $\fv{\phi} =
\{x_1,\ldots,x_n, \\ y_1,\ldots,y_m\}$, where the existentially
quantified variables $x_1,\ldots,x_n$ are left free.

First, we show that, contrary to $\bsr(\fol)$, the satisfiability of
$\bsr(\seplogk{k})$ is undecidable for $k\geq2$. Second, we carve two
nontrivial fragments of $\bsr(\seplogk{k})$, for which the infinite
and finite satisfiability problems are
both \pspace-complete. Technically, these fragments are defined based
on restrictions of \begin{inparaenum}[(i)]
\item polarities of the occurrences of the separating implication,
  and
\item occurrences of universally quantified variables in the scope of
  separating implications.
\end{inparaenum}

These results draw a rather precise chart of decidability within the
$\bsr(\seplogk{k})$ fragment, the only remaining gap being the
decidability for the case $k=1$. Indeed, it is known that the fragment
of $\seplogk{1}$ whose formulae do not contain the separating
implication is decidable but not elementary recursive
\cite{BrocheninDemriLozes11}. However, the decidability status for
$\bsr(\seplogk{1})$ with unrestricted uses of the separating
implication is still open.

\subsection{Undecidability of $\bsr(\seplogk{k})$}
\label{sec:bsr-sl-undec}

We consider the finite satisfiability problem of the
$[\forall,(0),(2)]_=$ fragment of $\fol$, which consists of sentences
of the form $\forall x ~.~ \phi(x)$, where $\phi$ is a quantifier-free
boolean combination of atomic propositions $t_1 \teq t_2$, and $t_1,
t_2$ are terms built using two function symbols $f$ and $g$, of arity
one, the variable $x$ and constant $c$. It is known\footnote{See
  e.g.\ \cite[Theorem 4.1.8]{BorgerGradelGurevich97}.} that finite
satisfiability is undecidable for $[\forall,(0),(2)]_=$

Given a sentence $\varphi = \forall x ~.~ \phi(x)$ in
$[\forall,(0),(2)]_=$, we proceed by first \emph{flattening} each term
in $\phi$ consisting of nested applications of $f$ and $g$. The result
is an equivalent sentence $\varphi_{\mathit{flat}} = \forall x_1
\ldots \forall x_n ~.~ \phi_{\mathit{flat}}$, in which the only terms
are $x_i$, $c$, $f(x_i)$, $g(x_i)$, $f(c)$ and $g(c)$, for $i \in
[1,n]$. For example, the formula $\forall x~.~ f(g(x)) \teq c$ is
flattened into $\forall x_1 \forall x_2~.~ g(x_1) \not \teq x_2 \vee
f(x_2) \teq c$. The formal construction is standard and thus
omitted. We define the following $\bsr(\seplogk{2})$ sentences, for
each $\gencard \in \set{\fincard,\infcard}$:
\begin{eqnarray}\label{eq:flat-seplog}
\varphi^\gencard_\tinyseplog \ifLongVersion\else\!\!\!\!\fi & \defequal & \ifLongVersion\else\!\!\!\!\fi \alpha^\gencard \wedge x_c \pto (y_c,z_c) \wedge \\[-2mm]
&& \ifLongVersion\else\!\!\!\!\fi \forall x_1 \ldots \forall x_n \forall y_1 \ldots \forall y_n \forall z_1 \ldots \forall z_n ~.~
\bigwedge_{i=1}^n (x_i \pto (y_i,z_i) \rightarrow \phi_{\tinyseplog}) \nonumber
\end{eqnarray} 
where\footnote{Note that the two definitions of
$\alpha^\fincard$ are equivalent.}:
\[\begin{array}{rcl}
\alpha^\fincard & \defequal & \forall x ~.~ \alloc(x) \text{ or } \alpha^\fincard \defequal \len{h} \geq \len{U} - 0 \\
\alpha^\infcard & \defequal & \forall x \forall y \forall z ~.~ x \pto (y,z) \rightarrow \alloc(y) \wedge \alloc(z)
\end{array}\]
and $\phi_{\tinyseplog}$ is obtained from $\phi_{\mathit{flat}}$ by
replacing each occurrence of $c$ by $x_c$, each term $f(c)$ [$g(c)$]
by $y_{c}$ [$z_c$] and each term $f(x_i)$ [$g(x_i)$] by $y_i$
[$z_i$]. Next, we show that $\varphi$ and $\varphi_\tinyseplog$ are
equisatisfiable, which permits to deduce that:

\begin{theorem}\label{thm:bsr-sl-undec}
  The finite and infinite satisfiability problems are undecidable for
  $\bsr(\seplogk{k})$.
\end{theorem}
\paragraph{Proof} 
Let $\varphi = \forall x ~.~ \phi$ be a sentence from
$[\forall,(0),(2)]_=$. We show that the following statements are
equivalent: \begin{compactenum}
  \item\label{it:fol-fin} $\varphi$ has a finite model
    $(\U,\astore,\afunc)$,
  \item\label{it:bsr-fin} $\varphi^\fincard_\tinyseplog$ has a finite
    model $(\U,\astore',\aheap)$, and
  \item\label{it:bsr-inf} $\varphi^\infcard_\tinyseplog$ has an
    infinite model $(\U^\infty,\astore',\aheap)$.
\end{compactenum}

\noindent ``(\ref{it:fol-fin}) $\Rightarrow$ (\ref{it:bsr-fin})'' We
define the store $\astore' \defequal \astore[x_c \leftarrow
  c^\afunc,y_c \leftarrow f^\afunc(c^\afunc),z_c \leftarrow
  g^\afunc(c^\afunc)]$ and the heap $\aheap$ such that $\dom(\aheap) =
\U$ and $\aheap(\ell) \defequal (f^\afunc(\ell), g^\afunc(\ell))$, for
all $\ell \in \U$. By construction, we have $(\U,\astore',\aheap)
\models \alpha^\fincard \wedge x_c \pto (y_c,z_c)$, because
$\dom(\aheap) = \U$ and $\aheap(c^\afunc) = (f^\afunc(c^\afunc),
g^\afunc(c^\afunc))$. Consider a store $\astore'' \defequal
\astore'[x_i \leftarrow \ell_i, y_i \leftarrow \ell'_i, z_i \leftarrow
  \ell''_i \mid i=1,\ldots,n]$, for an arbitrary set
$\{\ell_i,\ell'_i,\ell''_i \mid i \in [1,n]\} \subseteq \U$ and assume
that $(\U, \astore'', \aheap) \models \bigwedge_{i=1}^n x_i \pto
(y_i,z_i)$. Then by definition of $\aheap$, for all $i\in [1,n]$, we
have $\ell_i' = f^\afunc(\ell_i)$ and $\ell''_i = g^\afunc(\ell_i)$;
hence, $(\U, \astore'', \aheap)\models \phi_{\tinyseplog}$. Since
$\ell_i, \ell'_i$ and $\ell''_i$ are arbitrary, for $i \in [1,n]$,
this proves that $(\U, \astore', \aheap)$ is a finite model of
$\varphi^\fincard_\tinyseplog$.

\noindent``(\ref{it:bsr-fin}) $\Rightarrow$ (\ref{it:bsr-inf})'' We
define $\U^\infty \defequal \U \uplus L$, where $L$ is an infinite set
of locations not in $\U$. Clearly $(\U^\infty,\astore',\aheap) \models
\alpha^\infcard$, because $x \pto (y,z)$ is false for any extension of
$\astore'$ with a pair of the form $[x \leftarrow \ell]$, $[y
  \leftarrow \ell]$ or $[z \leftarrow \ell]$, where $\ell \in
L$. Furthermore, the valuation of $x_c \pto (y_c,z_c)$ is unchanged
between $(\U,\astore',\aheap)$ and
$(\U^\infty,\astore',\aheap)$. Consider a store $\astore'' \defequal
\astore'[x_i \leftarrow \ell_i, y_i \leftarrow \ell'_i, z_i \leftarrow
  \ell''_i \mid i=1,\ldots,n]$, for an arbitrary set
$\{\ell_i,\ell'_i,\ell''_i \mid i \in [1,n]\} \subseteq \U$ and assume
that $(\U, \astore'', \aheap)\models \bigwedge_{i=1}^n x_i \pto
(y_i,z_i)$. Then necessarily, $\set{\ell_i,\ell'_i,\ell''_i \mid i \in
  [1,n]} \cap L = \emptyset$.  Next, we show that
$(\U,\astore'',\aheap) \models \phi_{\tinyseplog} \iff
(\U^\infty,\astore'',\aheap) \models \phi_{\tinyseplog}$, by induction
on the structure of $\phi_{\tinyseplog}$. Since $(\U,\astore'',\aheap)
\models \phi_{\tinyseplog}$ by the hypothesis, we have $(\U^\infty,
\astore'', \aheap)\models \phi_{\tinyseplog}$, thus $(\U^\infty,
\astore, \aheap) \models \varphi^\infcard_\tinyseplog$.
  
\noindent``(\ref{it:bsr-inf}) $\Rightarrow$ (\ref{it:fol-fin})'' Let
$\U \defequal \dom(\aheap) \cup \set{\ell_1,\ell_2 \mid \exists \ell
  \in \U^\infty ~.~ \aheap(\ell) = (\ell_1,\ell_2)}$. Since $\aheap$
is finite, so is $\U$. Let $\astore$ be an arbitrary\footnote{The
  store is arbitrary because $\varphi$ contains no free variables.}
store on $\U$ and define $\afunc$ such that: \begin{compactitem}
  \item $c^\afunc = \astore'(x_c)$, and,
  \item for each $\ell \in \U$, such that $\aheap(\ell) =
    (\ell',\ell'')$, we have $f^\afunc(\ell) = \ell'$ and
    $g^\afunc(\ell) = \ell''$.
\end{compactitem}
Note that $c^\afunc\in \U$, because by hypothesis
$(\U^\infty,\astore',\aheap) \models x_c \pto (y_c,z_c)$, hence
$\astore'(x_c) \in \dom(\aheap)$.  Similarly,
$f^\afunc(\ell),g^\afunc(\ell) \in \U$, for each $\ell \in \U$, by the
definition of $\U$.  Moreover, since $(\U^\infty,\astore',\aheap)
\models \alpha^\infcard$ we obtain that $f^\afunc$ and $g^\afunc$ are
well-defined total functions.  For each set $\set{\ell_i
  \mid i = 1,\ldots,n} \subseteq \U$, the function $\astore'' =
\astore[x_i \leftarrow \ell_i, y_i \leftarrow f^{\afunc}(\ell_i), z_i
  \leftarrow g^\afunc(\ell_i) \mid i=1,\ldots,n]$ is a store on
$\U^\infty$ such that $(\U^{\infty},\astore'',\aheap) \models x_i \pto
(y_i,z_i)$ for every $i\in [1,n]$, hence
$(\U^{\infty},\astore'',\aheap) \models \phi_{\tinyseplog}$.  By
induction on the structure of $\phi$, one shows that
$(\U^\infty,\astore'',\aheap) \models \phi_{\tinyseplog} \iff
(\U,\astore'',\afunc) \models \phi_{\mathit{flat}}$. Since
$(\U^\infty,\astore'',\aheap) \models \phi_{\tinyseplog}$, we have
$(\U,\astore,\afunc) \models \phi_{\mathit{flat}}$. \qed

\subsection{Two Decidable Fragments of $\bsr(\seplogk{k})$}
\label{sec:bsr-sl-dec}

The reductions (\ref{eq:flat-seplog}) use either positive occurences
of $\alloc(x)$, where $x$ is universally quantified, or test formulae
$\len{h} \geq \len{U}-n$. We obtain decidable subsets of
$\bsr(\seplogk{k})$ by eliminating the positive occurrences of
both 
\begin{inparaenum}[(i)]
\item $\alloc(x)$, with $x$ universally quantified, and 
\item $\len{h} \geq \len{U}-n$,
\end{inparaenum}
from $\genmt{\phi}$, where $\gencard \in \set{\fincard,\infcard}$ and
$\forall y_1 \ldots \forall y_m ~.~ \phi$ is any $\bsr(\seplogk{k})$
formula.  Note that $\infmt{\phi}$ does not contain 
formulae of the form $\len{h} \geq \len{U}-n$ anyway, which explains why 
slightly less restrictive conditions are needed for infinite structures.

\begin{definition}\label{def:bsr-decidable-fragments}
  Given an integer $k \geq 1$, we define: \begin{compactenum}
  \item\label{it:bsr-inf-def} $\infbsrsl{k}$ as the set of sentences
    $\forall y_1 \ldots \forall y_m ~.~ \phi$ such that for all $i \in
    [1,m]$ and all formulae $\psi_1 \wand \psi_2$ occurring at
    polarity $1$ in $\phi$, we have $y_i \not\in \fv{\psi_1} \cup
    \fv{\psi_2}$,
  \item\label{it:bsr-fin-def} $\finbsrsl{k}$ as the set of sentences
    $\forall y_1 \ldots \forall y_m ~.~ \phi$ such that no formula
    $\psi_1 \wand \psi_2$ occurs at polarity $1$ in $\phi$.
  \end{compactenum}
\end{definition}
\noindent
Note that $\finbsrsl{k} \subsetneq \infbsrsl{k} \subsetneq
\bsr(\seplogk{k})$, for any $k \geq 1$.

We start by showing decidability, in \pspace, of the infinite
satisfiability problem for the $\infbsrsl{k}$ fragment. To this end,
given a $\bsr(\fol)$ formula $\varphi$, whose only function of a
strictly positive arity is the boolean function $\pfunc$, we provide
an axiom $\infaxiom{\varphi}$ that guarantees the existence of an
infinite model for $\varphi$.

\begin{definition}
\label{def:infaxiom}
Let $\varphi = \forall y_1 \ldots \forall y_m ~.~ \phi$ be a
$\bsr(\fol)$ formula, where $\set{x_1,\dots,x_n}$ is the set of
constants and free variables in $\varphi$ and $\phi$ is
quantifier-free. Let $c_1,\dots,c_m$ be pairwise distinct constants
such that $\set{c_1,\ldots,c_m} \cap \set{x_1,\ldots,x_n} =
\emptyset$. We define:
\begin{eqnarray*}
\varphi_\infty \ifLongVersion\else\!\!\!\!\fi & \defequal & 
\ifLongVersion\else\!\!\!\!\fi \bigwedge_{i=1}^m \forall z_1 \ldots \forall z_k ~.~ \neg \pfunc(c_i,z_1,\ldots,z_k) ~\wedge \\[-1mm]
&& \bigwedge_{\begin{array}{l}
    \scriptscriptstyle{i,j \in [1,m]} \\[-2mm]
    \scriptscriptstyle{i \neq j}
\end{array}} \!\!\!\! c_i \not \teq c_j \wedge 
\ifLongVersion\else\!\!\!\!\fi \bigwedge_{\begin{array}{l}
    \scriptscriptstyle{i \in [1,n]} \\[-2mm]
    \scriptscriptstyle{j \in [1,m]}
\end{array}} \!\!\!\! x_i \not \teq c_j ~\wedge \\[-1mm]
&& \forall x \forall z_1 \ldots \forall z_{k} ~. \!\!\!\!
\bigwedge_{\begin{array}{l}
    \scriptscriptstyle{\ell \in [1,k]} \\[-2mm]
    \scriptscriptstyle{j \in [1,m]}
\end{array}} 
\ifLongVersion\else\!\!\!\!\fi \neg \pfunc(x,z_1,\dots,z_{\ell-1}\,c_j,z_{\ell+1},\dots,z_k)
\end{eqnarray*}
\end{definition}

\begin{proposition}\label{prop:infinite}
Let $\varphi = \forall y_1 \ldots \forall y_m ~.~ \phi$ be a
$\bsr(\fol)$ formula, where $\phi$ is a quantifier-free formula built
on a signature containing only constants and the boolean function
symbol $\pfunc$, of arity $k+1$. The formula $\varphi \wedge
\infaxiom{\varphi}$ is satisfiable iff $\varphi$ has an infinite model
$(\U, \astore, \afunc)$, such that $\card{\pfunc^\afunc} \in \nat$.
\end{proposition}
\proof{ Let $\astruct = (\U,\astore,\afunc)$ be a model of $\varphi
  \wedge \infaxiom{\varphi}$.  Since $\varphi \wedge
  \infaxiom{\varphi}$ is in $\bsr(\fol)$, we may assume that $\U$ is
  finite by \cite[Proposition 6.2.17]{BorgerGradelGurevich97}, hence
  $\card{\pfunc^{\afunc}} \in \nat$ as well. Consider an extension
  $\U'$ of $\U$ obtained by adding infinitely many new elements. Let
  $\vec{a} = (a_1,\dots,a_m)$ be a vector of elements in $\U'$ and
  $\vec{b} = (b_1,\dots,b_m)$ be a vector of elements in $\U$ such
  that for all $i,j$: \begin{compactitem}
    \item $b_i = b_j$ iff $a_i = a_j$, 
    \item if $a_i \in \U$ then $a_i = b_i$, and 
    \item if $a_i \not \in \U$ then $b_i \in
      \set{\astore(c_j) \mid j \in [1,m]}$. 
  \end{compactitem}
  It is straightforward to verify that such a sequence always exists.
  Furthermore, by definition of $\infaxiom{\varphi}$, $c_i^\afunc$
  does not occur in $\pfunc^\afunc$, hence $(\U,\astore[y_1 \leftarrow
    b_1] \ldots [y_m \leftarrow b_m],\afunc)$ and $(\U',\astore[y_1
    \leftarrow a_1] \ldots [y_m \leftarrow a_m],\afunc)$ coincide on
  every atom in $\phi$.  Since $(\U,\astore,\afunc) \models \varphi$
  we deduce that $(\U,\astore[y_1 \leftarrow b_1] \ldots [y_m
    \leftarrow b_m],\afunc) \models \phi$, hence $(\U',\astore[y_1
    \leftarrow a_1] \ldots [y_m \leftarrow a_m], \afunc) \models
  \phi$.  Since $\vec{a}$ is arbitrary, we deduce that
  $(\U',\astore,\afunc) \models \varphi$.  The converse is
  immediate. \qed}

The first decidability result of this paper is stated below: 

\begin{theorem}\label{thm:pspaceBSR}
  For any integer $k\geq1$ not depending on the input, the infinite
  satisfiability problem for $\infbsrsl{k}$ is \pspace-complete.
\end{theorem}
\paragraph{Proof}
  \pspace-hardness is an immediate consequence of the fact that the
  quan\-ti\-fier-free fragment of $\seplogk{k}$, without the
  separating implication, but with the separating conjunction and
  negation, is \pspace-hard \cite[Proposition
    5]{CalcagnoYangOHearn01}.

  To show membership in \pspace, let $\varphi = \forall y_1 \ldots
  \forall y_m ~.~ \phi$ be a sentence in $\infbsrsl{k}$, where $\phi$
  is quantifier-free and $\fv{\phi} =
  \set{x_1,\ldots,x_n,y_1,\ldots,y_m}$. Let $\varphi' \defequal
  \forall y_1 \ldots \forall y_m ~.~ \bigvee_{M \in \infmt{\phi}}
  M$. If $\alloc(x)$ occurs in a minterm in $\infmt{\phi}$, then by
  Lemma \ref{lemma:polarity}, $x$ necessarily occurs in the scope of a
  positive occurrence of $\wand$, which entails by definition of
  $\infbsrsl{k}$ that $x \not \in \{ y_1,\dots,y_n\}$.  Therefore, by
  Lemma \ref{lem:bsrfoltrans}, there exists a $\bsr(\fol)$ formula
  equivalent to $\foltrans{\varphi'}$, with the same constants and
  free variables as $\foltrans{\varphi'}$. Consequently, the same
  holds for the formula $\psi \defequal \foltrans{\varphi'} \wedge
  \axioms{\varphi'}$ (Definition \ref{def:maxn-axioms}). Let
  $\infaxiom{\psi}$ be the formula defined in Definition
  \ref{def:infaxiom}.  By definition, $\infaxiom{\psi}$ is in
  $\bsr(\fol)$ and contains exactly $m$ constants not occurring in
  $\psi$. Thus $\psi \wedge \infaxiom{\psi}$ has a model iff it has a
  model $(\U,\astore,\afunc)$, with $\card{\U} \leq \max(1,p+m+n)$,
  where $p$ denotes the number of constants and free variables in
  $\axioms{\varphi'}$ \cite[Proposition
    6.2.17]{BorgerGradelGurevich97}. We have $p =
  \bigO(\maxn{\varphi'})$ and, by Corollary \ref{cor:sl-mnf-size},
  $p=\bigO(\size{\varphi}^2)$, so that $\card{\U} =
  \bigO(\size{\varphi}^2)$. Then we can guess a $\fol$-structure
  $(\U,\astore,\afunc)$ such that $\card{\U} =
  \bigO(\size{\varphi}^2)$ and check that $(\U,\astore,\afunc) \models
  \foltrans{\varphi'} \wedge \axioms{\varphi'} \wedge
  \infaxiom{\psi}$. This test is feasible
  in \pspace:\begin{compactitem}
    \item the problem $(\U,\astore,\afunc) \models
      \foltrans{\varphi'}$ is in \pspace\ by Lemma \ref{cor:pspacemc},
  \item the problem $(\U,\astore,\afunc) \models \axioms{\varphi'}$ is
    in $\mathsf{P}$, because $\axioms{\varphi'}$ is a conjunction of
    $\bigO(p)$ universally quantified formulae, each having a constant
    number of universal quantifiers, because $k$ does not depend on
    the input, and
  \item the problem $(\U,\astore,\afunc) \models \infaxiom{\psi}$ is
    also in $\ptime$, since it suffices to check that the
    $c_1^\afunc,\dots,c_m^\afunc$ are pairwise distinct, distinct from
    $\astore(x_1),\dots,\astore(x_n)$ and, moreover, do not occur in
    $\pfunc^\afunc$.
    \end{compactitem}
  Finally, by Proposition \ref{prop:infinite}, $\psi \wedge
  \infaxiom{\psi}$ is satisfiable iff $\psi$ admits an infinite model
  for which $\card{\pfunc^\afunc} \in \nat$.  By Lemma
  \ref{lemma:test-fo-sat}, the latter property holds iff $\varphi$ has
  an infinite model. \qed

In the remainder, we prove that finite satisfiability
is \pspace-complete for the class $\finbsrsl{k}$, defined as the set
of formulae with no positive occurrence of separating
implications. Even with this stronger restriction, the previous proof
based on a translation to first-order logic requires an additional
argument. The problem is that, in the case of a finite universe,
$\alloc(x)$ test formulae may occur at a positive polarity, even if
every $\phi_1 \wand \phi_2$ subformula occurs at a negative polarity,
due to the positive occurrences of $\alloc(x)$ within
$\lambda^\fincard$ (\ref{eq:septraction-elim-lambda}) in the
definition of $\finelim_\septraction(M_1,M_2)$, used for the
elimination of separating implications. As previously discussed
(Remark \ref{rem:test-fol-alloc}), positive occurrences of $\alloc(x)$
hinder the translation into $\bsr(\fol)$, because of the existential
quantifiers that may occur in the scope of a universal quantifier.

\begin{figure}[thb]
  \centerline{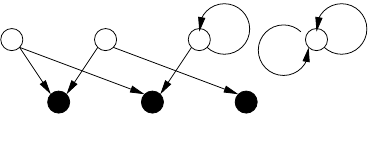}
  \caption{A finite \controlledk{3}\ $\seplogk{2}$
    structure.}\label{fig:controlled}
  \vspace*{-\baselineskip}
\end{figure}

The solution is to distinguish a class of finite structures
$(\U,\astore,\aheap)$, so-called \emph{{\controlled} structures}, for
some $\maxi \in \nat$, for which there exists a set of locations
$\ell_1,\ldots,\ell_\maxi$, such that every location $\ell \in \U$ is
either $\ell_i$ or points to a tuple from the set $\set{\ell_1,
  \ldots, \ell_\maxi,\ell}$. An example of a \controlledk{3} structure
is given in Figure \ref{fig:controlled}.

\begin{definition}\label{def:controlled}
  A structure $\I$ is \controlled\ iff $\I \models \contAx$, where
  \ifLongVersion
  \[\contAx \defequal \exists \kconst_1 \ldots \exists \kconst_n
  \forall x ~.~ \bigvee_{i=1}^\maxi x \teq \kconst_i \vee
  \bigvee_{\vec{y} \in \allvect{\kconst_1,\ldots,\kconst_\maxi,x}} x
  \pto \vec{y}\] \else \(\contAx \defequal \exists \kconst_1 \ldots
  \exists \kconst_n \forall x ~.~ \bigvee_{i=1}^\maxi x \teq \kconst_i
  \vee \bigvee_{\vec{y} \in
    \allvect{\kconst_1,\ldots,\kconst_\maxi,x}} x \pto \vec{y}\)\fi
  and $\allvect{S}$ is the set of $k$-tuples of symbols in $S$.
\end{definition}
Note that any \controlled\ structure is finite, since $\U =
\dom(\aheap) \cup \set{\astore(\kconst_1), \ldots,
  \astore(\kconst_\maxi)}$, but its cardinality is not necessarily
bounded. Furthermore, if $\card{\U} \leq \maxi$, then
$(\U,\astore,\aheap)$ is necessarily \controlled, since we can
extend $\astore$ to a store $\astore'$ such that $\U \subseteq
\set{\astore'(\kconst_1), \ldots, \astore'(\kconst_\maxi)}$.

For a sentence $\varphi = \forall y_1 \ldots \forall y_m ~.~ \phi$ in
$\finbsrsl{k}$, we distinguish the following cases:\begin{compactenum}
\item If $\varphi$ has an \controlled\ model $\I$, the formula
  obtained by replacing each occurrence of an $\alloc(x)$ with
  $\bigwedge_{i=1}^\maxi (x \teq \kconst_i \rightarrow
  \alloc(\kconst_i))$ in $\forall y_1 \ldots \forall y_m$ $\bigvee_{M
    \in \finmt{\phi}} M$ is satisfied by $\I$,\footnote{If $\I$ is
    \controlled, then $\I \models \alloc(x) \leftrightarrow
    \bigwedge_{i=1}^\maxi (x \teq \kconst_i \rightarrow
    \alloc(\kconst_i))$.}
\item Otherwise, each finite model of $\varphi$ is \ncontrolled\ and
  we can build a model $\I$, with a sufficiently large universe, such
  that each test formula $\theta \in \set{\len{U} \geq n, \len{h} <
    \len{U} - n \mid n \in \nat}$ becomes true in $\I$. Because each
  positive occurrence of $\alloc(x)$ in a $\lambda^\fincard$
  (\ref{eq:septraction-elim-lambda}) subformula of some
  $\finelim_\septraction(M_1,M_2)$ formula generated by the
  elimination of the separating implication from $\phi$ occurs in
  disjunction with a formula $\len{h} < \len{U}-n_1 \wedge \len{U}\geq
  n_2$, which is satisfied by $\I$, its truth value in $\I$ can be
  ignored.
\end{compactenum}
In both cases, we obtain an equisatisfiable universally quantified
boolean combination of test formulae with no positive occurrence of
$\alloc(y_i)$ formulae, for any universally quantified variable
$y_i$. We translate this into an equisatisfiable $\bsr(\fol)$
sentence, for which finite satisfiability is decidable and apply a
similar argument to that for the infinite case, to obtain
the \pspace\ upper bound.

\begin{lemma}\label{lemma:pspace-m-controlled-model}
Given a formula $\varphi \in \finbsrsl{k}$ and a number $\maxi\in\nat$
encoded in unary, the problem whether $\varphi$ has an
\controlled\ model is in \pspace.
\end{lemma}
\proof{ Let $x_1,\ldots,x_n$ be the existentially quantified variables
  occurring in $\varphi$ and let $\varphi'$ be the prenex form of
  $\varphi \wedge \contAx$. It is clear that $\varphi'$ is of the form
  $\forall y_1 \ldots \forall y_m ~.~ \phi$, where $\phi$ is
  quantifier-free. Moreover, by definition, $\varphi$ has an
  \controlled\ model iff $\varphi'$ has a model, and this model is
  necessarily finite.

  We denote by $\Tcont{\phi}$ the formula obtained from $\bigvee_{M\in
    \finmt{\phi}} M$ by replacing every formula $\alloc(x)$ with
  $\bigwedge_{i=1}^\maxi (x \teq \kconst_i \rightarrow
  \alloc(\kconst_i))$. Let $\psi \defequal \forall y_1 \ldots \forall
  y_m ~.~ \Tcont{\phi}$ and $\psi' \defequal \forall y_1 \ldots
  \forall y_m ~.~\bigvee_{M\in \finmt{\phi}} M$. We have $\psi \equiv
  \psi'$ because, in each structure $(\U,\astore,\aheap) \models
  \contAx$, all locations must be allocated, except possibly for
  $\astore(\kconst_1), \ldots, \astore(\kconst_\maxi)$, thus
  $(\U,\astore,\aheap) \models \alloc(x) \leftrightarrow
  \bigwedge_{i=1}^\maxi (x \teq \kconst_i \rightarrow
  \alloc(\kconst_i))$.

  The formula $\psi$ contains no occurrence of $\alloc(y_i)$, since by
  definition the only test formulae $\alloc(x)$ occurring in
  $\Tcont{\phi}$ are such that $x \in
  \set{\kconst_1,\ldots,\kconst_\maxi}$. Thus, by Lemma
  \ref{lem:bsrfoltrans}, $\foltrans{\psi}$ is equivalent to a formula
  in $\bsr(\fol)$ with the same free variables and
  constants. Therefore, the same holds for $\foltrans{\psi} \wedge
  \axioms{\psi}$, and $\foltrans{\psi} \wedge \axioms{\psi}$ has a
  model iff it has a model $(\U,\astore,\afunc)$, with $\card{\U} =
  n+\maxi+\bigO(\maxn{\Tcont{\phi}})$, since $\foltrans{\psi}$
  contains $n+\maxi$ free variables and some constants from
  $\axioms{\psi}$, and the number of constants and free variables in
  $\axioms{\psi}$ is bounded by $\bigO(\maxn{\Tcont{\phi}})$.  We have
  $\maxn{\Tcont{\phi}} = \maxn{\psi'}$, thus by Corollary
  \ref{cor:sl-mnf-size}, $\maxn{\Tcont{\phi}}$ is bounded by
  $\bigO(\size{\varphi'}^2)$, and $\card{\U} =
  \bigO(\size{\varphi'}^2) = \bigO((\size{\varphi}+\maxi)^2)$.

  Therefore $\foltrans{\psi} \wedge \axioms{\psi}$ has a model iff it
  has a model $(\U,\astore,\afunc)$, with $\card{\U} =
  \bigO((\size{\varphi}+\maxi)^2)$. Then we can guess a structure
  $(\U,\astore,\afunc)$ such that $\card{\U} =
  \bigO((\size{\varphi}+\maxi)^2)$ and check in polynomial space
  whether $(\U,\astore,\afunc) \models \foltrans{\psi}$, by Corollary
  \ref{cor:pspacemc}, and whether $(\U,\astore,\afunc) \models
  \axioms{\psi}$, as it is done in the proof of Theorem
  \ref{thm:pspaceBSR}. By Lemma \ref{lemma:test-fo-sat},
  $\foltrans{\psi} \wedge \axioms{\psi}$ has a model iff $\psi$ has a
  finite model. Since $\psi \equiv \psi'$ we deduce by Lemma
  \ref{lemma:sl-mnf} that $\foltrans{\psi} \wedge \axioms{\psi}$ has a
  finite model iff $\varphi'$ has a finite model iff $\varphi$ has an
  \controlled\ model.\qed}

\begin{lemma}\label{lemma:non-controlled-model}
  Let $m\geq1$ be an integer and $(\U,\astore,\aheap)$ be a finite
  \ncontrolled\ structure, where $\maxi > (k+1) \times (\card{\D}+m)$
  for a set $\D \subseteq \U$. Then for any universe $\U' \supseteq
  \U$ and each tuple $(a_1,\ldots,a_m) \in {(\U')}^m$ there exists a
  tuple $(b_1,\ldots,b_m) \in \U^m$ such that, for all $i,i'\in[1,m]$:
  \begin{compactenum}
  \item\label{b:condeqx} if $a_i \in \U$ then $b_i = a_i$,
  \item\label{b:condeqy} $b_i = b_{i'}$ iff $a_i = a_{i'}$,
  \item\label{b:condx} if $a_i \not \in \U$ then for no $\ell \in \D
    \cap \dom(\aheap)$ does $b_i$ occur in $\aheap(\ell)$,
  \item\label{b:condy} if $a_i \not \in \U$ then $\aheap(b_i)$ is
    either undefined or contains a location that does not occur in $\{
    b_1,\dots,b_m \} \cup \D$.
  \end{compactenum}
\end{lemma}
\proof{
  Let $\aset \defequal \set{p \in \interval{1}{m} \mid
    a_p \in \U}$ and $\aset' \isdef \set{j_1, \ldots, j_{m'}} \isdef
  \interval{1}{m} \setminus \aset$. For all $p \in \aset$, we set $b_p
  \isdef a_p$. The sequence $b_{j_i}$ for $1\leq i\leq m'$ is
  constructed inductively as follows. Assume that $b_{j_1},\ldots,
  b_{j_i}$ is constructed.  If $a_{j_{i+1}} = a_{j_p}$, for some $p \in
  \interval{1}{i}$ then we set $b_{j_{i+1}} \isdef b_{j_p}$.  Otherwise,
  we define the sets $S_i$ and $S_i'$ as follows:
  \[\begin{array}{rcl}
  S_i & \defequal & \{b_p \ \mid \ p \in \aset\} \cup \{b_{j_1},\ldots,b_{j_i}\} \cup \D \\ 
  S_i' & \defequal & \set{\alocation_1, \ldots, \alocation_k \mid 
    \alocation \in \dom(\aheap) \cap S_i \text{ and } \aheap(\alocation) =
    \sequence{\alocation_1}{\alocation_k}}.
  \end{array}\]
  By definition, $T \isdef S_i \cup S_i'$ contains less than
  $(k+1)\times(n+m)$ elements: indeed, there are at most $n+m$ elements
  in $S_i$, and for each element $\alocation$ in $S_i$, there are at
  most $k$ elements in $\aheap(\alocation)$. By hypothesis,
  $(k+1)\times(n+m) < \maxi$.  Since $(\U,\astore,\aheap)$ is
  \ncontrolled, this means that there exists a location $\ell \in \U
  \setminus T$, such that $\aheap(\ell)$ is either undefined or contains
  a location not occurring in $T \cup \set{\ell}$. We then let
  $b_{j_{i+1}} \isdef \ell$. 

  Conditions \ref{b:condeqx}-\ref{b:condx} are a straightforward
  check. We prove that condition~\ref{b:condy} also holds, by
  contradiction. Suppose that $a_{j_i} \not \in \U$ and $b_{j_i} \in
  \dom(\aheap)$, where $\aheap(b_{j_i}) =
  \sequence{\alocation_1}{\alocation_k}$ and $\set{\alocation_1,
    \ldots, \alocation_k} \subseteq \set{ b_1,\dots,b_m} \cup \D$. By
  construction of $b_{j_i}$, $\aheap(b_{j_i})$ contains an element,
  say $\alocation_{s}$, for some $s \in \interval{1}{k}$, that does
  not belong to $S_{i-1} \cup S_{i-1}' \cup \{b_{j_i}\}$. Since
  $S_{i-1} = \{b_p \ \mid \ p \in \aset\} \cup \{b_{j_1},\ldots,
  b_{j_{i-1}}\} \cup \D$, necessarily, $\alocation_{s} = b_{j_{i'}}$
  for some $i < i' \leq m'$, because $\alocation_s \in \set{
    b_1,\dots,b_m} \cup \D$ by hypothesis. However, $b_{j_{i}} \in
  S_{i'-1}$, because $i < i'$, and therefore $\alocation_{s} \in
  S_{i'-1}'$, and we cannot have $b_{j_{i'}} = \alocation_{s}$,
  contradiction. \qed}

We state below the second decidability result of the paper, concerning
the decidability of the finite satisfiability for $\finbsrsl{k}$:

\begin{theorem}\label{thm:pspaceBSRfin}
For any integer $k\geq1$, not depending on the input, the finite
satisfiability problem for $\finbsrsl{k}$ is $\pspace$-complete.
\end{theorem}
\paragraph{Proof} 
$\pspace$-hardness is proved using the same argument as in the proof
of Theorem \ref{thm:pspaceBSR}, which does not rely on the finiteness
of the universe.

Let $\varphi = \forall y_1,\dots,y_m~.~ \phi$ be a sentence in
$\finbsrsl{k}$, where $\phi$ is quantifier-free and $\fv{\phi} =
\{x_1,\ldots,x_n,y_1,\ldots,y_m\}$. Let $\maxi
\defequal(\max((k+1)\times(n+m),\Nmaxi)+1$, with $\Nmaxi \defequal
\maxn{\bigvee_{M \in \finmt{\phi}} M}$. We first check whether
$\varphi$ admits an \controlled\ model, which can be done in \pspace,
by Lemma \ref{lemma:pspace-m-controlled-model} since $\Nmaxi =
\bigO(\size{\varphi}^2)$, thus $\maxi=\bigO(\size{\varphi}^2)$. In
this case, $\varphi$ has a finite model, and otherwise $\varphi$ has a
finite model iff it has a \ncontrolled\ finite model. We now assume
that $\varphi$ does not have an {\controlled} model.

Let $\varphi' \defequal \forall y_1,\dots,y_m~.~ \Tnocont{\phi}$,
where $\Tnocont{\phi}$ is obtained by replacing all positive
occurrences of $\alloc(x)$, with $x \in \{x_1, \ldots, x_n,
y_1,\ldots, y_m\}$, by $\bot$, in $\bigvee_{M \in \finmt{\phi}} M$. We
shall prove that $\varphi'$ has a finite model iff $\varphi$ has a
finite \ncontrolled\ model. By Lemma \ref{lemma:sl-mnf}, $\varphi$ has
a finite model iff $\forall y_1,\dots,y_m~. \bigvee_{M \in
  \finmt{\phi}} M$ has a finite model. Because the replaced
occurrences of $\alloc(x)$ are all positive, it is clear that
$\varphi' \models \forall y_1,\dots,y_m~. \bigvee_{M \in \finmt{\phi}}
M$, thus the direct implication holds. Now, assume that $\forall
y_1,\dots,y_m~. \bigvee_{M \in \finmt{\phi}} M$ admits a
\ncontrolled\ finite model $(\U,\astore,\aheap)$. In particular, we
have $\card{\U} > \maxi$.

We build a structure $(\U',\astore,\aheap)$, where $\U'$ is obtained
by adding $\max(0, \card{\aheap}+\Nmaxi-\card{\U}+1)$ fresh locations
to $\U$, so that $\card{\U'} > \card{\aheap}+\Nmaxi$. Let $\vec{a} =
(a_1,\dots,a_m) \in {(\U')}^m$ be an arbitrary tuple of locations. We
show that $(\U',\astore[y_i \leftarrow a_i \mid i=1,\ldots,n], \aheap)
\models \Tnocont{\phi}$. By Lemma \ref{lemma:non-controlled-model},
taking $\D = \set{\astore(x_i) \mid i \in [1,n]} \subseteq \U$, there
exists a tuple $\vec{b} = (b_1,\ldots,b_m) \in \U^m$ satisying the
conditions (\ref{b:condeqx}), (\ref{b:condeqy}), (\ref{b:condx}) and
(\ref{b:condy}) of Lemma \ref{lemma:non-controlled-model}. Let
$\astore_a \defequal \astore[y_i \leftarrow a_i \mid i=1,\ldots,m]$
and $\astore_b \defequal \astore[y_i \leftarrow b_i \mid
  i=1,\ldots,m]$. Since $(\U,\astore,\aheap) \models \forall y_1
\ldots \forall y_m ~.~\bigvee_{M \in \finmt{\phi}} M$, we have
$(\U,\astore_b,\aheap) \models \bigvee_{M \in \finmt{\phi}} M$, hence
there exists a minterm $M \in \finmt{\phi}$ such that
$(\U,\astore_b,\aheap) \models M$. We show that for all literals
$\ell$ occurring in $M$: $(\U,\astore_b,\aheap) \models \ell
\Rightarrow (\U',\astore_a,\aheap) \models \ell$, by distinguishing
the following cases:

\noindent
$\underline{\ell \in \set{x \teq y, \neg x \teq y}}$. By conditions
(\ref{b:condeqx}) and (\ref{b:condeqy}) of Lemma
\ref{lemma:non-controlled-model}, $(\U,\astore_b,\aheap)$ and
$(\U',\astore_a,\aheap)$ coincide on every equational atom $x \teq y$.

\noindent
$\underline{\ell \in \set{t_0 \pto (t_1, \ldots, t_k), \neg t_0 \pto
    (t_1, \ldots, t_k)}}$. We show that $(\U,\astore_b,\aheap) \models
t_0 \pto (t_1, \ldots, t_k)$ iff $(\U',\astore_a,\aheap) \models t_0
\pto (t_1, \ldots, t_k)$. Note that we have $t_0,\ldots,t_k \in
\set{x_1,\ldots,x_n,y_1,\ldots,y_m}$. Let $Y \defequal \{i \in [1,m]
\mid y_i \in \set{t_0,\ldots,t_k}\}$. We distinguish several
cases:  \begin{compactenum}
\item $\set{a_i \ \mid \ i \in \asetbis} \subseteq \U$. In this case,
  for all $j \in \interval{0}{k}$, we have $\astore_b(t_j) =
  \astore_a(t_j)$. This is immediate when $t_j \in
  \set{x_1,\ldots,x_n}$, and when $t_j = y_{j'}$ for some $j' \in
  \asetbis$, this is due to the fact that $b_{j'} = a_{j'}$ by
  condition (\ref{b:condeqx}) of Lemma
  \ref{lemma:non-controlled-model}. Consequently,
  $(\U,\astore_b,\aheap) \models t_0 \pto (t_1, \ldots, t_k)$ iff
  $(\U',\astore_a, \aheap) \models t_0 \pto (t_1, \ldots, t_k)$.
\item There is an $i \in \asetbis$ such that $a_i \not \in \U$. Then
  $(\U',\astore_a,\aheap) \not\models t_0 \pto (t_1, \ldots, t_k)$,
  since $a_i \not \in \U$. We show that $(\U,\astore_b,\aheap) \not
  \models t_0 \pto (t_1, \ldots, t_k)$ by distinguishing the following
  cases:

  \begin{compactenum}
  \item $t_0 = y_i$. By condition (\ref{b:condy}) of Lemma
    \ref{lemma:non-controlled-model}, either $\aheap(b_i)$ is
    undefined or it contains an element that does not occur in $\{
    b_1,\dots,b_m \} \cup D$. In the first case,
    $(\U,\astore_b,\aheap) \not \models t_0 \pto (t_1, \ldots, t_k)$.
    In the second case, assuming that $\aheap(b_i) =
    \sequence{\alocation_1}{\alocation_k}$, there is an $s \in
    \interval{1}{k}$ such that $\alocation_s \not \in \{ b_1,\dots,b_m
    \} \\ \cup \D$.  Since $\astore_b(t_s) \in \set{b_1,\ldots, b_m}
    \cup \D$, we obtain $(\U,\astore_b,\aheap) \not \models t_0 \pto
    (t_1, \ldots, t_k)$.
  \item $t_0 \in \set{x_1,\ldots,x_n}$ and $t_s = y_i$ for some $s \in
    \interval{1}{k}$. By condition (\ref{b:condx}) of Lemma
    \ref{lemma:non-controlled-model}, $b_i$ does not occur in the
    tuple $\aheap(\alocation)$ for any $\alocation \in \D \cap
    \dom(\aheap)$. Consequently, $(\U,\astore_b,\aheap) \not \models
    t_0 \pto (t_1, \ldots, t_k)$.
  \item $t_0 = y_j$ with $a_j = b_j \in \U$ and $t_s = y_i$ for some
    $s \in \interval{1}{k}$. Then, by condition (\ref{b:condx}) of
    Lemma \ref{lemma:non-controlled-model}, $b_i$ is not an element of
    $\aheap(b_j)$, thus $(\U,\astore_b,\aheap) \not \models t_0 \pto
    (t_1, \ldots, t_k)$.
  \end{compactenum}
\end{compactenum}

\noindent
$\underline{\ell \in \setof{\len{h} \geq \len{U}-i}{i\in \nat}}$. This
case is impossible, because $\varphi$ contains no positive occurrence
of $\wand$ and by Lemma \ref{lemma:polarity}, $\bigvee_{M \in
  \finmt{\phi}} M$ contains no positive literal $\len{h} \geq
\len{U}-i$.

\noindent
$\underline{\ell \in \set{\len{U} \geq i, \len{h} < \len{U}-i \mid i
    \in \nat}}$. We obtain $(\U',\astore_a,\aheap) \models \ell$,
because $\card{\U'} > \Nmaxi \geq i$ by definition of $\Nmaxi$.

\noindent
$\underline{\ell \in \setof{\len{U} < i}{i\in\nat}}$.  These formulae
are false in $(\U,\astore_b,\aheap)$, since $\card{\U} > \maxi >
\Nmaxi \geq i$ by definition of $\Nmaxi$.

\noindent
$\underline{\ell = \neg\alloc(x)}$. If $\astore_a(x) \in \U$ then
$(\U,\astore_b,\aheap)$ and $(\U',\astore_a,\aheap)$ coincide on
$\alloc(x)$. Otherwise, $\astore_a(x) \not\in \dom(\aheap)$ and
$(\U',\astore_a,\aheap) \models \neg\alloc(x)$.

Consequently, if $M$ contains no literal of the form $\alloc(x)$ then
$(\U',\astore_a,\aheap) \models M$, hence $(\U',\astore_a,\aheap)
\models \Tnocont{\phi}$. Otherwise, let $\{ \alloc(x_1), \dots,
\alloc(x_p) \}$ be the set of literals $\alloc(x)$ occurring in $M$.
Since all occurrences of $\wand$ in $\varphi$ are negative or neutral,
by Lemma \ref{lemma:polarity} (point \ref{it:polarity3}), every
literal $\alloc(x_i)$ occurs within a subformula $\lambda^\fincard$ of
some formula $\elim^\fincard_\septraction(M_1,M_2)$, hence inside a
formula of the form $\alloc(x_i) \vee (\len{h} < \len{U} - q_i \wedge
\len{U} \geq r_i)$, for some $q_i,r_j \leq \Nmaxi$.  Thus
$\finmt{\phi}$ necessarily contains a minterm $M'$ identical to $M$,
except that each literal $\alloc(x_i)$ is replaced by $(\len{h} <
\len{U} - q_i \wedge \len{U} \geq r_i)$.  But, by definition, $\U'$
contains more than $\Nmaxi$ unallocated elements, hence necessarily
$(\U',\astore_a,\aheap) \models (\len{h} < \len{U} - q_i \wedge
\len{U} \geq r_j)$.  Therefore, $(\U',\astore_a,\aheap) \models M'
\models \Tnocont{\phi}$.

We have proved that $(\U',\astore_a, \aheap) \models \Tnocont{\phi}$.
Since $\vec{a}$ is arbitrary, we deduce that $(\U',\astore,\aheap)
\models \forall y_1,\dots,y_m~.~ \Tnocont{\phi} = \varphi'$.

Consequently, it is sufficient to test that $\varphi'$ has a finite
model.
%%%%%%%%%%%%%%%%%%%%%%%%%%%%%%%%%%%%%%%%%%%%%%%%%%%%%%%%%%%%%%%%%%%%%%%%%%%%%%%%
\ifLongVersion
%%%%%%%%%%%%%%%%%%%%%%%%%%%%%%%%%%%%%%%%%%%%%%%%%%%%%%%%%%%%%%%%%%%%%%%%%%%%%%%%
We show that the finite satisfiability problem for $\varphi'$ is
in \pspace. By Lemma \ref{lem:bsrfoltrans}, since by definition
$\Tnocont{\phi}$ contains no positive occurrence of $\alloc(x)$, the
formula $\foltrans{\varphi'} \wedge \axioms{\varphi'}$ is equivalent
to a formula in $\bsr(\fol)$ with the same free variables and
constants, and $\foltrans{\varphi'} \wedge \axioms{\varphi'}$ has a
finite model iff it has a model $(\U,\astore,\afunc)$, with $\card{\U}
= n+\bigO(\maxn{\Tnocont{\phi}})$, since the number of constants and
free variables in $\foltrans{\varphi'} \wedge \axioms{\varphi'}$ is
$n+\bigO(\maxn{\Tnocont{\phi}})$.  It is clear that
$\maxn{\Tnocont{\phi}} = \maxn{\bigvee_{M \in \finmt{\phi}} M}$, hence
by Corollary \ref{cor:sl-mnf-size},
$\maxn{\Tnocont{\phi}}=\bigO(\size{\varphi}^2)$, thus $\card{\U} =
\bigO(\size{\varphi}^2)$.
  
Then we can guess a $\fol$-structure $(\U,\astore,\afunc)$ such that
$\card{\U} = \bigO(\size{\varphi}^2)$ and check in polynomial space
that $(\U,\astore,\afunc) \models \foltrans{\varphi'}$ (this is done
as in Lemma \ref{cor:pspacemc}, except that the minterms containing a
test formula $\alloc(x)$ are discarded) and $(\U,\astore,\afunc)
\models \axioms{\varphi'}$ (as in the proof of Theorem
\ref{thm:pspaceBSR}).  By Lemma \ref{lemma:test-fo-sat}, the formula
$\foltrans{\varphi'} \wedge \axioms{\varphi'}$ has a finite model iff
$\varphi'$ has a finite model, i.e., iff $\varphi$ has an
\ncontrolled\ finite model (since we know at this point that $\varphi$
has no finite \controlled\ model).
%%%%%%%%%%%%%%%%%%%%%%%%%%%%%%%%%%%%%%%%%%%%%%%%%%%%%%%%%%%%%%%%%%%%%%%%%%%%%%%%
\else
%%%%%%%%%%%%%%%%%%%%%%%%%%%%%%%%%%%%%%%%%%%%%%%%%%%%%%%%%%%%%%%%%%%%%%%%%%%%%%%%
Since by construction $\varphi'$ contains no positive occurrences of
$\alloc(x)$, this can be done by following the same steps as in the
proof of Theorem \ref{thm:pspaceBSR}, except that the axiom
$\infaxiom{\varphi}$ is not added (Appendix \ref{sect:endofproof}).
%%%%%%%%%%%%%%%%%%%%%%%%%%%%%%%%%%%%%%%%%%%%%%%%%%%%%%%%%%%%%%%%%%%%%%%%%%%%%%%%
\fi
%%%%%%%%%%%%%%%%%%%%%%%%%%%%%%%%%%%%%%%%%%%%%%%%%%%%%%%%%%%%%%%%%%%%%%%%%%%%%%%%
\qed

Note that the $\pspace$-completeness results for $\finbsrsl{k}$ and
$\infbsrsl{k}$ allow us to re-establish the \pspace-completeness of
the satisfiability problem for quantifier-free formulae of
$\seplogk{k}$, both in finite and infinite domains.  Indeed, every
quantifier-free formula $\phi$ is sat-equivalent to a formula $\phi
\septraction \top$ that is both in $\finbsrsl{k}$ and $\infbsrsl{k}$,
since the antecedent of $\wand$ has neutral polarity.

\section{Conclusions and Future Work}

We have studied the decidability problem for the class of Separation
Logic formulae with quantifier prefix in the language
$\exists^*\forall^*$, denoted as $\bsr(\seplogk{k})$. Although the
fragment was found to be undecidable, we identified two non-trivial
subfragments for which the infinite and finite satisfiability
are \pspace-complete. These fragments are defined by restricting the
use of universally quantified variables within the scope of separating
implications that occur at positive polarity. Since, in practice,
solving most Separation Logic entailments that arise as verification
conditions in programs or inductive solvers do not involve considering
separating implications that contain universally quantified variables,
the decidable classes found in this work are of practical interest.

Future work involves using the techniques for proving decidability,
namely the translation of quantifier-free $\seplogk{k}$ formulae into
boolean combinations of test formulae, to solve other logical
problems, such as frame inference, abduction and possibly
interpolation.

\vspace*{\baselineskip}
\paragraph{Acknowledgements}
The authors wish to acknowledge the contributions of St\'e\-phane Demri
and \'Etienne Lozes to the insightful discussions during the early
stages of this work. \hfill$\blacksquare$

\bibliography{refs}
\bibliographystyle{abbrv}
%%%%%%%%%%%%%%%%%%%%%%%%%%%%%%%%%%%%%%%%%%%%%%%%%%%%%%%%%%%%%%%%%%%%%%%%%%%%%%%
\end{document}